\documentstyle[12pt,epsf]{article}
\newcommand{\sbz}{{\mbox{\scriptsize            \bf             {Z}}}}

\newcommand{\bunit}{\mbox{\bf{1}}}

\newcommand{\link}{\mbox{\begin{picture}(4.15,10)
\put(0,3){\circle*{2}}                     \put(0,2.75){\line(1,0){7}}
\put(7.3,3){\circle*{2}}\mbox{  }  \end{picture}  }  }

\newcommand{\site}{\mbox{\begin{picture}(4.15,4)
\put(2,2){\circle*{2}} \end{picture}  }  }
\newcommand{\on}{U_{sign\;U(\Box)}(\Box)}
\newcommand{\B}{U_{BIO}(\Box)}
\newcommand{\sitex}{\site^{\!\!\!x^{\mu}}}

\newcommand{\linkxy}{\;\link^{\!\!\!\!\!\!\!x^{\mu}\;\;y^{\mu}}}

\newcommand{\linkx}{\;\link^{\!\!\!\!\!\!\!x^{\mu}\;\;}}

\newcommand{\beq}{\begin{equation}}
\newcommand{\eeq}{\end{equation}}
\newcommand{\nin}{\noindent}
\newcommand{\mc}{\multicolumn}

\newcommand{\dof}{degrees of freedom}
\newcommand{\dofx}{degrees of freedom }
\newcommand{\pcp}{partially       confining       phase }
\newcommand{\pcps}{partially        confining         phases }

\newcommand{\bz}{\mbox{\bf{Z}}}
\newcommand{\br}{\mbox{\bf{R}}}
\newcommand{\bu}{\mbox{\bf{U}}}

\begin{document}
\bibliographystyle{unsrt}
\pagenumbering{roman}

\begin{flushright}
\begin{small}
NBI-HE-96-29 \\

31 May 1996
\end{small}
\end{flushright}

\begin{center}

{\Large \bf
Gauge Couplings calculated from

\vspace{.3cm}

Multiple Point Criticality yield  $\alpha^{-1}=136.8\pm 9$: \\

\vspace{.3cm}

At Last the Elusive Case of $U(1)$}

\vspace{50pt}

{\sl D.L. Bennett}

\vspace{6pt}

\begin{small}

The Niels Bohr Institute, Blegdamsvej 17 DK-2100 Copenhagen {\O} \\
{\em \tiny and } \\

The Quality-of-Life Research Centre,  St. Kongensgade 70, DK-1264 Copenhagen K
\\

\end{small}

\vspace{18pt}

{\sl H.B.  Nielsen}

\vspace{6pt}

\begin{small}

The  Niels  Bohr Institute,
Blegdamsvej 17, DK-2100 Copenhagen {\O}, Denmark \\

\end{small}

\end{center}

\tableofcontents
\newpage
\pagenumbering{arabic}
\setcounter{page}{1}

\begin{footnotesize}
\begin{center}
{\bf ABSTRACT}
\end{center}

We calculate the $U(1)$ continuum gauge coupling using the values of
action parameters coinciding with the multiple point. This is a point in the
phase diagram of a lattice gauge theory where a maximum number of phases
convene. We obtain for the running inverse finestructure constant the values
$\alpha_1^{-1}=56\pm 5$ and $\alpha_1^{-1}=99\pm 5$ at respectively the Planck
scale and the $M_Z$ scale.
The gauge group underlying the phase diagram in which we seek multiple point
parameters is what we call the Anti Grand Unified Theory (AGUT) gauge group
$SMG^3$ which is the
Cartesian product of 3 standard model groups (SMGs). There is one SMG
factor for
each of the $N_{gen}=3 $ generations of quarks and leptons. In our model, this
gauge group $SMG^3$ is the predecessor to the usual standard model group. The
latter
arises as the diagonal subgroup surviving the
Planck scale breakdown of $SMG^3$. This breakdown leads to a weakening of the
$U(1)$ coupling by a $N_{gen}$-related factor. For $N_{gen}=3$, this factor
would be
$N_{gen}(N_{gen}+1)/2=6$
if phase transitions between all the phases convening at the multiple point
were purely second order. The factor $N_{gen}(N_{gen}+1)/2=6$ corresponds
to the six
gauge invariant combinations of the $N_{gen}=3$ different $U(1)$s that
give action contributions that are second order in $F_{\mu\nu}$. The factor
analogous to this $N_{gen}(N_{gen}+1)/2=6$ in the case of the earlier
considered
non-Abelian
couplings reduced to the factor $N_{gen}=3$ because action terms quadratic
in $F_{\mu\nu}$ that arise as contributions from two different
of the $N_{gen}=3$ SMG factors of $SMG^3$ are forbidden by the requirement of
gauge symmetry.

Actually we seek the multiple point in the phase diagram of the gauge group
$U(1)^3$ as a simplifying approximation to the desired gauge group $SMG^3$.
The most important correction obtained from using
multiple point parameter values (in a multi-parameter phase diagram instead
of the single critical parameter value
obtained say in the 1-dimensional phase diagram of a Wilson action) comes from
the effect of
including the influence of also having at this point phases confined
solely w.r.t. discrete subgroups. In particular, what matters is that the
degree of first orderness is taken into account in making the transition
from these latter phases at the multiple point to the totally Coulomb-like
phase. This gives rise to a discontinuity $\Delta \gamma_{eff}$ in an effective
parameter $\gamma_{eff}$. Using our calculated value of the quantity
$\Delta \gamma_{eff}$, we calculate the above-mentioned weakening factor to
be more like 6.5 instead of the $N_{gen}(N_{gen}+1)/2=6$ as would be the
case if all
multiple point transitions were purely second order. Using this same
$\Delta \gamma_{eff}$, we also calculate the
continuum $U(1)$ coupling corresponding to the multiple point of a single
$U(1)$. The product of this latter and the weakening factor of about
6.5 yields our Planck scale prediction for the continuum $U(1)$ gauge coupling:
i.e., the multiple point
critical coupling of the diagonal subgroup of
$U(1)^3\in SMG^3$. Combining this with the results of earlier work
on the non-Abelian gauge
couplings leads to our prediction of
$\alpha^{-1}=136._8\pm9$ as the value for the fine-structure  constant at low
energies.

\end{footnotesize}

\section{Introduction}

Over a period of a number of  years  we
have put forth\cite{van,gosen,long,nonabel,lap1,ngen1} the observation that
the actual values of the standard model running coupling constants
$\alpha_i(\mu)$ $(i\in \{U(1), SU(2), SU(3) \})$
at
the  Planck  energy  scale  $\mu_{Pl}$  (i.e.,   experimental   values
extrapolated to the Planck scale using the assumption of a minimal standard
model) depart from critical  values  for  a
standard model group lattice gauge theory by a factor that is close to
three for the non-Abelian $SU(2)$ and $SU(3)$ couplings (strictly speaking,
this applies rather to the simple groups $SU(2)/\bz_2$ and $SU(3)/\bz_3$ having
the same Lie algebra as $SU(2)$ and $SU(3)$).

We  proposed  the  so-called  $AGUT$  gauge  group  at  the
``fundamental scale'' $\mu_{Pl}$ consisting of  the  3-fold  Cartesian
product of the standard model group (sometimes referred to by use of the
acronym ``$AGUT$'': {\bf A}nti {\bf G}rand {\bf U}nified {\bf T}heory):

\beq SMG\times SMG\times SMG\stackrel{def}{=}SMG^3.\eeq

\nin as way of ``explaining'' the phenomenologically indicated factors of
three. Here we are anticipating that these factors,
up to minor, controllable
corrections, really have the {\em integer} value three.

Actually, rather than taking  the  $SMG$  as  the  group
$U(1)\times SU(2)\times SU(3)$, we  use  instead  a  different
group\footnote{We define the standard model {\em group} $(SMG)$ as the
factor  group  obtained  from  the  $SMG$  covering  group  $\br\times
SU(2)\times SU(3)$ by identifying the elements of the centre belonging
to      the      discrete      subgroup      $\{(2\pi,-\bunit^{2\times
2},e^{i\frac{2\pi}{3}}\bunit^{3\times3})^n|n \in \bz\}$:

\beq
SMG   \stackrel{def}{=}   S(U(2)\times   U(3))\stackrel{def}{=}
(\br\times   SU(2)\times
SU(3))/\{(2\pi,\bunit^{2\times   2},e^{i\frac{2\pi}{3}}\bunit^{3\times
3})^n|n \in \bz\}  \eeq

\nin The defining representation of $S(U(2)\times U(3))$  is  the
set of $5\times 5$ matrices

\beq     S(U(2)\times      U(3))_{def}\stackrel{def}{=}\left\{\left.\left(
\begin{array}{cc} \bu_2 & \begin{array}{ccc} 0 & 0 & 0 \\ 0 &  0  &  0
\end{array} \\ \begin{array}{cc} 0 & 0 \\ 0 & 0 \\ 0 & 0 \end{array} &
\bu_3 \end{array} \right ) \right| \begin{array}{l} \bu_2\in U(2),  \\
\bu_3\in U(3), \\ \mbox{det}\bu_2\cdot  \mbox{det}\bu_3=1  \end{array}
\right \} \eeq

\nin This representation is  suggested  by  the
spectrum of representations in the standard model.}
denoted as $S(U(2)\times U(3))$. However,
both groups have the same Lie  algebra.  From
the beginning we have assumed that the  number  of  Cartesian  product
factors of the $SMG$ in $SMG^3$ is equal to the number  of  quark  and
lepton generations $N_{gen}$ so that each generation has its  own
set of gauge degrees of freedom. In our early\cite{ngen1,ngen2,kk3} work in
this
series, we used a fitting procedure that lead us to conclude that  the
number of generations $N_{gen}$ was  3  prior  to  the  subsequent
experimental confirmation of $N_{gen}=3$ in $LEP$ measurements. We might
therefore claim to have predicted the number of generations  as  being
three at a time when the  number  $N_{gen}$  was  essentially  unknown
except from notoriously unreliable cosmological fits.

In our more recent work\cite{gosen,van,long,nonabel} we venture what  can
be said to be a more eloquent formulation in which fitting  parameters
are  avoided  and  a  number  of  assumptions  made   previously   are
essentially reduced to the postulate that the running gauge  couplings
at the Planck scale assume {\em multiple point} critical  values.
The multiple point is  defined  as
the point in plaquette action parameter  space  where  the  largest
possible number of ``phases''  in  a  lattice
gauge theory come together. This requires that the functional form  of
the plaquette action - as defined by  the  parameters  that  span  the
space in which we seek the multiple point - is sufficiently general to
``provoke'' all the ``phases'' that we  seek  to  bring  together  at  the
multiple point.

Although we suspect that it may be possible to avoid the assumption of
a fundamental,  truly  existing  lattice
\footnote{The  possibility  of
avoiding the assumption of a fundamental lattice lies in the  speculation
that the all important multiple point critical coupling in  our  model
may, in a purely continuum theory, play  the  role  of  the  strongest
coupling for which a pure continuum field is meaningful in the  sense  of
having a well defined $A^{\mu}$ field. Recall that
at the multiple point, there will be a phase boundary where there is a
jump from a finite value of $\langle U(\link) \rangle$  to  the  value
$\langle U(\link) \rangle=0$ for each degree of  freedom.  Going  into
the phase for which $\langle U(\link) \rangle=0$  corresponds  to
going  into  confinement.  Thinking  of  the  lattice  theory  in  the
traditional way as a regularization of a  continuum  theory,  we  know
that a lattice gauge theory that is  in  confinement  already  at  the
lattice scale cannot meaningfully be brought into correspondence  with
a continuum field theory because of the impossibility  of  defining  parallel
transport (i.e., $A^{\mu}$) when $\langle U(\link)  \rangle=0$.

In a more general context, we suspect that the consistency of any quantum
field theory requires the assumption of a fundamental regulator in some form.
A lattice is just one way to implement this assumption. The hope would be that
any implementation of the fundamental regulator of Nature would demonstrate
critical behaviour corresponding to the same values of couplings at
what corresponds to the multiple point.}
,  this
assumption is the most straight forward approach to our model.

To get an idea of what is meant by the ``phases'' that convene at  the
multiple point, recall first that the gauge field  $U(\link)$  assigns
an element of the gauge group $G$ to each link  of  the  lattice:  $U:
\{\linkxy \} \rightarrow G$. For  complicated  gauge  groups
such as  $SMG$  and  $SMG^3$  having  many  subgroups  (and  invariant
subgroups), it will be seen  that  \dofx corresponding to different subgroups
can have qualitatively different
   fluctuation patterns. In other words,
the region of action parameter  space
corresponding to some ``phase'' corresponds, in general, to  different
fluctuation patterns along different subgroups. The various phases
can be classified using the different subgroups of the gauge group. We
shall see that a ``phase'' is labelled  by  a  subgroup  $K  \subseteq
SMG^{N_{gen}}$ and an invariant subgroup  $H  \lhd  K$.  These  labels
$(K,H)$ characterise a possible qualitative physical behaviour  of  the
vacuum that could be perceived using a small band of wavelengths at  a
given energy scale (the Planck scale in this  case).  More  precisely,
the elaboration of the  ``phases''  $(K,H)$  is  a  classification  of
qualitatively different physical behaviours of the vacuum of a  lattice
gauge theory at the lattice scale\cite{frogniel} according to  whether
or not there is spontaneous breakdown of the gauge symmetry  remaining
after  making  a  choice  of  gauge  that  we  here  take  to  be  the
(latticized)  Lorentz  gauge   condition   (i.e., for all sites
$\sitex$,   $\prod_{\;\;\linkx\;\;\; \mbox{\tiny at  } \sitex}
U\mbox{(  $\linkx$  )}=1$
where $\;\;\linkx\;\;\; \mbox{\tiny at } \sitex$ denotes a link that
emanate from $\sitex$~). This choice of gauge still allows the  freedom  to
perform  gauge  transformations  of  the  types  $\Lambda_{Const.}(x)=
e^{i\alpha^at^a}$              and               $\Lambda_{Linear}(x)=
e^{i\alpha^at^aa_{\mu}x^{\mu}}$ respectively having constant and linear
gauge functions ($a$ is a ``colour'' index labelling
components $\alpha^a$ and generators $t^a$ of the Lie algebra).
Such gauge transformations are used in  defining  the
``phases'' $(K,H)$ of the vacuum.  Here  we  use  the  idea  of different
degrees of spontaneous symmetry breaking as a way to classify phases.
This sort of classification  can depend on the scale
at which the classification is made. To illustrate this, one can think of
two different regions of action parameter space with which are associated
different {\em finite} correlation lengths. If the physics of these two
regions is probed at a scale intermediate to these two correlation lengths,
it would appear as though one region were Coulomb-like and the other
confining and therefore separated by a phase transition. However, an
examination of these regions at a length scale short compared to
both correlation lengths would not detect a phase transition because both
regions would appear to be in a Coulomb-like phase. The phase
transition would also go undetected if the physics of the two regions were
probed at a length scale large compared to both correlation lengths because
both regions would appear to be confining. This situation is well known
(e.g., in non-Abelian groups):
a transition between Coulomb-like and confining lattice scale phases
 - ``lattice artifacts'' - often ``disappear'' in going to long wavelengths
because phase diagram regions on both sides of the lattice scale
transition are perceived as being in confinement at long distances.

Although we classify these phases with reference to a certain
scale of wavelengths, there are nevertheless typically true first order
phase transitions that are sharply defined in a specific regularization.
In our model, we take the view that the regulator is not arbitrary but rather
an ontological attribute of fundamental scale physics. From this point of
view, ``lattice artifact'' phases assume a role that has physical
implications.

In Section~\ref{modell} we state the principle of multiple point criticality
first in terms of the ``lattice artifact'' phases of a lattice gauge theory
and then, in a more general context, as the prototype of a fine-tuning
mechanism that results from arguing that even randomly
fixed values of some extensive quantities of some sort or another
will with a finite probability
enforce the coexistence of two or more phases if the transition between
these phases are first order. The presence of two or more
phases separated by first order transitions fine-tunes ``coupling
constants'' in a manner  reminiscent of the way that temperature  and
pressure are fine-tuned at the triple point of water. Section~\ref{interest}
considers problems encountered in implementing the principle of multiple point
criticality in the case of $U(1)$ as compared to the simpler case of the
non-Abelian subgroups of the standard model. These problems, related to the
``Abelian-ness'' of $U(1)$ include the problem of charge normalisation and
also the interactions between the $N_{gen}=3$ replicas of $U(1)$ in the AGUT
gauge
group $SMG^3$. In the roughest approximation, these interactions result
in a weakening of the diagonal
subgroup coupling of $U(1)^3\in SMG^3$ by a factor of $N_{gen}(N_{gen}+1)/2=6$
instead of the weakening factor $N_{gen}=3$ that applies to  the non-Abelian
subgroups (for which such
interactions are not gauge invariant). Section~\ref{phasediagram} deals with
approximate methods for constructing phase diagrams for the gauge group
$U(1)^3$
in which we can seek out multiple point parameter values. After a brief
discussion
of how phases at the scale of a lattice regulator are classified, we develop
a formalism that allows us to seek multiple point parameter values by
adjusting the metric (which amounts to adjusting the parameters of
a Manton action) in a $N_{gen}$-dimensional space upon which
is superimposed an hexagonally symmetric lattice of points identified
with the identity
of $U(1)^3$. The hexagonal symmetry takes into account the allowed
interactions between the $N_{gen}=3$ $U(1)$ factors of $U(1)^3$. Using this
formalism, two approximative methods of determining
phase boundaries are developed: the independent monopole and the group
volume approximations.
These describe respectively phase transitions that are purely
second order and strongly first order. Calculations are done in
Section~\ref{calculation} where we interpolate between the
extreme situations described by the group volume and independent monopole
approximations. This interpolation is done by calculating the discontinuity
$\Delta \gamma_{eff}$ in an effective coupling $\gamma_{eff}$ at the multiple
point. The dominant contributions to $\Delta \gamma_{eff}$ are due to
multiple point transitions between phases that differ by the confinement
of discrete subgroups (rather than continuous subgroups). The
calculated $\Delta \gamma_{eff}$ reflects the degree of first-orderness
of these transitions. As a result of including this effect
 the weakening factor $N_{gen}(N_{gen}+1)/2=6$
increases to about 6.5. The quantity $\Delta \gamma_{eff}$ is also used
(together with $\gamma_{eff}$) to calculate the continuum $U(1)$ coupling
corresponding to the multiple point of a single $U(1)$ which is then
divided by the square root of the
weakening factor of about 6.5 to get our prediction for
the value of the running $U(1)$ coupling at the Planck scale. We also give
values of the $U(1)$ coupling at scale of $M_Z$ obtained using the
assumption of the minimal standard model in doing the renormalization group
extrapolation. We present a number of values reflecting various
approximations.
In presenting what we take to be the ``most correct'' result, we compute the
uncertainty from the deviations arising from  plausibly correct
ways of making distinctions in how different discrete subgroups
enter into the calculation of $\Delta \gamma_{eff}$. The paper ends with
concluding remarks in Section~\ref{conclusion}.

\section{The Model}\label{modell}

\subsection{Multiple Point Criticality}

The $SMG$ and, to an even greater extent, the $AGUT$ gauge
group  $SMG^3$
are non-simple groups with many subgroups and invariant  subgroups. As
already mentioned  above,  there  can  be  a  distinct
``phase'' for each pair of subgroups $(K,H)$ such that $H\triangleleft
K \subseteq SMG^3$ (the symbol``$\triangleleft$~'' means  ``invariant
subgroup  of'').  Such  ``phases'',  which are often referred to as lattice
artifacts, characterise qualitative physical  behaviours  that  can  be
distinguished at least at the scale of the lattice. The principle  of  multiple
point criticality states that Nature seeks out a point  (the  multiple
point) in the phase diagram of a  lattice  gauge  theory  (with  gauge
group $SMG^3$) where a maximum  number  of ``phases''  come
together.

\subsubsection{Multiple point criticality: a very  general  model  for
fine-tuning}

While the validity of the  multiple  point  criticality  principle  is,
in the context of our model,
suggested alone on phenomenological  grounds,  we  suspect
that in a more general context,  this  principle
                           is the consequence of having fixed  amounts
of some - presumably a multitude of - extensive quantities in spacetime .
Universally fixed amounts of these extensive  quantities  could  quite
plausibly impose constraints that can only be fulfilled by having  the
coexistence of several ``phases''.  The  idea  is  that  having  fixed
amounts of certain extensive quantities can enforce the coexistence of
several phases and in  so  doing place  constraints  on
intensive  parameters  (e.g.,  gauge   couplings,   the   cosmological
constant).  This  idea  provides  a  possible  explanation  for  the
``fine-tuned'' parameters found in Nature in a manner suggested by  the
following analogous situation. Think
of an equilibrium system enclosed by  a container  within
which there is water in all three phases: solid, liquid, and  ice.  If
the container is  rigid  and  also  impenetrable  to  heat  and  water
molecules,  we  have  accordingly  fixed  amounts  of  the   extensive
quantities: energy, mole number of water, and volume. There is a whole
range of values at which the average energy and average volume
per molecule can be fixed
such that the system is forced to maintain the presence of all  three  phases.
The
enforced coexistence of all three  phases  in  the  presence  of rapid
changes in energy (i.e., non-vanishing heats  of   fusion,   sublimation   or
vaporisation) as a function of the intensive parameters temperature and
pressure in going from one
phase  to  another  ``fine-tunes''  the
values of these parameters to those of
the triple point of water.
This mechanism provides a mechanism for
fine-tuning that is effective if the transition is first
order so that there is a ``hop'' in energy
(and  entropy) as a function of e.g. temperature.

Having (globally) fixed amounts of extensive quantities corresponding to
4-dimensional path integrals implies, strictly
speaking, that
locality is broken. This could,  for  example,  come
about in statistical mechanics where the use of a microcanonical
ensemble can, taken in a stringent  sense,  imply  correlations  over
long distances. This last remark is to be understood in the following
way: having knowledge of the fixed amounts of the extensive quantities
and thereby the amounts of the various phases means that if we at some
place find a given phase, we can immediately conclude that the probability
for finding more of this phase elsewhere - even at far-removed places -
is smaller.

However  it  has  been  shown  \cite{frogniel}  that  the  non-locality
introduced in this manner is harmless insofar as it does not  lead  to
experimentally observable violations of locality. This  mild  form  of
non-locality is manifested in spacetime  in  a  completely  homogeneous
way and it is tolerable physically precisely because its  omni-presence
in spacetime allows it to be  incorporated  into  universal  intensive
constants of Nature (coupling constants).

The idea\cite{glas94,nl1,nl2} of fixed extensive quantities is suggested by
examining  the
limitations on the form that a nonlocal  action  can  have  if  it  is
required to be diffeomorphism invariant (as  a  natural  extension  of
translational invariance in general relativity). A simple but, for the
purposes of long wavelength limits, very general form for a  nonlocal
Lagrangian    action\cite{frogniel}    is    a    general     function
$S_{nl}(I_1,I_2,\cdots  )$  of  all  the   (diffeomorphism   invariant)
spacetime integrals

\beq    I_{j}     \stackrel{def}{=}     \int     \sqrt{g(x)}     {\cal
L}_j(\phi(x),\partial_{\mu} \phi(x))d^4x \eeq

\nin where $g=|\det g_{\mu\nu}|$ and  the  index  $j$  enumerates  the
possible local field functions $ {\cal L}_j$ with symmetry  properties
allowing them as Lagrange density terms.

With this functional form of the action, it  follows  that  there  are
fixed amounts (i.e., given alone by the form of the  action  $S_{nl}$)
of extensive quantities $I_j$ equal to the values of  these  spacetime
integrals $I_j$ which yield the extremum of $S_{nl}$
(classical approximation). The action terms
$S_{nl}$ consist typically of nonlinear (and thereby nonlocal) functions
of the reparameterization invariant integrals $I_j$. Such action terms
$S_{nl}(\{I_j\})$ are then also reparameterization invariant and
the functional derivatives of $S_{nl}$ w.r.t. the fields become essentially
local in the sense that these derivatives have the same value as seen from
all spacetime points. Non-locality
that comes about in this phenomenologically un-offensive way is manifested
as the omnipresent values of constants of Nature.
It is interesting that the potential paradoxes inherent to a theory with
non-locality
(e.g., of the
``matricide'' type naively encountered in ``time machines'')
are averted by a unique compromise
that exists generically with a finite probability. We can show\cite{bomb} that
this
unique solution coincides with multiple point values of intensive
quantities such as fine structure constants and the cosmological constant.

\subsection{AGUT with the gauge group $SMG^3$ and its  breakdown
to the $SMG$}

In the context of our model, the experimental values
for the fine-structure  constants
for the weak $SU(2)$ and the $QCD$ $SU(3)$ (after extrapolation to the
Planck scale) are, to within the bounds of our uncertainty  in  making
continuum corrections and other uncertainties, three times weaker
than the values corresponding to the triple point in an $SU(N)$ lattice
gauge     theory     using     results      from      Monte      Carlo
calculations\cite{bachas,bhanot1,bhanot2,drouffe}. However, to claim that the
weakening factor has the {\em integer} value three would be somewhat
presumptuous at this stage because of our lack of complete
understanding of possible singularities (at the multiple point)
in the continuum coupling as a function of bare parameters. Our concern comes
about because such singularities are observed in numerical
simulations\cite{jersak}
of the $U(1)$ continuum coupling. There is the
hope that the effect of analogous singularities on the values of the
non-Abelian couplings would be mitigated by a factor $1/(N^2-1)$ for
$SU(N)$ groups. This is suggested if it is assumed that monopoles are
responsible for the singularity.
If however, we proceed using the assumption of an integer 3 relationship,
we propose as an explanation that the gauge
subgroups $SU(2)$, $SU(3)$, and $U(1)$ are embedded  as  the  diagonal
subgroup   in   the   $AGUT$   gauge   group   $SMG^3$.   The
$AGUT$ gauge group $SMG^3$ is the Cartesian product  of
$N_{gen}=3$ group factors $SMG$:

\beq      SMG^3\stackrel{def}{=}(U(1)_{Peter}\times      U(1)_{Paul}\times
U(1)_{Maria})\times \eeq

$$  \times    (SU(2)_{Peter}\times      SU(2)_{Paul}\times
SU(2)_{Maria})\times (SU(3)_{Peter}\times      SU(3)_{Paul}\times
SU(3)_{Maria})  $$

\nin where the labels ``$Peter"$, ``$Paul"$, $etc.$  distinguish  the
$N_{gen}$ different isomorphic Cartesian product group  $SMG$  factors.
                                     The  quantity  $N_{gen}$ denotes the
number of quark and lepton generations and is taken as  3  in  accord
with experimental results.

Before dealing with the gauge subgroup  of  primary  interest  in  the
present  article  -  namely  $U(1)^3\subseteq  SMG^3$  -  it  will  be
instructive to first consider the subgroup $SU(N)^3 \subseteq SMG^3 $ $
(N=2,3)$. In our proposal, the $SU(2)$ and $SU(3)$  subgroups  of  the
standard model  group  are  realized  as  the  diagonal  subgroups  of
$SU(2)^{N_{gen}}$ and $SU(3)^{N_{gen}}$ respectively.  The  breakdown
of the non-Abelian Cartesian product subgroups of  the  $AGUT$
gauge group $SMG^3$  to  the  diagonal  subgroups  $SU(2)_{diag}$  and
$SU(3)_{diag}$ can come about by different mechanisms. One breakdown
mechanism, referred to as ``confusion'' \cite{conf1,conf2,frogniel,gaeta},
comes about due to ambiguities that arise under group automorphic symmetry
operations. Such ambiguities are also present in gauge groups with duplicated
factors in Cartesian product groups such as our $AGUT$ with gauge group
$SMG^{N_{gen}}$. Having the confusion mechanism of breakdown provides
therefore a
natural explanation for having a diagonal pattern of symmetry breakdown;
i.e., gauge group couplings corresponding to
groups embedded diagonally in groups with repeated Cartesian product
factors.

The diagonal  subgroup  by
definition consists of the  elements  $(U_{Peter},U_{Paul},  \cdots  ,
U_{N_{gen}})$ of $SU(N)^{N_{gen}}$ having identical excitations of the
``$Peter$'',   ``$Paul$'',   $etc.$   Cartesian    product    factors:
$U_{Peter}=U_{Paul}=\cdots = U_{N_{gen}}=U_{common}$. That is,

\beq SU(N)\stackrel{-}{\simeq} SU(N)_{diag}^{N_{gen}}\stackrel{def}{=} \eeq

\[ \{(U_{common},U_{common},\cdots ,U_{common}) |U_{common}\in SU(N)\}
\subseteq \]

\[ \subseteq
\underbrace{SU(N)\times SU(N)\times \cdots \times SU(N)}_{\tiny N_{gen}
\mbox{  factors}}\;\;\;(N=2,3). \]

When the gauge group is realized in this way, it is readily shown that
the inverse fine-structure constant for $SU(N)_{diag}^{N_{gen}}$
is  additive  in
those  of  the  $N_{gen}$   Cartesian   product   group   factors   of
$SU(N)^{N_{gen}}$ (at least to lowest order perturbatively in weak coupling):

\beq
\frac{1}{\alpha_{diag}(\mu_{Planck})}=\frac{1}{\alpha_{Peter}(\mu_{Planck})}+
\frac{1}{\alpha_{Paul}(\mu_{Planck})}+ \cdots + \frac{1}{\alpha_{N_{gen}}
(\mu_{Planck})} \eeq

This follows because the diagonal subgroup of $SU(N)^3$ corresponds by
definition to identical excitations of the $N_{gen}$ isomorphic  gauge
fields  (with  the  gauge  couplings   absorbed)\footnote{As   it   is
$gA_{\mu}$ rather than $A_{\mu}$ that appears in  the  (group  valued)
link  variables  $U(\link)\propto  e^{iagA_{\mu}}$,  it  is  the   quantities
$(gA_{\mu})_{Peter}$, $(gA_{\mu})_{Paul}$, etc. which are equal in the
diagonal subgroup.}:

\begin{equation} (g{\cal A}_{\mu}(x))_{Peter} =(g{\cal A}_{\mu}(x))_{Paul} =
\cdots  =   (g{\cal   A}_{\mu}(x))_{N_{gen.}}   \stackrel{def.}{=}(   g{\cal
A}_{\mu}(x))_{diag.}.\label{diagconfig} \end{equation}

\nin This has the consequence that the common $(gF_{\mu \nu})^2_{diag}$
in each term of the Lagrangian density
for $SU(N)^{N_{gen.}}$ can be factored out:

\begin{equation}  {  \cal
L}=-1/(4g_{Peter}^2)(gF_{\mu                      \nu}^a(x))^2_{Peter}
-1/(4g_{Paul}^2)(gF_{\mu      \nu}^a(x))^2_{Paul}       -       \cdots
\end{equation}

\[ \cdots -1/(4g_{N_{gen.}}^2)(gF_{\mu \nu}^a(x))^2_{N_{gen.}} \]

\[ =(-1/(4g_{Peter}^2)  -1/(4g_{Paul}^2)-\cdots
-1/(4g_{N_{gen.}}^2)) \cdot (F_{\mu \nu}^a(x))^2_{diag}  =\]
\[ =-1/(4g_{diag}^2)
\cdot (gF_{\mu\nu}^a(x))^2_{diag}. \]

The inverse squared couplings for the
diagonal subgroup is indeed  just  the  sum  of  the  inverse  squared
couplings for each of  the  $N_{gen}$  isomorphic  Cartesian  product
factors of $SU(N)^3$. Additivity in the inverse squared  couplings  in
going to the diagonal subgroup applies  separately  for  each  of  the
invariant   Lie   subgroup types\footnote{Strictly   speaking    this is   an
approximation; it will be  further  elaborated  upon  later.}  $i  \in
\{SU(2), SU(3)\}\subset SMG$. So for the non-Abelian couplings we  have
that

\begin{equation}
\frac{1}{g_{i,diag}^2}=\frac{1}{g_{i,Peter}^2}+\frac{1}{g_{i,Paul}^2}+\cdots
+\frac{1}{g_{i,N_{gen}}^2}     \;\;(i\in     \{SU(2),\;SU(3)\})      .
\label{confadd} \end{equation}

Assuming that the inverse squared
couplings for a given $i$ but different labels  $\{
Peter,$ $ Paul, \cdots,N_{gen.} \}$ are all driven to the  multiple
point in accord with the principle of multiple point criticality,
these $\{ Peter, Paul, \cdots,N_{gen.} \}$ couplings
all become equal to the multiple point value $g_{i,multi.\; point}$; i.e.,:

\begin{equation}
\frac{1}{g_{i,Peter}^2}=\frac{1}{g_{i,Paul}^2}=\cdots
=\frac{1}{g_{i,N_{gen}}^2}=\frac{1}{g_{i,\;multi.\; point }^2}.
\end{equation}

We see that the inverse squared coupling $1/g_{i, \;  diag}^2$  for  the
$i$th subgroup of the diagonal subgroup, i.e.,

\beq i\in
\{SU(2)\overline{\simeq}SU(2)^{N_{gen}}_{diag},
SU(3)\overline{\simeq}SU(3)^{N_{gen}}_{diag}\}, \eeq
is enhanced  by  a
factor $N_{gen}$ relative to the corresponding subgroup type $i$ of each of
the $N_{gen}$ individual Cartesian product  factors  $Peter$,  $Paul,
\cdots, N_{gen.}$ of $SMG^{N_{gen.}}$:

\begin{equation}
\frac{1}{g_{i,diag}^2}=\frac{N_{gen}}{g_{i,\;multi.\; point }^2}.
\end{equation}

In the context of our model,  $N_{gen}=3$ yields values for
$\frac{1}{g_{i,diag}^2}$ that agree (within anticipated uncertainties)  with
the  experimental   values   of   the   non-Abelian   couplings   after
extrapolation to the Planck scale using the assumption of a ``desert''.

However, we shall see that the simple additivity rule  that  works  so
well  for  the  non-Abelian  couplings  yields  poor   agreement   with
experiment for the $U(1)$ couplings. The explanation for this, in the context
of our model, is in part that, for $U(1)$,  there  can  be  ``mixed''
action terms of the type $F^{Peter}_{\mu\nu}F^{\mu\nu\;Paul}$ even  in
the continuum Lagrangian as opposed to the case  for  the  non-Abelian
degrees     of     freedom     where     only     quadratic      terms
$F^{Peter}_{\mu\nu}F^{\mu\nu\;Peter}$ with the same $Peter$, $Paul$  or
$Maria$ label are gauge invariant.

\section{Gauge Group of Interest: $U(1)^3$}\label{interest}

The gauge group to which we ultimately  want  to  apply  the {\bf M}multiple
{\bf P}oint {\bf C}riticality {\bf P}rinciple (MPCP) is the
{\bf A}nti {\bf G}rand {\bf U}nified {\bf T}heory (AGUT) gauge
group $SMG^3$ or some
group in which the latter is embedded in such a way that $SMG^3$ dominates
as the group to be considered. However
for the purpose of finding the multiple point
$U(1)$ coupling, it can be argued that we can approximately ignore the
interaction between  the
Abelian and non-Abelian subgroups  provided  we  identify  the  $U(1)_i$
factors in  $U(1)^3$  with  the  factor  groups  $SMG_i/(SU(2)  \times
SU(3))_i$ $ (i\in\{Peter, Paul, Maria\})$. In  this  approximation, we
essentially treat $SU(3)^3$, $SU(2)^3$ and $U(1)^3$ separately.
We shall now address the $U(1)$ \dofx by
endeavouring the construction of  some rather rough approximations
to  the  phase diagram for a lattice gauge theory with the gauge group
$U(1)^3$.  In
order to provoke the many possible phases $(K,H)$,
including in principle  the
denumerable infinity of  ``phases'' involving  the  discrete
subgroups of $U(1)^3$, it is necessary to use a  functional  form  for
the plaquette action that is quite general.

\subsection{Special problems with $U(1)^3$}

In the case of the non-Abelian subgroups of the $SMG$
that  we  have  dealt  with  in  earlier
work\cite{nonabel,albu}, the correction
factor in going from the multiple point couplings of $SMG^3$ to the
diagonal subgroup couplings is 3 corresponding to the value of the
number of generations $N_{gen}$. Recall that the diagonal subgroup couplings
are in our model predicted to coincide
with the experimental $U(1)$ coupling after extrapolation to the Planck scale.

However, the relation of the diagonal subgroup couplings to the multiple
point critical  couplings in the case of $U(1)^3$
turns out to be more complicated than for the
non-Abelian $SMG$  couplings.
The resolution of these  complications  helps  us  to
understand the  phenomenological  disagreement  found  when  a  naively
expected correction factor of $N_{gen}=3$ is used in going from the $U(1)$
couplings at  the multiple point  of $U(1)^3$ to the couplings
for the diagonal subgroup of $U(1)^3$.

For the fine-structure  constants  of  the  non-Abelian
groups $SU(2)$ and $SU(3)$, it  was  found  that  experimental  values
extrapolated to the Planck scale agree to within the uncertainties  of
our calculation with the predicted values
$1/\alpha_{diag\;multicr}=3/\alpha_{multicr.}$  (i.e.
the inverse fine-structure constants for the diagonal  subgroups  of the
non-Abelian subgroups of
$SMG^3$). While the factor 3 correction to the  multiple point  inverse
squared coupling values obtained for a  lattice  gauge  theory  yields
rather noteworthy agreement with the experimental values of  non-Abelian
fine-structure constants, the analogous relation does not  hold for
the $U(1)$ gauge algebra (weak hyper-charge). For $U(1)$  a  correction
factor  of  roughly  6 (or 7) is   indicated
phenomenologically. This would naively  suggest  that  at  the  Planck
scale we should postulate something like

\beq U(1)^{6\;or\;7}=\overbrace{U(1)\times \cdots \times
U(1)}^{6\;or\;7\;factors} \eeq

\nin rather than $U(1)^3$ as suggested by our preferred ``fundamental''
gauge group $SMG^{N_{gen}}$ with $N_{gen}=3$.

An explanation for this disparity when
we use $U(1)^3$ as the gauge group (rather than the naively indicated
$U(1)^6$ or $U(1)^7$)
can be sought by considering how
the ``Abelian-ness'' of $U(1)$ distinguishes it from the non-Abelian subgroups.

\subsubsection{The normalisation problem for $U(1)$ \label{onlyz6}}

For $U(1)$, there is no natural unit of charge in contrast to the
non-Abelian groups $SU(2)$ and $SU(3)$. For these latter,  there is a
way to normalise the fine-structure constants by  means  of  the  commutators.
The commutation algebra  provides  a  means of unambiguously  fixing  a
convention for the gauge couplings that alone pertains to the Yang-Mills
fields without reference to the charge of, for example, a  matter  field;  the
Yang-Mills fields are themselves charged in  the  non-Abelian
case and can therefore be used to define a charge convention.
Essentially this is because the Lie algebra commutator relations
are non-linear and are therefore not invariant under re-scalings of the
gauge potential $g{\bf A}_{\mu}$. Such scalings,  if  not
forbidden, would  of  course  deprive  gauge  couplings  of  physical
significance.

Because such a rescaling is possible in the case of $U(1)$, the weak
hyper-charge fine-structure constant is only normalizable  by  reference
to some quantum of charge. This immediately raises the question of  which
particle should  be  declared  as  having  the  unit  quantum of charge
as its hyper-charge. An equivalent way to address this question is to ask
which $U(1)$-isomorphic factor group of $SMG$ should be identified with the
$U(1)$ on the lattice to give us the critical coupling.

It is only when - on the lattice - the group of real numbers $\br$
(in the covering group $\br\times SU(2)\times SU(3)$ of the $SMG$) is
compactified  to  a  $U(1)$
that a normalisation becomes possible and thereby that the idea  of  a
critical  coupling  acquires  a  meaning.  The  only  remnant  in  the
continuum of having chosen a specific group on the lattice is  the
quantisation rule of the charges (more generally, a constraint on  the
allowed representations) and the  lattice  artifact
monopoles. This suggests that we should take the length of the  $U(1)$
in such a way as to enforce empirical charge quantisation rules.  When
we state that the critical coupling for a $U(1)$ lattice gauge  theory
is given by

\beq  \alpha_{crit}\propto\frac{1}{4\pi\beta_{crit}}=\frac{1}{4\pi\cdot 1.01},
\eeq

\nin the meaning is that this $\alpha_{crit}$ is the fine-structure constant
at the phase transition {\em corresponding to the coupling to the smallest
charge quantum allowed on the lattice}. For the $SMG$ as we define it:

\beq
SMG   \stackrel{def}{=}   S(U(2)\times   U(3))\stackrel{def}{=}
(\br\times   SU(2)\times
SU(3))/\{(2\pi,\bunit^{2\times   2},e^{i\frac{2\pi}{3}}\bunit^{3\times
3})^n|n \in \bz\},  \eeq

\nin the charge quantisation rule
for weak hyper-charge is very sophisticated\cite{skew1,skew2}:

\beq y/2+d/2+t/3=0\;\; (\mbox{mod }1). \label{rule}\eeq
This means that depending on whether the non-Abelian subgroups are
represented trivially or non-trivially, the smallest allowed quantum for the
weak hyper-charge is respectively $y/2=1$ and $y/2=1/6$. This complicated
quantisation rule can be regarded as a consequence of Nature having
chosen the gauge {\em group}\cite{michel,oraifear}
$S(U(2)\times U(3))$. In spite of the fact that the global structure of this
group imposes the severe restriction (\ref{rule}) on the possible
representations, it still allows all representations that are seen
phenomenologically.

The $U(1)$ centre of $SMG$
is embedded in the latter in a complicated way.
In order to determine the non-Abelian coupling of the $SMG$,
one must relate
the $U(1)$ centre of the $SMG$ and the simple $U(1)$ studied using Monte Carlo
methods on a lattice. Our earlier work
suggests that the disconnected $\bz_2$ and $\bz_3$ centres of respectively
the non-Abelian $SMG$ subgroups $SU(2)$ and $SU(3)$ should both
alone be confined in phases that convene at the multiple point. In
order to respect this requirement in the present work, it is necessary to
require that the class of $\bz_N$ discrete subgroups $\bz_N$
for which there can be phases convening
at the multiple point that are
solely confined along $\bz_N$ must be as follows: when $\bz_K$ is in this
class, then so are the groups $\bz_K+\bz_2=\bz_{K^{\prime}}$
(where $K^{\prime}$ is the smallest integer multi-plum of $K$ that is
divisible by 2) and the groups $\bz_K + \bz_3=\bz_{K^{\prime\prime}}$ (where
$K^{\prime\prime}$ is the smallest integer multi-plum of $K$ divisible by 3).
Hence, for the phases that convene at the multiple point, the greatest $N$
of a phase that is solely confined w.r.t a  subgroup $\bz_N$
must be such that $N$ is divisible by 2 and 3 and thus
also by 6.

A rule\footnote{In calculating the continuum coupling for a
continuous
Lie (sub)group, the effect on this continuum coupling due to having
discrete subgroups that convene at the multiple point can be
taken into account by calculating as if these discrete subgroups were
{\em totally} confined (instead of being critical as is
the case at the multiple point).}
from our earlier work\cite{long} states that the coupling for a continuous Lie
(sub)group $L$ at the multiple
point is given - to a good approximation - by the critical coupling
for a the factor group $L/\bz_{N_{max}}$ anywhere along the phase border where
the Coulomb-like \dofx corresponding to this factor group are critical. Here
$\bz_{n_{max}}$ denotes the largest discrete subgroup that alone confines in
a phase that convenes at the multiple point. We shall refer to this rule as
the $\bz_{N_{max}}$ factor group rule.

We shall argue below that the largest  discrete
subgroup of the $U(1)$ centre of $SMG$ that is solely confined in a phase
that convenes at the multiple point does not result in a  $U(1)$-isomorphic
factor group of length shorter than that corresponding to the
identification
of $SU(2)\times SU(3)$ with the identity. This corresponds to dividing  the
largest possible non-Abelian subgroup out of the $SMG$; the result
is a factor group isomorphic with $U(1)/\bz_6$:

\beq U(1)/\bz_6\stackrel{-}{\simeq}SMG/(SU(2)\times SU(3)).\label{facmax}\eeq

Consequently, we shall also argue that the
$U(1)$
critical coupling $\sqrt{4\pi\alpha_{crit}}$ obtained using Monte Carlo
simulations of a $U(1)$ lattice gauge
theory is to be identified with the charge quantum of the
factor group $SMG/(SU(2)\times SU(3))$. Subsequently we shall substantiate
that it is reasonable to take this charge quantum as the weak hyper-charge
of the left-handed positron (i.e., $y/2=1$).
The arguments for this choice
are indeed pivotal for the credibility of the proposed  model.  Had  we
for example taken the lattice critical coupling
$\sqrt{4\pi\alpha_{crit}}$ as the hyper-charge of
the left-handed quarks - which are assigned to the
$\underline{2}\otimes\underline{3}$ representation of $SU(2)\times SU(3)$:

\beq \left( \begin{array}{ccc} u_r & u_b & u_y \\ d_r^c & d_b^c & d_y^c
\end{array} \right), \eeq

\nin this would lead  to  an  $\alpha_{crit}(\mu_{Pl})$ that  was  a
factor $6^2=36$ times larger than that obtained
the left-handed positron.

\nin We return to these matters in Section~\ref{resolve}.

\subsubsection{The infinity of discrete subgroups of $U(1)^3$}

Recall that at the multiple point, there are, in addition to phases confined
w.r.t. continuous subgroups,  also phases that are confined solely
w.r.t.  discrete subgroups. We use as the definition of  confinement that
Bianchi identities can be disregarded in the sense that plaquette variables
can be treated as independent variables. We define
Bianchi variables to be the group product of the plaquette variables
enclosing a 3-volume. The simplest Bianchi variable on a hyper-cubic lattice
are the 3-cubes enclosed by six plaquettes. Bianchi variables are identically
equal to the group identity.
This constraint introduces in general
correlations between the values taken by
plaquettes forming the boundary of a 3-volume. In the case of a first
order phase transition, there is a ``jump'' in the
width of the distribution of plaquette variables in going from a Coulomb to a
confining phase. Our claim is that this ``jump'' is explained
by a change in how effective Bianchi identities are in enforcing correlations
between plaquette variable distributions for different plaquettes forming
the closed surface of a 3-volume. In the Coulomb phase, Bianchi identities
can presumably only be satisfied by having the sum of phases (thinking now
of $U(1)$) of the plaquettes bounding a 3-volume add up to zero. At the
transition to a confining phase, the width of plaquette variable
distributions is large enough so that Bianchi identities are readily fulfilled
in any of a  large number of ways in which the values of plaquette variables
can sum to a non-zero multiple of $2\pi$. This greater ease (energetically)
with which
Bianchi identities can be satisfied for a variety of configurations
of values of boundary plaquette variables means that Bianchi
identities are less
effective in causing correlations between plaquette variables which in turn
allows even greater fluctuations in plaquette variables in a sort of chain
reaction that we claim is the explanation for the sudden decrease in the
Wilson loop operator at the Coulomb to confining phase transition.

Were it not for Bianchi identities, the distributions of values taken by
Bianchi variables would correspond (for a simple 6-sided cube) to the 6-fold
convolution of an independent plaquette variable distribution (i.e.,
uncorrelated with the distribution on other plaquettes). For such a
distribution, it turns out that the critical value of the inverse squared
coupling coincides with a change from a distribution centred at the group
identity to
an essentially ``flat'' (i.e.,
Haar measure) distribution. That the 6-fold convolution of independent
plaquette variable distributions becomes rather``flat'' at the critical value
of the coupling concurs nicely with our characterisation of
confinement as the condition that prevails when
the fulfilment of
Bianchi identities has become almost ``infinitely easy''
energetically and can therefore be neglected in the sense that plaquette
variable distributions for different plaquttes can be taken as
approximately independent.

If it is a discrete subgroup that is confined,
there will be subsidiary peaks in the exponentiated plaquette action
$e^{S_{\Box}}$ at nontrivial elements of this discrete subgroup. Confinement
occurs just when the subsidiary peaks are accessed with sufficient probability
so that the 6-fold convolution of the plaquette distribution over elements
of the discrete subgroup leads to comparable probabilities for accessing
all of these discrete subgroup elements
(i.e., when the 6-fold convolution
of a plaquette variable distribution takes values at all elements
of the discrete subgroup with roughly the same probability).

Having in the plaquette distribution
the presence of subsidiary  peaks (i.e., maxima of the distribution of
group elements) at nontrivial elements of discrete subgroups
affects the value of the critical coupling
of the {\em continuous} (i.e., Lie) group degrees of freedom
at the Coulomb to confinement phase transition.
However, once the discrete subgroup is in the confining phase, the dependence
of the Lie group critical coupling on the relative heights of the peaks has
essentially reached a plateau. This is so because fluctuations along the
discrete subgroup are by definition large enough so that
the transition-relevant distribution obtained as the 6-plaquette
convolution of the plaquette
distribution over the discrete group is essentially already flat
so that going deeper into confinement will hardly access more elements of the
Lie group.
So the Lie group coupling is essentially unchanged in going from
the multiple point to
where the discrete subgroup is deeply confined (meaning parameter values
for which the discrete peaks are equally high). Here the fluctuations
along the discrete subgroup and the cosets that are translations of it
are maximal
(i.e., equal probabilities for all the elements in a coset) and one therefore
needs effectively only to consider
the factor group obtained by dividing out the discrete
subgroup. This is the reasoning underlying the $\bz_{N_{max}}$ factor group
rule discussed above. The rule states that
to a good approximation, the multiple point continuous group
coupling equals the critical coupling for this factor group.

\subsubsection{Resolving the $U(1)$ normalisation problem}\label{resolve}

There is the problem with $U(1)$ that the principle of multiple point
criticality suggests that there should even be phases convening at the
multiple in which there is solely confinement of $\bz_N$  subgroups
of arbitrarily large $N$.
This would result in couplings that vanish. However,
if we also give the matter fields some arbitrarily large
number of the charge quanta of the $U(1)$ that corresponds to the lattice
compactification of $\br$, the coupling of these matter particles need not be
zero. But then our prediction would (only) be that the matter
coupling is a rational number times the multiple point critical coupling.

In order to suggest the manner in which this rational factor might arise,
let us speculate in terms of a model for how our universe came about.
First we describe the model; then we formulate two concise statements from
which the model follows. We end this Section by arguing for the validity of
the two statements.

Assume that at high temperatures (e.g.
immediately following the``Big Bang"), the phase that dominates is that having
the largest number of light particles. Recalling that the various phases
convening at the multiple point have the same vacuum energy density
(in Minkowski language), such a phase would constitute
the "highest pressure" phase that could be expected to expand at
the expense of
other phases.  We speculate that such a phase
has an optimal balance of
unconfined fermions and unconfined monopoles. However, unconfined monopoles are
present in phases
that are {\em confined} w.r.t. discrete subgroups (i.e., $\bz_N$ subgroups). So
in terms of our speculative
picture, we do not expect the high temperature dominant phase to be a totally
Coulomb-like phase but
rather a phase confined w.r.t. some discrete subgroups.
In this scenario, we would  claim that the phase in which we live
- ``our'' cold-universe phase - has the maximal number of
monopole charges consistent with having the phenomenologically known
electrically charged particles (quarks and leptons). This leads us to a
system of monopoles (in ``our'' cold-universe phase) causing confinement
for any fraction of the electric charges known to exist phenomenologically.
The picture to have in mind is that ``our'' cold-universe phase is but one
of many degenerate phases that can convene at the multiple point of a cold
universe. We speculate that the reason that only our phase is realized
is because ``our'' phase dominated so effectively at the high temperatures
following the ``Big Bang'' that all other phases disappeared with the result
that these phases are non-existent in the present low-temperature universe.
Had there existed ``seeds'' of these  phases in the present universe, they
could have competed more or less successfully with ``our'' phase.

Let us examine this proposal for ``our'' universe in the context of a
$U(1)$ lattice gauge theory. We denote by the symbol $U(1)_{fund}$
the $U(1)$ gauge group that is associated with the compactification
that establishes the Abelian degrees of freedom on the fundamental
lattice. Let us furthermore assume that there is some integer
$N_{max}$ such that $\bz_{N_{max}}$ is the largest discrete subgroup
of $U(1)_{fund}$ that can confine alone in one of the phases convening at
the cold-universe multiple point.
This corresponds to having Coulomb-like behaviour for the coset-\dofx
of the factor group $U(1)_{fund}/\bz_{N_{max}}$. This means that if a
$\bz_N$ with $N>N_{max}$ confines in a phase that convenes at the
multiple point, it does so not alone but because the continuous $U(1)$
\dofx also confine. Finally, let $N_{our}$ be defined such that $\bz_{N_{our}}$
is the largest
discrete subgroup that alone is confined in ``our'' phase (which is
assumed to be among the phases that meet at the multiple point).

With the assumption of an $N_{max}$, we can
immediately conclude that the $U(1)_{fund}/\bz_{N_{max}}$
representation of $U(1)_{fund}$ has the largest minimum allowed  charge
quantum. Let us denote this as $Q_{max}$. Furthermore, we can conclude
that the smallest allowed charge quantum - namely that of
$U(1)_{fund}$ - is $Q_{max}/N_{max}$.

In terms of monopoles, we have of course the dual situation: denoting
the smallest
allowed monopole charge for $U(1)_{fund}$ as  $m_{fund}$, the factor
group $U(1)_{fund}/\bz_{N_{max}}$ allows monopoles of fractional charge
the only restriction being that these must be multiples of $m_{fund}/N_{max}$.

The above proposal for ``our'' cold-universe phase as a vacuum that allows
monopoles causing confinement
for any fraction of the electric charges (measured in charge quanta of
$U(1)_{fund}$) known to exist phenomenologically
follows as a consequence of the validity of {\bf two statements:}

\begin{enumerate}

\item $N_{our}$ and $N_{max}$ are such that:

\[ N_{max}=6\cdot N_{our} \]

\[ N_{our} \mbox{ not divisible by 2 or 3.} \]

\item The critical coupling $e_{crit}=\sqrt{4\pi\alpha_{crit}}$ for a
$U(1)$ lattice gauge theory determined
using Monte Carlo methods should be identified with the charge quantum
$Q_{max}$ of the factor group $U(1)_{fund}/\bz_{N_{max}}$.

\end{enumerate}

Before substantiating these statements, we first discuss some conclusions
that that follow from assuming the validity of them.

As long as the conditions of {\bf  statement 1} are fulfilled,
$N_{max}$ can be arbitrarily
large without making the coupling at the multiple point vanish (see first
paragraph of (this) Section~\ref{resolve}).
The smallest allowed charge quantum in ``our'' phase is
$N_{our}(Q_{max}/N_{max})\stackrel{def}{=}Q_{our}$; the discrete subgroups
$\bz_2$ and $\bz_3$ are not confined in ``our'' phase. These
discrete subgroups
$\bz_2$ and $\bz_3$ - which are only found once as subgroups of
$\bz_{N_{max}}$ - are confined (alone) only in phases to which are associated
minimum allowed charge quanta larger than $Q_{our}$.
Using the {\bf statement 2}, we can fix the value of the smallest allowed
charge quantum in the phase with $/bz_{N_{max}}$ alone confined
as $\sqrt{4\pi\alpha_{crit}}$ and thus in ``our'' phase as
$Q_{our}=N_{our}\cdot(\sqrt{4\pi\alpha_{crit}}/N_{max})$.

It is now necessary to give an argument for which physical particles should
have $Q_{max}=\sqrt{4\pi\alpha_{crit}}$ as its charge quantum.
As stated above, earlier work leads us to  expect the $\bz_2$ and $\bz_3$
centres  of respectively
$SU(2)$ and $SU(3)$ to confine alone in phases convening at the multiple
point. The phase with $\bz_2\times \bz_3$ confined alone coincides
with the phase with Coulomb-like behaviour for the coset \dofx of the
factor group $SMG/(SU(2)\times SU(3))\stackrel{-}{\simeq} U(1)/\bz_6$
corresponding to the trivial representation of the $SU(2)\times SU(3)$
\dof. The left-handed positron $e^+_L$ is the singlet under
$SU(2)\times SU(3)$ that has the smallest charge.

At the end of this Section, we shall give a speculative argument for why
it is natural that the phase in which there  alone
is confinement of $SMG/(SU(2)\times SU(3))\stackrel{-}{\sim} U(1)/\bz_6$
should be identified with the phase in which there is confinement
solely of the discrete subgroup $\bz_{N_{max}}$ corresponding to
Coulomb-like \dofx for the cosets of
$\frac{U(1)_{fund}/\sbz_{N_{our}}}{\bz_6}\
=U(1)_{fund}/\bz_{N_{max}}$. This
identification puts the hyper-charge of the left-handed positron into
correspondence with the
factor group $U(1)_{fund}/\bz_{N_{max}}$ charge quantum
$\sqrt{4\pi\alpha_{crit}}$.

Use now the usual convention for hyper-charge: $y/2=Q/6Q_L$ (for particles
of hyper-charge $Q$) and associate $(y/2)_{e^+_L}=1$ with $Q=Q_{max}=
\sqrt{4\pi\alpha_{crit}}$ (the $U(1)$ lattice gauge critical coupling).
This determines the hyper-charge
quantum $Q_L$ of ``our'' phase (which has unconfined quarks and leptons
at the Planck scale) as $Q_L=\frac{\sqrt{4\pi\alpha_{crit}}}{6}$.
This is the charge quantum of the
$\underline{2}\otimes\underline{3}$ representation of $SU(2)\times SU(3)$.

The properties ascribed to ``our'' cold-universe phase are contingent
upon the validity of {\bf statements 1} and {\bf 2} above.  Let us now argue
for
the validity of these statements (in reverse order).

{\bf Statement 2} follows basically from  the $Z_{N_{max}}$
factor group
rule for the multiple point coupling of continuous \dofx as discussed on
page \ref{resolve}. This rule states that if the multiple
point for  $U(1)_{fund}$ has contact with a phase in which a discrete
subgroup $\bz_N\in U(1)_{fund}$ is alone confined, then to a very good
approximation,
the {\em multiple point} value of the coupling for the continuous \dofx (i.e.,
the coupling values that reflect the effect of also having a
phase confined alone w.r.t $\bz_N$ that convenes at the multiple point)
is obtained by assuming that this discrete
subgroup is {\em totally} confined (instead of having the multiple point
(i.e., critical) coupling value). This is tantamount to identifying
the multiple point value of the coupling of the continuous \dofx of
$U(1)_{fund}$ with the value of the critical coupling for the factor
group $U(1)_{fund}/\bz_N$. If there are more than one phase convening
at the multiple point that is confined solely w.r.t. some discrete
subgroup, then the best approximation to the multiple point coupling
of the continuous \dofx of $U(1)_{fund}$ is given by the critical value of
the coupling of the factor group with the largest discrete subgroup
$\bz_{N_{max}}$ divided out: i.e., the critical coupling value of
$U(1)_{fund}/\bz_{N_{max}}$. We referred to this approximation as the
$\bz_{N_{max}}$ factor group rule.

The approximate validity of
statement 2) follows using results from Monte Carlo simulations of
lattice gauge theories. From these results the critical value
$e_{crit}=\sqrt{4\pi\alpha_{crit}}$ of the coupling for factor groups
groups of the type $U(1)_{fund}/\bz_N$ with $N=2$ or 3 can be deduced.
As the identification of the critical coupling for
$U(1)_{fund}/\bz_{N_{max}}$ with the critical coupling for
$U(1)_{fund}/\bz_N$ ($N=2$ or 3) is good even for $N<<N_{max}$, the approximate
validity of statement 2) follows.

To establish the validity of {\bf statement 1},
write as above $N_{max}=pN_{our}$ where $p\in \bz$ and
$N_{our}$ is such that $\bz_{N_{our}}$ is the largest discrete
subgroup of $U(1)_{fund}$ that is confined in ``our'' phase. We
note first that $N_{our}$ cannot be divisible by 2 or 3.
Had this been
the case, we would have respectively the subgroups $\bz_2$ and $\bz_3$
confined in ``our'' phase. This would correspond to a restriction of
the possible Coulomb-like \dofx to those having the charge quantum of
a factor group isomorphic to $SMG/(SU(2)\times SU(3))$. The latter is
a singlet w.r.t $SU(2)\times SU(3)$ and accordingly has a charge
quantum too large to allow the
$\underline{2}\otimes\underline{3}$ representation of
$SU(2)\times SU(3)$ needed for having the phenomenologically observed
left-handed quarks and leptons. Phenomenologically at least, our phase
does not have confinement of quarks and leptons at the Planck scale.

However, in order to have the (unrealized) phases with $\bz_2$ and
$\bz_3$ alone confined among the degenerate cold-universe phases that
convene at the multiple point, it is necessary that $p$ be divisible
by 2 and 3: $p=q\cdot 6$. To establish statement 1) however,
we need to
argue that $q=1$. This somewhat speculative
argument goes as follows. Let us imagine that there are extra \dofx
that are hidden from us but which also tend to go into different
phases. Let us speculate that the extra hidden \dofx influence the
form of our ``fundamental'' Lagrangian. So really our ``fundamental''
Lagrangian is an effective Lagrangian; which effective Lagrangian is
realized as our ``fundamental'' Lagrangian can depend on which phases
that hidden \dofx are in. It is important for the argument that the
difference that these extra \dofx can make as to which effective
Lagrangian is realized as our ``fundamental'' Lagrangian can even be
manifested as different numbers of quanta of $U(1)_{fund}$ for quarks
and leptons for different effective Lagrangians. From this point of
view, figuring out which phase would have maximum pressure
immediately following
the
``Big Bang'' also requires looking at different
``possible'' effective Lagrangians (corresponding to
hidden \dofx being in
different phases and even perhaps having quarks and leptons
made up of
different numbers
of quanta of $U(1)_{fund}$)  before
``deciding'' on what our
``fundamental'' Lagrangian should be. These
different ``fundamental ''
Lagrangians (i.e., different effective
Lagrangians among which ours is found)
are different because the extra
to us hidden \dofx of other fundamental
theories can be in phases
having various
different minima. Using as input that observed quarks and
leptons must not be confined, this picture favours a choice for our
``effective''
Lagrangian that corresponds to quarks and
leptons having the largest possible
number
of the charge quanta of
$U(1)_{fund}$; i.e., the largest possible number of
the quanta
$Q_{max}/N_{max}$. This allows the largest possible discrete
subgroup
to be confined in ``our'' phase and accordingly the greatest
number of monopoles
consistent with having observed fermions.

Another
way of putting this is that
phenomenology tells us that $\bz_2$ and
$\bz_3$ cannot be confined in our
phase. So the corresponding
monopoles are not available for helping to have
a high pressure at the
high temperatures immediately following
the Big Bang. However, all
possible other monopoles can help create high
pressure at high
temperatures; the corresponding discrete subgroups are expected to be confined
in ``our''
phase.
The argument is that when the hidden \dofx can go into one or
another phase
that
lead to one or another ``effective'' Lagrangian
for us, the effective
Lagrangian that can be expected to become our
``fundamental'' Lagrangian
is one
that
doesn't ``waste'' monopoles in
the sense that the charge quanta of ``our''
phase (i.e., of the factor-group $U(1)_{fund}/\bz_{N_{our}}$)
do not consist of a smaller number of  fundamental
quanta $Q_{max}/N_{max}$
than absolutely necessary
in order to have
the phenomenologically forbidden $\bz_2$ and $\bz_3$
monopoles
convene in (unrealized) cold-universe degenerate phases
convening at
the multiple point\footnote{E.g., if there were two effective Lagrangians
${\cal L}_{eff\;1}$ and ${\cal L}_{eff\;2}$  - one leading to
$N_{max}=42\cdot N_{our_{1}}$ and the other to
$N_{max}=6\cdot N_{our_{2}}$ (assuming $N_{max}$ the same in both cases) -
we would expect ${\cal L}_{eff\;2}$ to be be realized as {\em the} ``our''
effective Lagrangian because
$Q_{our_{2}}= N_{our_{2}}\frac{Q_{max}}{N_{max}}=\frac{Q_{max}}{6}$ is
larger than $Q_{our_{1}}=\frac{Q_{max}}{42}$. Relative to the Lagrangian
${\cal L}_{eff\;2}$, the Lagrangian ${\cal L}_{eff\;1}$ lacks a
confined $\bz_7$ subgroup and therefore the pressure contribution from the
corresponding monopoles.}.
This dictates that $N_{max}$ is just a single
factor 6 larger than $N_{our}$ so that $q=1$ above as we set out to show.

\subsubsection{Cartesian product gauge groups, additive actions and
factorizable subgroups}

The fact that the ``fundamental''  gauge  group  $SMG\times  SMG\times
SMG$ is a Cartesian product group means that it is possible to  have
an action that is additive in contributions from  each  of  the  three
group factors in the Cartesian product:

\beq S=S_{Peter}+S_{Paul}+S_{Maria}.\label{addact} \eeq

We have used such an additive action in connection with the calculation
of the non-Abelian gauge couplings in previous work. In
this section we  explain why, from the standpoint
of the $MPCP$, it is necessary to use a more general action in the
case of the Abelian gauge coupling.

With such an action, we are restricted to  bringing  together,  at  an
approximative multiple point, the confining phases that correspond
to {\em factorizable} invariant subgroups which  means  invariant  subgroups
that are Cartesian products of  invariant  subgroup  factors  each  of
which can be identified as coming from  just  one  of  the  isomorphic
$SMG$ factors (labelled by ``$Peter$'', ``$Paul$'', $\cdots$)
of $SMG^{N_{gen}}$. So if  we  restrict  ourselves
to an additive action of the type (\ref{addact}),
the phase diagram for  the  ``fundamental''
gauge group $SMG^{N_{gen}}$ is completely determined from a  knowledge
of  the  phase  diagram  for  just  one  of  the  group   factors   of
(e.g. $SMG_{Peter}$) of $SMG^{N_{gen}}$.
The additive action approximation yields the same value of the coupling for
the  $U(1)$
subgroup of each of the $SMG$ factors (labelled by the $N_{gen}$ indices
``$Peter$'', ``$Paul$'', $\cdots$). The same applies for the  three
$SU(2)$'s and $SU(3)$'s. In going to the diagonal subgroup, all three
$SMG$ fine-structure constants (i.e. for $U(1)$, $SU(2)$ and $SU(3)$) are
each enhanced by the same factor $N_{gen}=3$:

\beq \frac{1}{\alpha_{diag}}(\mu_{Pl})=\frac{1}{\alpha_{Peter}}(\mu_{Pl})+
\frac{1}{\alpha_{Paul}}(\mu_{Pl})+\frac{1}{\alpha_{Maria}}(\mu_{Pl})=
\label{diagcoup} \eeq

\[ =\frac{1}{\alpha_{multicr.}}+\frac{1}{\alpha_{multicr.}}+
\frac{1}{\alpha_{multicr.}}=\frac{3}{\alpha_{multicr.}}. \]

For the non-Abelian subgroups, it turns out that the  approximate
multiple point found in this way lacks contact with relatively few  of
the possible partially confining\footnote{In the case of a non-simple group
such as the $SMG$, it is possible to have confinement w.r.t. some but not
all gauge \dofx. Such phases are referred to as partially confining phases.}
phases  whereas  such an approximate
multiple point
lacks contact with  an infinity of partially confining phases of
$U(1)^3$.
Accordingly, we have found that the approximate  multiple point
critical couplings obtained using an  additive  action  (\ref{addact})
yield excellent predictions for the non-Abelian fine-structure constants
whereas the analogous prediction for the $U(1)$ fine-structure constant
is off by about 100 \%.

The phases that are lacking when the action is restricted to
being additive - i.e., phases corresponding to confinement along
non-factorizable subgroups - are present unless all the
group factors of a Cartesian product group are
without common nontrivial isomorphic subgroups of the centre.
In the  case of the Cartesian product group $SMG^3$,  the centre
(which itself is a Cartesian product)
has nontrivial repeated subgroup factors that are in different $SMG$
factors of $SMG^3$. Diagonal subgroups of such repeated subgroup factors are
non-factorizable in the sense that they cannot be factorized into
parts that each
are unambiguously  associated with just one $SMG$ factor of $SMG^3$.
With an additive action, it is not possible to  have  confinement alone along
the diagonal subgroups of such repeated factors.

Getting the phases  that  are  confined  w.r.t.  non-factorizable
invariant subgroups to convene at the  multiple point  (together
with phases for factorizable invariant subgroups) requires interaction
terms in the action that obviously are incommensurate with having  an
additive action. Having such interaction terms means that it does  not
suffice to consider just one
$SMG$ factor at a time as  was  the  case  for  the
additive  action  (\ref{addact}).  In   general,   the   presence   of
interaction  terms  means  that  it   is   necessary   to   seek   the
multiple point  for  the  whole  $SMG^3$.  For  simplicity,  we  might
approximate  the  problem  by  considering  $U(1)^3$,  $SU(2)^3$   and
$SU(3)^3$ separately - but even this may ignore  some  non-factorizable
subgroups that could confine by  having  appropriate  interaction
terms in the action. However, for  the  non-Abelian  groups,  an  even
rougher approximation  is  rather  good:  finding the multiple point
couplings for  $SU(2)$  and  $SU(3)$
instead of respectively for $SU(2)^3$ and $SU(3)^3$ corresponds to finding
the multiple point using the approximation of an  additive
action (\ref{addact}).

Having non-factorizable subgroups requires
having invariant (and therefore necessarily central)
                                   ``diagonal-like''
subgroups (i.e., diagonal subgroups or subgroups that are diagonal up to
automorphisms within subgroups of the centre).
The centre of $SMG^3$ is the Cartesian product

\beq [(U(1)\times \bz_2\times \bz_3)/``\bz_6"]_{Peter} \times
     [(U(1)\times \bz_2\times \bz_3)/``\bz_6"]_{Paul} \times
     [(U(1)\times \bz_2\times \bz_3)/``\bz_6"]_{Maria} \eeq

In the case of the non-Abelian subgroups $SU(2)^3$ and $SU(3)^3$,
the possibility
for non-factorizable subgroups is limited to the finite number of
``diagonal-like'' subgroups that can be
formed from $\bz_2^3$ and $\bz_3^3$ (i.e., the respective centres of
$SU(2)$ and $SU(3)$). An examples is

\beq \{(g,g)|g\in \bz_3 \} \stackrel{-}{\simeq} \bz_3 \eeq

\nin where the element $(g,g)$ is the special (diagonal) case of say
an element $(g_{Peter},g_{Paul})\subset SMG^3$ for which
$g_{Peter}=g_{Paul}\stackrel{def}{=}g$. Other examples are

\beq \{(g,g^{-1})|g\in \bz_3 \} \stackrel{-}{\simeq} \bz_3, \eeq

\beq \{(g,g,g)|g \in \bz_3 \} \stackrel{-}{\simeq} \bz_3, \eeq

\beq \{(h,h^{\prime},h^{\prime\prime})|h,h^{\prime},h^{\prime\prime}\in \bz_2,
\mbox{two out of three of the }h,h^{\prime},h^{\prime\prime} \mbox{odd}\}\eeq
\[ =\{(1,1,1), (1,-1,-1), (-1,1,-1), (-1,-1,1)\}\stackrel{-}{\simeq}\bz_2
\times \bz_2 \]
and
\beq \{(h,h,g,g)|h\in \bz_2, g\in \bz_3 \}\stackrel{-}{\simeq}
\bz_2\times \bz_3 \}. \eeq

In the case of $U(1)^3\subset SMG^3$, any subgroup is invariant
(because $U(1)^3$ lies entirely in the centre of $SMG^3$). In particular,
any diagonal-like  subgroup
is invariant and constitutes
therefore a
non-factorizable subgroup along which there separately can be confinement.
While the non-factorizable (invariant) subgroups for $SU(2)^3$ and $SU(3)^3$
are exclusively of dimension $0$, such
subgroups can occur for $U(1)^3$ with
dimension 0, 1, 2 and 3.
For $U(1)$, non-factorizable subgroups occur as diagonal-like subgroups of
all possible  Cartesian products  having two or three repeated subgroup
factors (with different labels ``$Peter$'', ``$Paul$'', $\cdots$). These
repeated factors can be
discrete $\bz_N$ subgroups (for all $N\in \bz$) and  also $U(1)$ subgroups.
The latter are of
importance as regards plaquette action terms that are bilinear
in gauge fields:
unlike the case for continuous non-Abelian subgroups, it is possible to have
{\em gauge invariant} quadratic action terms of, for example, the type
$F_{\mu\nu\;Peter}F^{\mu\nu}_{Paul}$ defined on, for example,
$U(1)_{Peter}\times U(1)_{Paul}\subset U(1)^3$. Because the subgroup
$U(1)_{Peter}\times U(1)_{Paul}$ lies
in the centre of $SMG^3$, diagonal-like subgroups are {\em invariant} and
it is therefore meaningful to consider the transition between phases that
are confining and Coulomb-like for such diagonal-like subgroups. By
introducing terms in the action  of the type
$F_{\mu\nu\;Peter}F^{\mu\nu}_{Paul}$, we can extend the space of parameters and
thereby find additional phases that we subsequently can try to make accessible
at the multiple point.
In fact,
such terms can explain the factor ``6'' enhancement of Abelian inverse
squared couplings in going to the diagonal subgroup of $U(1)^3$.
The analogous factor for the non-Abelian diagonal subgroup couplings
is recalled as being only three - i.e., $N_{gen}=3$.

\section{Phase Diagram}\label{phasediagram}

\subsection{``Phase'' classification according to symmetry  properties
of vacuum}

We classify the lattice artifact phases of  the  vacuum  according  to
whether or not there is spontaneous breakdown of symmetry under  gauge
transformations  corresponding to the sets of gauge functions
$\Lambda_{Const}$   and   $\Lambda_{Linear}$ that are
respectively  constant  and  linear  in  the
coordinates:

\beq \Lambda_{Const}\in
\{\Lambda:\br^4\rightarrow G|\exists\alpha[\forall x\in \br^4 [\Lambda(x)=
e^{i\alpha}]] \}\eeq

\nin   and

\beq\Lambda_{Linear}\in
\{\Lambda:\br^4\rightarrow G|\exists\alpha_{\mu}[\forall x\in \br^4[\Lambda(x)=
e^{i\alpha_{\mu}x^{\mu}}]]\}. \label{lin} \eeq

\nin
Here $\alpha=\alpha^at^a$ and $\alpha_{\mu}=\alpha^a_{\mu}t^a$ where
$a$ is a ``colour'' index in the case of non-Abelian subgroups.
Spontaneous  symmetry   breakdown   is   manifested   as
non-invariant values for gauge variant quantities.  However,  according
to Elitzur's theorem, such quantities cannot survive  under  the  full
gauge symmetry. Hence a partial fixing of the gauge  is  necessary  before  it
makes sense to talk about the spontaneous breaking of these types of
symmetry.  We
choose the Lorentz gauge for the reason that  this  still  allows  the
freedom of making gauge transformations of the types $\Lambda_{Const}$
and $\Lambda_{Linear}$ to be used in classifying the lattice  artifact
``phases'' of the vacuum.

When the gauge field $U(\linkxy)$ takes values in  a  non-simple  gauge
group such as $SMG^3$ having many subgroups and  invariant  subgroups
(including discrete subgroups),
it is possible for \dofx corresponding say to  different  subgroups  to
take  group  values  according  to  distributions  that   characterise
qualitatively  different  physical  behaviours  along   the   different
subgroups. Some \dofx can have a fluctuation pattern  characteristic
of a Higgsed phase; some  of  the  \dofx  having  fluctuation  patterns
characteristic  of  an  un-Higgsed  phase  can  be  further  classified
according  to  whether  they  have  Coulomb-like  or
confinement-like  patterns  of  fluctuation.  The  point  is   that   a
``phase'', which of course corresponds to  a region  in  the  action
parameter space, can, for a non-simple gauge group,  be described in terms of
characteristica that differ along different subgroups.
The fluctuation patterns for the various \dofx corresponding to these
subgroups  can be classified according to the
transformation properties of the vacuum under the two classes of gauge
transformations $\Lambda_{Const}$ and $\Lambda_{Linear}$.
We shall see that the set of possible ``phases'' corresponds one-to-one to
the set of  all  possible  subgroup  pairs\footnote{In this classification
scheme it has been assumed that the action
energetically favours $U(\Box)\approx \bunit$; however, a vacuum also having
fluxes corresponding to nontrivial elements of the centre could be
favoured if for instance there were negative values for
coefficients of plaquette terms in the action. Such terms would lead to new
partially confining phases that were Coulomb-like
but for which fluctuations in the  \dofx are centred at a
nontrivial element of the centre instead of the identity.}
$(K,H)$  consisting  of  a
subgroup $K\subseteq G$ (where the gauge group $G$ of interest here is $SMG^3$)
and an invariant subgroup $H\triangleleft K$.
Each ``phase'' $(K,H)$ in general corresponds to a partitioning of the
\dofx (these latter can be labelled by a Lie algebra basis) - some  are
Higgsed, others that are un-Higgsed; of the latter, some  \dofx  can
be confining, others Coulomb-like. It is therefore useful to think of a
group element $U$ of the gauge group as being parameterised  in  terms
of  three  sets  of  coordinates  corresponding  to  three   different
structures that are appropriate to the symmetry properties  used
to define a given phase $(K,H)$ of the vacuum.  These  three  sets  of
coordinates,
which are definable in terms of the gauge group $SMG^3$, the  subgroup
$K$, and the invariant subgroup $H\triangleleft K$,
are the {\em homogeneous space}  $SMG^3/K$,  the  {\em  factor  group}
$K/H$, and $H$ itself:

\beq U=U(g,k,h) \;\;\mbox{with }\;g\in SMG^3/K, k\in K/H, h\in H. \eeq

\nin The coordinates $g\in SMG^3/K$ will  be  seen  to  correspond  to
Higgsed \dof, the coordinates $k\in  K/H$  to  un-Higgsed,  Coulomblike
\dof, and the coordinates $h\in H$ to un-Higgsed, confined \dof.

For each phase $(K,H)$, the \dofx taking values in the subgroup $K$
(after having fixed the gauge by the choice of say the Lorentz gauge condition
-
see above))
are  said
to exhibit ``un-Higgsed'' behaviour which by definition means  that  $K$
is the maximal subgroup of gauge transformations having constant gauge
transformations $\Lambda_{Const.}$ that leaves the  vacuum  invariant.
The gauge symmetry of the vacuum for the \dofx that take as values  the
cosets of the homogeneous  space  $SMG^3/K$  is  spontaneously  broken
under gauge transformations  with  constant  gauge  functions  and  is
accordingly taken as the defining feature of a Higgsed phase.

Lattice \dofx that take values in the invariant subgroup $H\lhd K$  are
said to have ``confinement-like'' behaviour which  by  definition  means
that $H$ is the maximal  invariant  subgroup  $H\lhd  K$  of  elements
$h=\exp\{i\alpha^1_at_a\}$ such that the  gauge  transformations  with
linear gauge functions $\Lambda_{Linear}$  exemplified  by\footnote{In
the quantity $x^1/a$, $a$ denotes the lattice constant; modulo lattice
artifacts, rotational invariance  allows  the  (arbitrary)  choice  of
$x^1$ as the axis $x^{\mu}$  that  we  use.} $\Lambda_{Linear}\stackrel
{def.}{=}h^{x^1/a}$ leave the vacuum invariant. For the \dofx that take
values in the factor group $K/H$, there is invariance  of  the  vacuum
under  gauge  transformations  for  which  the  (exponentiated)  gauge
function is constant and takes values in $K$
while there is spontaneous breakdown of  the  vacuum under
the gauge transformations with linear gauge functions.
Degrees of  freedom  for which the vacuum (in the Lorentz gauge) has  these
transformation properties are by definition said to
demonstrate ``Coulomb-like'' behaviour.

In implementing the multiple point criticality principle ($MPCP$) in practice,
we  seek  a  multiple
point in some restriction to  a  finite  dimensional
subspace  of  the  in  principle  infinite  dimensional  action
parameter space. This just amounts to making an action ansatz.
Consider an action parameter space that has  been
chosen so that we can realize a given phase $(K   \subseteq   G,   H
\triangleleft K)$.
In this paper, we consider only the special case $K=G$ corresponding to
not having \dofx that are Higgsed. However, we want to include a suggestion of
the manner in which one - at least in a discretized gauge theory - could
also have convening phases at the
multiple point that are Higgsed w.r.t. to various \dofx even though we
shall not make use of Higgsed phases in the sequel.

In order to bring about a Higgsing of the gauge group $G$ down to the
subgroup $K$, one could use
action terms  defined  on  gauge  invariant
combinations  of  site-defined  fields  $\phi(\sitex)$  and  the  link
variables $U(\link)$. The fields $\phi(\sitex)$  take values on homogeneous
spaces $G/K$ of
the gauge group $G$ where $K\subseteq G$. Such action terms
 are designed so that for
sufficiently large values
of a coefficient $\kappa$,  the  field  $\phi(\sitex)$  acquires  a
non-vanishing vacuum expectation value\footnote{Even if we in some natural
manner succeeded in embedding a
homogeneous space in an affine space, it would not in general be convex.
Therefore one needs to
construct the convex closure (e.g. in a  vector space) if we want to talk
about averages of field variables.
 As an example, think of the homogeneous space $SO(3)/SO(2)$
which is metrically equivalent with an $S_2$ sphere. In this case, one
could obtain the complex closure as a ball in the linear embedding space
$\br^3$. Alternatively, we can imagine supplementing the $SO(3)/SO(2)$ manifold
with the necessary (strictly speaking non-existent) points needed in order to
render averages on the $S_2$ meaningful.}: $\langle  \phi(\sitex)  \rangle
\neq 0$ with the result  that  the  gauge  symmetry  is  spontaneously
Higgsed from that of the gauge group $G$ down to that of the  subgroup
$K$. Then degrees of freedom corresponding to the cosets of  $G/K$  are
Higgsed and \dofx corresponding to elements of $K$ are un-Higgsed.  We
have seen that the defining feature of the subgroup $K$ is that it is the
maximal  subgroup  of  gauge  transformations  having  constant  gauge
functions that leave the vacuum invariant.

Other coefficients - call them $\beta$ and  $\xi$  -  multiply  action
terms defined on factor groups $K/H$ of the un-Higgsed subgroup
$K$ where $H \triangleleft K$. Two types of coefficients $\beta$
and $\xi$ having to  do  with  respectively  continuous  and  discrete
invariant subgroups $H$ are distinguished.  For  sufficiently  large
values of the parameters $\beta$  and/or  $\xi$,  the  gauge  symmetry
under  gauge  transformations  having  linear   gauge   functions   is
spontaneously broken from that of $K$ down to that of the  invariant
subgroup $H$. The \dofx corresponding to the factor  group  $K/H$
behave by definition Coulomb-like;  elements  of  the  invariant  subgroup
$H$ correspond to ``confined \dof''.  By  definition,  $H\triangleleft
K$ is the maximal invariant subgroup of gauge transformations having
linear gauge functions that leave the vacuum invariant.

Were we to include the
possibility of Higgsed phases, an extra interaction between the Higgs
field and the gauge field (in addition to the one implemented by the use of
covariant derivatives in the kinetic term for the Higgs field) would be
needed in order to make the various phases meet at the multiple point.
Otherwise there is the risk that the fine-structure constant changes (e.g.,
does not remain equal to $\alpha_{crit}$)
in going from
$\langle \phi(\sitex) \rangle=0$ to $\langle \phi(\sitex) \rangle\neq 0$.
A suitable interaction term might be of a rather explicit form; for
example, it could be implemented by replacing the parameters
$\beta$ and $\xi$ by functions of the Higgs fields so that the interaction
effectively (i.e., via the Higgs fields) will depend on the subgroup
$K \subseteq SMG^3$ of un-Higgsed \dof.
This could be accomplished using a term in the action of the form

\beq   c|\phi(\sitex)|^2    \mbox{Tr}(U(\Box)). \label{higgsint}\eeq

\nin A term such as (\ref{higgsint}) comes into play when the gauge symmetry
is spontaneously broken by Higgsing from $G$ down to $K\subseteq G$. It
could
compensate changes in the critical coupling that accompany such a
spontaneous breakdown inasmuch as it is obvious that

\beq \langle \phi(\sitex) \rangle \left \{ \begin{array}{l} = 0 \mbox{ in
phase  }(G,\bunit)  \\ \neq  0  \mbox{  in  phase  }  (K,\bunit),\;\;\;
\phi(\sitex)\in G/K  \end{array} \right. . \eeq

In other words, a term such as (\ref{higgsint}) vanishes in the phase
$(G,\bunit)$ where
$\langle \phi(\sitex) \rangle = 0$ but can,  in  going  into  the  phase
$(K,\bunit)$  where  $\langle   \phi(\sitex)   \rangle   \neq   0$,
make a contribution to the inverse  squared  coupling
for $K$.

\subsection{Portraying $U(1)^3$ and its subgroups}\label{sec-portray}

The  phase  diagram  for the  group   $U(1)^3\stackrel{def}{=}
U(1)_{Peter}\times
U(1)_{Paul}\times U(1)_{Maria} \subset SMG^3$ can be  expected  to  be
rather  complicated  because  of  its  many  subgroups.  There   is   a
denumerable infinity of compact subgroups of $U(1)^3$ (discrete  as  well  as
continuous subgroups ranging in dimension  from  zero  to  three).  We
shall seek an approximate $U(1)^3$ phase diagram in the context  of  a
Lattice gauge theory with a Manton action.

As mentioned above, even a continuum action term of  for  example  the
form $\int d^4x F_{\mu\nu}^{Peter}F^{\mu\nu\;Paul}$ is invariant under
gauge transformations in the case of Abelian groups such  as  $U(1)^3$
simply  because  $F_{\mu\nu}^{Peter}$   and   $F^{\mu\nu\;Paul}$   are
separately gauge invariant
\footnote{Under a gauge transformation, we have
\beq Tr[F_{\mu\nu}^{Peter}F^{\mu\nu\;Paul}]\rightarrow Tr[\Lambda^{-1\;Peter}
F_{\mu\nu}^{Peter}\Lambda^{Peter}\Lambda^{-1\;Paul}F^{\mu\nu\;Paul}
\Lambda^{Paul}] \neq Tr[F_{\mu\nu}^{Peter}F^{\mu\nu\;Paul}] \eeq
\nin unless gauge transformations commute with the $F_{\mu\nu}^{I}$'s
 $I \in \{``Peter", ``Paul",\cdots \}$.}.
In particular, a Manton action can have  a
term of this type and therefore a general Manton action can be written

\beq        S_{\Box;,Man}(\theta^{Peter},\theta^{Paul},\theta^{Maria})
=min\{\hat{\theta}^ig_{ik}\hat{\theta}^k|
\hat{\theta}^j=\theta^j   \mbox{   mod   }(2\pi)\}\eeq
\nin where $i,k\in
\{Peter, Paul,Maria\} \label{firstc}$ and $g_{ik}$ is the metric tensor.

We may choose more general coordinates  by  defining  new  coordinates
$\theta^i$ as linear combinations of the old  ones  $\tilde{\theta}^j$:
$\theta^i \rightarrow K^i_{\;k}\tilde{\theta}^k$. Under such a transformation,
an action term of for example the  type  $(F^{Peter}_{\mu\nu})^2$  may
transform into a linear combination involving also
terms of the type  $F^{\mu\nu\;Peter}F^{Paul}_{\mu\nu}$  and  vice
versa. Also, the identification $mod$ $2\pi$  is  transformed  into  a
more general identification modulo a lattice $L$ in the covering group
${\bf R}^3$:

\beq \vec{\theta}\;\;\;\;\;\widetilde{\mbox{\tiny   identified}}
\;\;\;\;\;\vec{\theta}+\vec{l} \;\;\mbox{where}\;\; \vec{l}\in L
\label{secondc}
\eeq

\nin The meaning of  (\ref{secondc}) is that $\vec{\theta}$ and
$\vec{\theta}+\vec{l}$  corresponds
to the same group element of $U(1)^3$.

Because the requirement of
gauge invariance for an action defined  on  the  Abelian  gauge  group
$U(1)_{Peter}\times U(1)_{Paul}\times U(1)_{Maria}$ does not  prohibit
linear combinations of  $F^{\mu\nu}_{Peter}$,  $F^{\mu\nu}_{Paul}$  and
$F^{\mu\nu}_{Maria}$ that can lead to bilinear terms of the type
$F^{\mu\nu}_{Peter}F_{\mu\nu\;Paul}$, there are many possible formulations
corresponding to the same physics (this assumes  of  course  that  the
functional form of the action and the quantisation rules  are  changed
appropriately in going from one formulation to  another).
So points in  the  phase  diagram  should  correspond  to  equivalence
classes of formulations having the same physics.

The gauge group $U(1)^3$ is a (compact) factor group of  the  covering
group ${\bf R}^3$ obtained by dividing out  a  discrete  subgroup  $L$
isomorphic to ${\bf Z}^3$ that we refer to as  the
identification lattice $L$. This is just  the  3-dimensional  lattice  of
elements of ${\bf R}^3$ that are identified with the unit  element  in
going to $U(1)^3$. If we assume that ${\bf R}^3$  is  provided
with an inner product, there will  be  a
recipe for constructing a unique Manton action

\begin{equation}
S_{\Box}(\vec{\theta})=min\{
                  \vec{\theta}^{\prime T}{\bf  g}\vec{\theta}^{\prime}
|\vec{\theta}^{\prime}\in     \vec{\theta}+L      \}.      \label{min}
\end{equation}

\noindent where ${\bf g}$ denotes the metric tensor. The point is that
we construct the metric ${\bf g}$ so that it describes the Manton action.
The  expression
(\ref{min}) is just the generalisation of (\ref{firstc}) to the case of
an arbitrary  choice  of  coordinates  instead  of  the  special  case
in (\ref{firstc}) where coordinates  are  referred  to  basis  vectors
$\vec{l}\in L$.

For ease of exposition, it is useful to consider $U(1)^2$ as a representative
prototype for $U(1)^3$.
Physically different Manton actions correspond to different classes of
isometric-ally  related  embeddings  of  the   identification   lattice
into the Euclidean plane (i.e., $\br^2$ provided with the action-related
metric).
A  pair  of
embeddings where  one  is  rotated  w.r.t.  the  other  correspond  to
physically the same Manton action. Such rotations could be implemented
by coordinate transformations that transfers the coordinate  set  from
one embedding into being the coordinate set of the rotated  embedding.
Obviously  the  two  lattice  constants  (call  them  $a_{Peter}$  and
$a_{Paul}$) and the angle (call it $\phi$)  between  the  two  lattice
directions  are  isometric-ally  invariant   (i.e.,   invariant   under
rotations).
Hence the specification of the properties of a  physically
distinct Manton action (for $U(1)^2$) requires three parameters. These
can be taken as the
three independent matrix elements of the metric  tensor.  We  re-obtain
the coordinate choice (\ref{firstc}) by  adopting  as  our  coordinate
choice  the  requirement  that  the  identification  lattice  has  the
coordinates\footnote{We require of  this  coordinate  system  that  it
allows the group composition  rule  (denoted  with  ``$+$'')  for  two
elements             $(\theta_{Peter},\theta_{Paul})$              and
$(\theta_{Peter}^{\prime},\theta_{Paul}         ^{\prime})$:
$(\theta_{Peter},\theta_{Paul})+(\theta_{Peter}^{\prime},\theta_{Paul}
^{\prime})         =          (\theta_{Peter}+\theta_{Peter}^{\prime},
\theta_{Paul}+\theta_{Paul}^{\prime})$.}

\beq 2\pi(n_{Peter},n_{Paul}) \;\;\;\mbox{with }\;\;\
n_{Peter}, n_{Paul} \in {\bf  Z}.
\label{coord2} \eeq

We give now a concrete example. Using the coordinates (\ref{coord2})
for the identification lattice,
the class of  embeddings  corresponding  to  a
given Manton  action  $S_{\Box}(\vec{\theta}$)  given  by  (\ref{min})  is
specified by the metric tensor

\begin{equation} {\bf g}=\left( \begin{array}{cc} g_{11} &  g_{12}  \\
g_{21}  &  g_{22}   \end{array}   \right)   =\left(   \begin{array}{cc}
\frac{\beta_{Peter}}{2}                                              &
\sqrt{\frac{\beta_{Peter}\beta_{Paul}}{4}}\cos\phi                \\
\sqrt{\frac{\beta_{Peter}\beta_{Paul}}{4}}\cos\phi &
\frac{\beta_{Paul}}{2}      \end{array}
\right). \end{equation}

\nin In particular, for $\vec{\theta}=(2\pi,0)$ it follows that

\beq      S_{\Box\;      Man}(\vec{\theta})=(2\pi,0){\bf       g}\left(
\begin{array}{c}    2\pi    \\     0     \end{array}     \right)     =
\frac{\beta_{Peter}}{2}(2\pi)^2. \label{metten}\eeq

\nin We define

\beq \frac{\beta_{Peter}}{2}(2\pi)^2 \stackrel{def}{=}a^2_{Peter}
\label{dist1dim}. \eeq

\[ \frac{\beta_{Paul}}{2}(2\pi)^2 \stackrel{def}{=}a^2_{Paul}. \]

\nin  where  $a_{Peter}$  and  $a_{Paul}$  denote   respectively   the
identification lattice constants in the respectively the  $Peter$  and
$Paul$ directions along the lattice.

Strictly speaking, two different metric tensors (\ref{metten})  may
correspond to the same physical action because there are different
ways of representing the same physics that are related by
(discrete)  isomorphic  mappings of  the  identification  lattice
into itself. But these discrete ambiguities do not affect  the  number
of (continuous) parameters needed - namely three for $U(1)^2$.

Using the covering group  $\br^3$  with  the  Manton-action  metric  and  the
embedded identification lattice, it is possible to depict, among other
things, the denumerable infinity of  compact  subgroups  of  $U(1)^3$.
Starting at the identity of the covering group  $\br^3$,  it  is  seen
that the identification lattice  induces  a  $U(1)$  subgroup  on  any
direction along which a lattice  point  is  encountered  at  a  finite
distance from the unit element of $\br^3$. Recall from above that  the
lattice constant $a_i$ is  inversely  proportional  to  the  coupling:
$a_i=2\pi \sqrt{\frac{\beta_i}{2}}$ $(i\in \{Peter, Paul, Maria \})$ .
So the larger the distance from the identity to the first  encountered
lattice point along some one-dimensional  subgroup  of  $U(1)^3$,  the
weaker is the coupling for this subgroup. In particular,  if  we  have
$a_i=a_{i\;crit}=2\pi   \sqrt{\frac{\beta_{i\;crit}}{2}}$   for    all
nearest neighbour lattice points, then all other  one-dimensional  subgroups
will be in a  Coulomb-like  phase  and  at  least
somewhat removed from the phase boundaries at which confinement  would
set in.

We want to let the $MPCP$ single out  the identification lattice $L$ -
which of course means a system of couplings - that
will bring the maximum number of phases together. We shall consider phases
corresponding to
subgroups of dimension ranging from 0 to 3  as  candidates  for  phases
that can meet at a multiple point.

If the $Peter$, $Paul$ and  $Maria$  directions  of  the  lattice  are
chosen to be mutually orthogonal (corresponding to a {\em cubic}
identification lattice), we have in this choice a proposal
for  a  multiple
point in the sense that, by  choosing  the  nearest  neighbour  lattice
constants to correspond to critical couplings, we have a Manton action
described by the geometry of this identification lattice such  that
various phases can be reached by  infinitesimal  changes  in
this lattice and  thereby in the  action  form.  By  such  infinitesimal
modifications, one can  reach  a  total  of  8  phases  with  confinement  of
8
subgroups. These subgroups are the ones  corresponding  to  directions
spanned by the 6 nearest neighbour
points  to,  for  example,  the  origin  (i.e.,  unit
element) of the orthogonal lattice: 1 zero-dimensional subgroup  (with
the Manton action, we do not  get  discrete  subgroups  confining),  3
one-dimensional  subgroups,  3  two-dimensional   subgroups   and   1
three-dimensional  ``subgroup''  (i.e.,  the  whole   $U(1)^3$).
For the choice of the orthogonal lattice, the action (\ref{firstc}) is
additive (i.e., without interactions)  in  the  $Peter$,  $Paul$,  and
$Maria$ terms and the diagonal coupling  is  multiplied  by  the  same
factor 3  as  for  the  non-Abelian  couplings  (see  (\ref{diagcoup})
above). However, as already mentioned, an additive action is without
interaction terms. These are important for
the $U(1)$ diagonal coupling.

It turns out that we can get a larger number of phases to convene at the
multiple point using a {\em hexagonal} lattice. Really this refers to
a special way of having interaction terms of the type $F_{\mu\nu}^{Peter}
F^{\mu\nu\;Paul}$ in such a way that there is an abstract symmetry similar
to that of a hexagonal lattice.
The hexagonal identification lattice results in a better implementation
of the $MPCP$. With the hexagonal  choice of lattice, it  is  possible
with infinitesimal departures from a lattice  with  critical  distance
to the nearest neighbours to provoke any one of
12  different  phases  in the ``volume'' approximation (after some slight
extra modifications; see Section (\ref{grvolapprox})
below) or 15 different phases in  the ``independent monopole'' approximation
(Section (\ref{moncon}) below):
{\em one} phase corresponding to confinement of the zero-dimensional subgroups,
{\em six} phases corresponding to confinement of one-dimensional subgroups,
{\em four} phases ({\em seven} in the ``independent monopole'' approximation)
corresponding to confinement of two-dimensional  subgroups
and {\em one}  phase  corresponding  to  confinement  of  the  whole
three-dimensional
$U(1)^3$.  The  choice  of  the  hexagonal  lattice  obviously  better
satisfies the $MPC$ principle. The fact  that  the  hexagonal  lattice
introduces  interactions  between  the  $Peter$,  $Paul$  and  $Maria$
degrees of freedom in the  Lagrangian  is  not  forbidden  for  $U(1)$
contrary to the situation for  the  non-Abelian  couplings  where  such
mixed terms in the Lagrangian would not be gauge invariant (unless they
were of fourth order or higher).

Originally the hexagonal identification lattice
was invented as a way of optimally realizing the multiple point criticality
idea for $U(1)^3$ and its {\em continuous} subgroups.
But we should also endeavour
to have phases confined alone w.r.t. {\em discrete} Abelian subgroups in
contact with the multiple point.
However, it is {\em a priori} not obvious that this hexagonal
identification lattice can be used for implementing the multiple
point criticality principle in the case of the discrete subgroups $\bz_N$
of $U(1)^3$
which,  according to the $MPCP$ should also be present at the multiple point.
For example, it seems unlikely that subgroups of $\bz_2^3$ can in analogy
to the $6+4+1+1=12$ continuous subgroups $U(1)^3$ (in the hexagonal scheme)
separately confine at the multiple point. The reason is that $\bz_2$
does not have sufficiently many
conjugacy classes so that the subgroups of $\bz_2^3$ can have a
generic multiple point at which 12 phases convene inasmuch as $\bz_2^3$
has only 8 elements and consequently only 8 conjugacy classes\footnote{
By including action terms involving several plaquettes it would in principle
be possible to have an action parameter space of dimension high enough to
have a generic confluence of the 12 phases each which is partially confined
w.r.t. a different discrete subgroup of $\bz_2^3$. However,
even assuming that our $MPCP$ were correct,
it might not be sufficiently favourable for Nature to implement it
to this extreme.}. Consequently, at most 8 phases can convene at a generic
multiple point if we restrict ourselves to  single
plaquette action terms and only allow confinement of $\bz_2^3$ and
subgroups thereof.

In general, having a phase for a gauge group $G$ that confines {\em alone}
along an (invariant) subgroup
$H$ requires that the distribution of elements along $H$ is rather
broad {\em and} that the cosets of the factor group $G/H$ {\em alone}
behave in a  Coulomb-like fashion which most often means that the distribution
of these cosets must
be more or less concentrated about the coset consisting of
elements identified with the identity.

Let us think of the hexagonal
identification lattice for $U(1)^2$ (the latter for the sake of
illustration instead of $U(1)^3$)
that is spanned by the variables $\theta_{Peter}$
and $\theta_{Paul}$ say. In the most general case, the action for a $U(1)^2$
gauge theory could be taken as an infinite sum of terms of the type

\beq  a_{nm}cos(n\theta_{Peter}+m\theta_{Paul}) \label{genact}  \eeq

Let us enquire as to what sort of terms could be used to attain criticality
for $\bz_{2\;Peter}\times \bz_{2\;Paul}\subset U(1)^2$ itself as well as for
subgroups
of $\bz_{2\;Peter}\times \bz_{2\;Paul}\subset U(1)^2$. Denote elements
of $U(1)^2$ as $(\theta_{Peter},\theta_{Paul})$ and use additivity in the
Lie algebra as the composition rule:

\beq (\theta_{1\;Peter},\theta_{1\;Paul})\circ
(\theta_{2\;Peter},\theta_{2\;Paul})
=(\theta_{1\; Peter}+\theta_{2\;Peter},\theta_{1\;Paul}+\theta_{2\;Paul}). \eeq

\nin Relative to the identity $(0,0)$, the elements of
$\bz_{2\;Peter}\times \bz_{2\;Paul}\subset U(1)^2$ (each of which
constitutes a conjugacy class) are  $(0,\pi)$, $(\pi,0)$, and $(\pi,\pi)$
(assuming a $2\pi$ normalisation).
Note that the terms in (\ref{genact}) having {\em even} values of both
$m$ and $n$ cannot be used to suppress the probability density at nontrivial
elements of $\bz_{2\;Peter}\times \bz_{2\;Paul}$ relative to the identity
element $(0,0)$; such {\em even} $n$ and {\em even} $m$ terms of (\ref{genact})
therefore
leave $\bz_{2\;Peter}\times \bz_{2\;Paul}$ and its subgroups totally confined.

Note however by way of example that {\em all} terms of (\ref{genact}) with {\em
odd} $n$
and {\em even} $m$ contribute to the suppression of the element
$(\pi,\theta_{Paul})\in \bz_{2\;Peter}\times \bz_{2\;Paul}$ relative to the
element
$(0,\theta_{Paul})\in \bz_{2\;Peter}\times \bz_{2\;Paul}$
(where $\theta_{Paul}\in \bz_{2\;Paul}$ can be anything) and can
therefore be used to render
the subgroup $\bz_{2\;Peter}$ critical (while the distribution over the
elements of the subgroup $\bz_{2\;Paul}$ is flat for any element of
$\bz_{2\;Peter}$ which means that $\bz_{2\;Paul}$ is left totally confined).
We observe that while all such odd-$n$ even-$m$ terms

\beq n=2p+1 \mbox{ for } p\in \bz \label{pee} \eeq

\nin suppress the probability density at
$(\pi,\theta_{Paul}\in \bz_{2\;Paul})$ relative to
$(0,\theta_{Paul}\in \bz_{2\;Paul})$, these odd-$n$ terms
also concentrate probability
density at $p$ different maxima along $U(1)_{Peter}
\setminus \bz_{2\;Peter}$; i.e., at
elements
$(0 <\theta_{Peter} < \pi, \theta_{Paul}\in \bz_{2\;Paul})$.
However these $p$ extra maxima in probability are not ``noticed''
by $\bz_{2\;Peter}\times \bz_{2\;Paul}$ and its subgroups because such
maxima are located at elements of
$U(1)_{Peter}\times U(1)_{Paul}$ that do not coincide with
elements of $\bz_{2\;Peter} \times \bz_{2\;Paul}$. The point to be gleaned
from this example is that for the purpose of rendering the $\bz_{2\;Peter}
\in U(1)_{Peter}\times U(1)_{Paul}$ \dofx critical, we can do the job with
any {\em one} representative from among the infinite number of terms of
(\ref{genact}) having coefficients $a_{nm}$ with $n$ odd and
$m$ anything.
We can therefore make the choice $n=1$ without loss of generality.
This choice will also be seen to be a convenient way to approximately
decouple the action parameters relevant to
\dofx corresponding to continuous subgroups of
$U(1)_{Peter}\times U(1)_{Paul}$
and the \dofx corresponding to discrete subgroups of
$U(1)_{Peter}\times U(1)_{Paul}$.

Generalising the above example, we can enumerate a choice for the
smallest set of parameters $a_{nm}$
in (\ref{genact}) that permits us maximal freedom in trying to get
partially confining phases w.r.t. subgroups of $U(1)_{Peter}\times U(1)_{Paul}$
(including $\bz_{2\;Peter} \times \bz_{2\;Paul}$ and subgroups thereof)
to convene at the multiple point. Such a choice is conveniently made as
follows:

\begin{itemize}
\item confinement alone along $\bz_{2\;Peter}$ and a peaked Coulomb-like
distribution of the cosets of the factor group
$(\bz_{2\;Peter}\times \bz_{2\;Paul})/\bz_{2\;Peter}$ is achieved using any
term $a_{nm}$ of (\ref{genact}) for which with $n$ is {\em even} and $m$ is
{\em odd}; we choose $a_{01}\stackrel{def}{=}\beta_{Paul}$ and set all other
$n$-even, $m$-odd terms equal to zero.

\item confinement alone along $\bz_{2\;Paul}$ and a peaked Coulomb-like
distribution of the cosets of the factor group
$(\bz_{2\;Peter}\times \bz_{2\;Paul})/\bz_{2\;Paul}$ is achieved using any
term $a_{nm}$ of (\ref{genact}) for which $m$ is {\em even} and $n$ is
{\em odd}; we choose $a_{10}\stackrel{def}{=}\beta_{Peter}$ and set all other
$m$-even, $n$-odd terms equal to zero.

\item confinement alone along\footnote{We want
the anti-diagonal subgroup if we want an analogy to the third direction
in the hexagonal identification lattice; however for $\bz_2$ the
anti-diagonal subgroup  coincides with the diagonal subgroup
$\{(1,1),(-1,-1)\}$. Here we have changed to a notation for the elements of
$U(1)_{Peter}\times U(1)_{Paul}$ corresponding to a multiplicative composition
of group elements.}
$\{(1,1),(-1,-1)\}
\subset \bz_{2\;Peter}\times \bz_{2\;Paul}$ and a peaked Coulomb-like
distribution of the cosets of the factor group
$(\bz_{2\;Peter}\times \bz_{2\;Paul})/\{(1,1),(-1,-1)\}$ is achieved using any
term $a_{nm}$ of (\ref{genact}) for which both with $n$ and $m$ is
{\em odd}; we choose $a_{11}\stackrel{def}{=}\beta_{interaction}$ and set all
other $n$-odd, $m$-odd terms equal to zero.

\end{itemize}

\nin This gives us effectively three free parameters with which we can try
to bring  {\em discrete} \pcps together at the multiple point. This choice
using

\beq a_{nm}=a_{10}\stackrel{def}{=}\beta_{Peter},\eeq

\[ a_{nm}=a_{01}\stackrel{def}{=}\beta_{Paul}\] and
\[ a_{nm}=a_{11}\stackrel{def}{=}\beta_{interaction}\] is the most
smooth choice. Other choices for action terms with
$n$ and/or $m$ odd could potentially result in additional maxima
in the probability density that are not centred at elements of
$\bz_{2\;Peter}\times \bz_{2\;Paul}\subset U(1)_{Peter}\times U(1)_{Paul}$
(e.g., for $p\neq 0$ in (\ref{pee})). But these additional maxima would
effectively not influence the distribution of continuum
\dofx as such additional maxima can easily  be suppressed by (dominant)
$n$-even, $m$-even action terms everywhere on $U(1)_{Peter}\times U(1)_{Paul}$
except at elements of $\bz_{2\;Peter}\times \bz_{2\;Paul}$ Representing these
dominant $n$-even, $m$-even action terms by the smoothest ones corresponds to
using just three non-vanishing
parameters to adjust the continuum \dofx along subsets of
$U(1)_{Peter}\times U(1)_{Paul}$:

\beq a_{20}\stackrel{def}{=}\gamma_{Peter}. \eeq
$$ a_{02}\stackrel{def}{=}\gamma_{Paul}$$ and
$$ a_{22}\stackrel{def}{=}\gamma_{interaction}.$$
So we end up with six parameters
where the three $n$-even, $m$-even ones can be used
to bring phases confined w.r.t.  continuous subgroups
of $U(1)_{Peter}\times U(1)_{Paul}$ together at the multiple point.
These parameters are approximately independent of
the parameters $\beta_{Peter}$, $\beta_{Paul}$ and $\beta_{interaction}$
than can be used to bring phases confined w.r.t.  discrete subgroups
of $U(1)_{Peter}\times U(1)_{Paul}$ together at the multiple point.
We end up with an action $S$

\beq S=\gamma_{Peter}\cos(2\theta_{Peter})+\beta_{Peter}\cos\theta_{Peter}+
       \gamma_{Paul}\cos(2\theta_{Paul})+\beta_{Paul}\cos\theta_{Paul} + \eeq
$$    +\beta_{interact}\cos(\theta_{Peter}+\theta_{Paul})
     +\gamma_{interaction}\cos (2(\theta_{Peter}+\theta_{Paul})).  $$

\nin Let us assume that $\gamma_{Peter}$, $\gamma_{Paul}$ and
$\gamma_{interaction}$ have been chosen so as to
bring $U(1)^2$ and the continuous subgroups
of $U(1)^2$ together at the multiple point.
This leaves three approximately independent parameters that can be used
as coefficients to plaquette action terms defined on
$\bz_2\times \bz_2$ and its subgroups. These parameters can be adjusted so as
to bring phases confined w.r.t. subgroups of $\bz_2\times \bz_2$
together at the multiple
point. That we have three (effectively) independent parameters up to a
constant action term
is in accord with $\bz_2\times \bz_2$ having just four elements (i.e., four
possible
conjugacy classes). With three parameters we can have a generic multiple point
at which four phases convene. However, the number of {\em possible} different
phases (regardless of whether they can all meet at the multiple point)
obtainable by varying the parameters
$\beta_{Peter}$, $\beta_{Paul}$, and $\beta_{interact}$ is five.
Two of the five possible phases
correspond to total confinement and
totally Coulomb-like behaviour for $\bz_{2\;Peter}\times \bz_{2\;Paul}$;
the remaining three possible phases correspond to confinement along
1-dimensional\footnote{Strictly speaking,
$\bz_{2\;Peter}\times \bz_{2\;Paul}$ and subgroups hereof are of course all
0-dimensional; when we talk about ``1-dimensional subgroups of discrete
groups'' we mean the (measure zero) sets that coincide with elements of,
e.g., the 1-dimensional subgroup $U(1)_{Peter}\in U(1)^3$.}
subgroups of $\bz_{2\;Peter}\times \bz_{2\;Paul}$ enumerated above in
connection with our procedure for choosing
$\beta_{Peter}$, $\beta_{Paul}$, and $\beta_{interact}$.
However, only two of these three phases with
confinement solely along 1-dimensional
subgroups can convene at a (generic) multiple point. This is different from
the
situation for $U(1)^2$ (i.e., for the continuum); it is
shown elsewhere that in this case, all three phases
that are confined solely along a
1-dimensional subgroups can convene at a single (generic) multiple point.

On the other hand,
for $\bz_N$ (with $N>3$) there are enough conjugacy
classes (and thereby potential action parameters) so that for any of the three
directions $\theta_{Peter}$, $\theta_{Paul}$ and
$\theta_{Paul}-\theta_{Peter}$ in $\bz_3$
we can independently choose to have a somewhat flat distribution
of group elements (corresponding to confinement-like behaviour)) along for
example the
$\theta_{Peter}$ direction
while at the same time having
a peaked distribution of the cosets of the factor group
$(\bz_{N\;Peter}\times \bz_{N\;Paul})/\bz_{N\;Peter}$ (corresponding to
Coulomb-like behaviour for these \dof).
This is of course just the \pcp confined w.r.t. $\bz_{N\;Peter}$. It turns
out that also for $\bz_3$, this is in principle at least just barely possible.

For $U(1)^3$,
an analogous difference between the subgroups $\bz_2^3$ and $\bz_N^3$
($N>3$) is found.
Of the six possible 1-dimensional subgroups of $\bz_2^3$, only three of the
corresponding \pcps
can convene at any (generic) multiple point
as compared to the situation for $U(1)^3$ where six such phases can convene
at the multiple point.

According to the multiple point criticality
principle, we should determine the critical $U(1)$ coupling corresponding to
the multiple point in a $U(1)^3$ phase diagram where a maximum number of
\pcps convene. This also
applies of course to the possible 1-dimensional
discrete subgroups. We deal with these latter subgroups by using an
appropriate correction to the continuum $U(1)$ coupling in a later Section.

Beforehand, it is not known whether it is even numerically possible
to have criticality for the discrete subgroups using the hexagonal symmetry
scheme for the couplings.
At least in the case of $\bz_2$, the subgroups in some directions are lacking
because there are not enough action parameters to bring them all to the
multiple point.
Hence the $\bz_2$ correction  should only have a weight
reflecting the contribution from the fraction of these 1-dimensional
discrete subgroups that (alone) can be confined at the
multiple point. For $\bz_2^3$,
it turns out that only one half (i.e., three out of six) of the hexagonal
nearest neighbour
1-dimensional  subgroups can convene at a (generic) multiple point.
In the
boundary case of $\bz_3^3$, it is not entirely clear as to whether
the contribution should also be
reduced by some factor.

On the other hand, for $\bz_N^3$ ($N>3$), it is not strictly excluded
to have the six 1-dimensional phases at a (generic) multiple point that
correspond to the six analogous phases of  $U(1)^3$.
This  reflects the fact that for $\bz_N^3$
with $N>3$,
there are sufficiently many conjugacy classes\footnote{Strictly speaking,
this is also true for $\bz_3$: there are eight conjugacy classes
corresponding to the eight elements of $\bz_3$. However, it can hardly be
useful to have
separate action terms for elements $g\in \bz_3^2$ and $-g\in \bz_3^2$. So for
the purpose of provoking different \pcps independently, there are effectively
only four conjugacy classes.  But four action parameters are in
principle at least just sufficient
to bring $1+3+1=5$ phases together at a generic multiple point.}
so that the hexagonal identification lattice that is so efficient in getting
phases corresponding to continuous
subgroups of $U(1)^3$ to convene at the multiple point can presumably also
bring the analogous phases of discrete subgroups  $\bz_N^3$ ($N>3$)
together at the multiple point.

When we talk about ``contributions'' of $\bz_N$ subgroups to $\frac{1}{g^2}$,
we are anticipating that in a later Section, we shall make approximate
corrections for our having initially
neglected that there should also be phases convening at the multiple point
for which the various discrete invariant subgroups are alone confining while
the corresponding continuous factor groups behave in a Coulomb-like fashion.
The correction
procedure that we use results in small corrections to the critical continuum
couplings that we loosely refer to as ``contributions'' to the inverse squared
couplings from $\bz_2$, $\bz_3$, etc.

In summary, it is possible for $\bz_N$ discrete subgroups of large enough $N$
to
realize all possible combinations of phases for the (nearest neighbour)
1~-~dimensional subgroups of the hexagonal identification lattice coupling
scheme. These \pcps should also convene at the multiple point; we deal with
this requirement in an approximate way in a later Section by making a
correction to $\frac{1}{g^2}$ for discrete subgroups $\bz_N$ with various
values of $N$.
The  result of the discussion above is that the approximate
correction that will be made to $(\frac{1}{g^2})_{mult\;point}$
coming from taking into account
that we also want to have \pcps w.r.t. $\bz_2^3$ at the multiple point is
reduced
by a factor $\frac{3}{6}=\frac{1}{2}$ relative to the analogous correction
for $\bz^3_N$ ($N> 3$). It may also well be that the contribution in the
marginal case of $\bz_3$ should also be reduced by some factor. These
considerations will be incorporated into the presentation
of our results.

\subsection{Mapping out the phase diagram for $U(1)^3$: approximative
techniques}

\subsubsection{Monopole condensate approximation - outline of procedure}
\label{moncon}
The philosophy of the first approximation to be used to estimate which
phase is obtained for given parameters is that the decisive factor  in
distinguishing the Coulomb-like phase (or Coulomb-like behaviour of some
of the degrees of freedom) from  the  confinement  phase  is  whether
quantum  fluctuations  are  such  that  the  Bianchi  identities   are
important or essentially irrelevant in introducing correlations between
plaquette variables.

That  is  to  say  we  imagine  that  the  phase  transition   between
a ``Coulomb'' and confining phase -  as  function  of  the
parameters $\beta$ - occur when the fluctuations of the  plaquette  variables
take such values that the fluctuation of the convolution of the number
of plaquette variable distributions (coinciding  with  the  number  of
plaquettes bounding a 3-cube - e.g., six  for  a  hyper-cubic  lattice)
become just large enough so as to be essentially spread out over the
whole  group
(or over the elements within the cosets of a factor group)
in question and thereby rendering Bianchi identities
essentially irrelevant.

The idea behind this philosophy is that when the fluctuations  are  so
large that a naive (i.e. neglecting Bianchi constraints) convolution
of  the  6  plaquettes  making  up  the
boundary  of  a  3-cube  fluctuates  over  the  whole  group  (leading
essentially to the Haar measure  distribution),  the  Bianchi-identity
is then assumed to be essentially  irrelevant  in  the  sense  that
each  plaquette
fluctuates approximately independently of the  other  plaquettes  that
form the boundary of a 3-cube. In this situation there is essentially
no (long range) correlations. This is  of  course  the  characteristic
feature of a confining phase.

If, however, fluctuations of the convolutions of  plaquettes  variable
distributions $e^{S_{\Box}}$ for the six plaquettes bounding a  3-cube
do not cover the whole group, the Bianchi identities are important in
the  sense  that  the  constraint  that  these  impose  leads   to   a
correlation  of  plaquette  variable   fluctuations   over   ``long''
distances  (i.e.,  at  length  scales   of at least   several   lattice
constants).  Such  ``long''  range  correlations  are  taken  as   the
characteristic feature of a Coulomb-like behaviour.

The idea of phase determination according to whether the  fluctuations
in plaquette variables are  small  enough  so  that  Bianchi  identity
constraints can introduce ``long'' range correlations or  not  can  be
translated into a  lattice  monopole  scenario:  a  Coulomb-like  phase
corresponds to a scarcity of monopoles while the vacuum of a confining
phase is copiously populated by monopoles. For a single  $U(1)$  gauge
group, a monopole (or rather the cross section in the time track of a
monopole)  is just  a  3-cube  for  which  the  values  of  the
bounding  plaquette  variables  -  defined say by  the  convention   that
Lie-algebra (angle) variables take values in the interval $[-\pi,\pi)$
- have fluctuations large enough so as to get back to the unit element
by first adding up to a circumnavigation  of  the  whole  group.
Such a traversal of the whole $2\pi$ length of the group  as  the  way
the Bianchi identity is realized is tantamount to  having  a  lattice  artifact
monopole. The  confinement  phase  is  characterised  by  the  copious
occurrence of such monopoles.

The  case  where  the  gauge  group  is  $U(1)^3$  is  slightly   more
complicated. As seen above, the group $U(1)^3$ can be  thought  of  as
the cosets of the group $\br^3$ modulo an identification lattice.  A
unique assignment of an  element  of  the  group  $\br^3$  to  each
$U(1)^3$-valued plaquette requires a convention which we  take  to  be
the choice of that element among the  coset  representatives having the
shortest distance to the  zero-element  of  $\br^3$.  With  such  a
convention, we can, for any 3-cube, now ask if the  sum  of  the  $\br^3$
representatives for the surrounding plaquette variables typically
add up to the unit element (as is characteristic  of  the  Coulomb-like
phase) or instead add up to one of  the  nontrivial  elements  of  the
identification-lattice (as is characteristic  of  a  confining  phase)
corresponding respectively to not having  a  monopole  or having  a  monopole
with  some   $N_{gen}$-tuple   of   magnetic    monopole    charges
$2\pi(n_{Peter},   n_{Paul},   n_{Maria})$    $(n_{Peter},    n_{Paul},
n_{Maria}\in \bz)$.

Monopoles come  about  when  the  Bianchi  identities  (one  for  each
of the  $N_{gen}$  $U(1)$  subgroups  labelled   by   names   $``Peter"$,
$``Paul"$ and $``Maria"$) are satisfied by having the values of  the
plaquette variables of a 3-cube add up to a lattice point  other  than
that corresponding to the  identity  element  of  $\br^3$.  In  other
words, a monopole is a  jump  from  the  origin  of  the  $\br^3$
identification lattice to another point of the identification  lattice
that takes place when values  of  the  variables  for  the  plaquettes
surrounding a 3-cube add up to a nonzero multiple  of  $2\pi$  for  at
least one of the $N_{gen}=3$  $U(1)$'s  of  $U(1)^3$  as  the  way  of
fulfilling the Bianchi identities.

Having a  phase  in  which  for  example a one-dimensional subgroup
 -  $U(1)_{Peter}$ say - is  confined
corresponds to having, statistically speaking, an abundance  of cubes
of  the  lattice  for  which   the  monopole   charge
$2\pi(n_{Peter},  n_{Paul},  n_{Maria})$  is  typically  $\pm 2\pi(1,0,0)$
but (depending on couplings) also with less frequent occurrences of the
monopole  charges  $\pm 2\pi(2,0,0)$,  $\pm 2\pi(3,0,0),\cdots$  as  well   as
only occasional monopoles with $n_{Paul}\neq 0$ and $n_{Maria}\neq 0$. Which
phase is realized is  determined  of  course  by  the  values  of  the
couplings. We recall that  the  information  about  the  couplings  is
``built  into''  the  distance   between   lattice   points   of   the
identification   lattice.   Confinement   along   for   example    the
$U(1)_{Peter}$ subgroup corresponds to having  a  less  than  critical
distance between nearest  neighbour  lattice  points  lying  along  the
$U(1)_{Peter}$ subgroup. It is also possible to have confinement along
two dimensional subgroups (including the orthogonal two-dimensional subgroups)
and the entire (three-dimensional) $U(1)^3$.

We want to use the monopole condensate  model  to  construct  a  phase
diagram for $U(1)^3$. A  confining
subgroup is generated in a direction along which the  spacing  between
nearest (identification lattice) neighbours is  smaller  than  that
corresponding  to  critical
coupling values. In general, the  critical
coupling for a given subgroup depends on which  phases  are
realized for the remaining $U(1)$ degrees of freedom.
For example, confinement for a given one dimensional subgroup of $U(1)^3$
occurs for
a  weaker  coupling  when
one or both of the other $U(1)$ \dofx are  confined than when both of these
other
\dofx are in Coulomb-like phases. In  the
roughest  monopole  approximation, these interactions
between phases is ignored. Accordingly, the critical distance in one
direction is taken to be independent  of
the distance between  neighbouring  identification  lattice  points  in
other directions. This approximation is appropriate if we take the transition
as being second order
because the fluctuation pattern then goes smoothly through
the transition so that the transition for one subgroup does not abruptly
change the fluctuation pattern significantly for another subgroup.

In this approximation, seeking the multiple point  is  easy.  Multiple
point criticality is achieved simply by having the  critical  distance
between identification lattice points in all nearest  neighbour  directions.
In  this
approximation, the number of phases convening at the multiple point is
maximised by having the largest possible number  of  nearest  neighbour
directions (i.e., maximum number of one-dimensional  subgroups).  This
just  corresponds  to  having  the  tightest   possible   packing   of
identification lattice points. In three dimensions  (corresponding  to
$N_{gen}=3$) tightest packing is attained using a  hexagonal  lattice.
The  generalisation  to   $U(1)^3$   for   the   coordinate choice  of
(\ref{coord2}) is that the points  to  be  identified  with  the  unit
element are

\beq 2\pi(n_{Peter},n_{Paul},n_{Maria}) \;\;\;
(n_{Peter}, n_{Paul}, n_{Maria} \in {\bf Z}). \label{coord3} \eeq

\nin and with this coordinate choice the value of the Manton action at
the multiple point is given by

\beq S_{\Box\;Man}(\vec{\theta}(\Box))= \theta^i(\Box)g_{ik}\theta^k(\Box)
\;\; (i,k \in \{Peter, Paul, Maria\}) \label{manmult} \eeq

\nin where

\beq {\bf g}= \frac{\beta_{crit}}{2}\left( \begin{array}{ccc}
1 & \frac{1}{2} & \frac{1}{2} \\
\frac{1}{2} & 1 & \frac{1}{2} \\
\frac{1}{2} & \frac{1}{2} & 1 \end{array} \right). \label{manten} \eeq

Here we review briefly the symmetry properties
of the hexagonal lattice in the metric of (\ref{manten}).
A point of the lattice has 12
nearest neighbours that define a cub-octahedron. Under an isometric
transformation that leaves the identification lattice invariant (as a set),
one of the 12 nearest neighbours be transformed into another one
in 12 ways.
Moreover, the 4 points adjacent to
any one of the 12 nearest neighbour points must be transformed into each other
in 4 ways.
In this way we account for the $4\times 12$ operations that exhaust
the allowed symmetry operations of the point group characterising the symmetry
of the hexagonal lattice.

\begin{figure}
\centerline{\epsfxsize=\textwidth \epsfbox{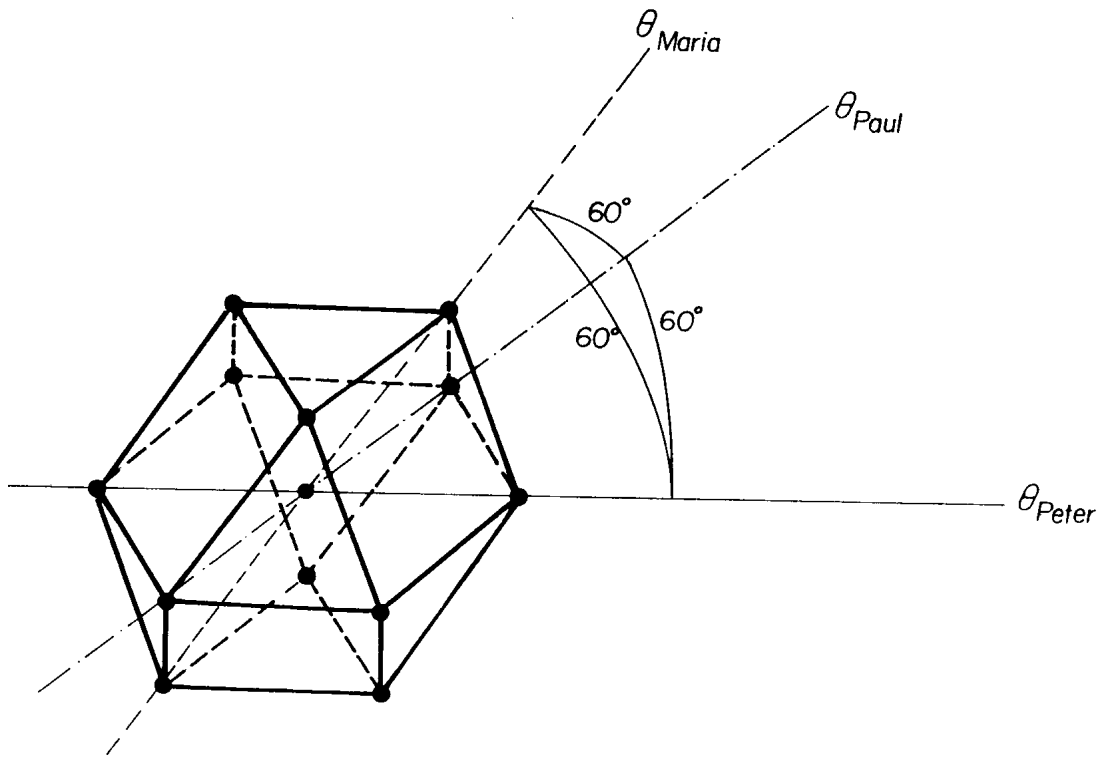}}
\caption[figcuboct]{\label{figcuboct} The nearest neighbours of a chosen point
in the identification lattice form a cub-octahedron. The metric used is that
which corresponds to taking the squared distance as the Manton action.}
\end{figure}

For the purpose of elucidating the symmetries of the hexagonal identification
lattice, it is useful to introduce an extra (superfluous) coordinate
$\theta_4$. First let us  rewrite $S_{\Box\;Manton}$
in (\ref{manmult})
as

\beq
\vec{\theta}^T{\bf g}\vec{\theta}=
\vec{\theta}^T\left\{\frac{\beta}{2}\left(\left(
\begin{array}{ccc} 1/2 & 0 & 0 \\ 0 & 1/2 & 0 \\ 0 & 0 & 1/2 \end{array}\right)
+\frac{1}{2} \left(\begin{array}{ccc} 1 & 1 & 1 \\ 1 & 1 & 1 \\ 1 & 1 & 1
\end{array}\right)\right)\right\}\vec{\theta} = \eeq

\[ = \frac{\beta}{4}(\theta_1^2+\theta_2^2+\theta_3^2 +
(\underbrace{-\theta_1-\theta_2-\theta_3}_{\stackrel{def}{=}\theta_4})^2)
=\frac{\beta}{4}\sum^4_{i=1} \theta_i^2
\]
 \nin where $\theta_1=\theta_{Peter},\cdots, \theta_3=\theta_{Maria};
\theta_4=-\sum_{i=1}^3\theta_i$.

In this coordinate system with the superfluous coordinate $\theta_4$, we
have the constraint

\beq \sum_{i=1}^4 \theta_i=0 \eeq

\nin and the hexagonal lattice is  characterised as the set of points with
coordinates

\beq (\theta_1,\theta_2,\theta_3,\theta_4)\in 2\pi{\bf Z}^4. \eeq

In this notation, it is apparent that the symmetry group
for the lattice and the
metric consists of the permutations combined with or without a
simultaneous sign shift of all four coordinates.

Each of the 12 nearest neighbours to any site of the identification lattice
(e.g. the group identity) have, in the 4-tuple coordinate notation, just
two non-vanishing coordinates (that sum to zero). The 1-dimensional subgroups
 correspond to the 6 co-linear pairs of these  12 nearest neighbours.

The
2-dimensional subgroups are of two types. One type, of which there are 4,
are spanned by the identity and any (non-co-linear) pair of the
12 nearest neighbour sites
that have a common non-vanishing coordinate.
A given subgroup of this type contains 6
nearest neighbour sites positioned at the corners of a hexagon; all 6 such
sites of a given 2-dimensional subgroup of this type have a vanishing
coordinate in common; e.g., the 6 nearest neighbours with a ``0''
for the first coordinate belong to the same 2-dimensional subgroup of this
type.
              That there are four such subgroups follows from the fact that
there are 4 possibilities for having a common vanishing coordinate in the
4-tuple notation. The other type of 2-dimensional subgroups - there
are 3 mutually orthogonal such subgroups - are each spanned by 2 pairs of
nearest neighbour sites where the two sites of each such pair have no common
non-vanishing coordinates. There are 3 such pairs:

\beq \begin{array}{c} (\pm 2\pi,0,\mp 2\pi,0) \\ (0,\pm 2\pi,0,\mp 2\pi)
\label{p1}\end{array}\eeq

\beq \begin{array}{c} (\pm 2\pi,\mp 2\pi,0,0) \\ (0,0,\pm 2\pi,\mp 2\pi)
\label{p2}\end{array}\eeq

\beq \begin{array}{c} (\pm 2\pi,0,0,\mp 2\pi) \\ (0,\pm 2\pi,\mp 2\pi,0).
\label{p3}\end{array}\eeq

\nin Any of the 3 pairs (\ref{p1}), (\ref{p2}), (\ref{p3}) span
one of the $\left(\begin{array}{c} 3 \\ 2 \end{array}\right)=3$ orthogonal
2-dimensional subgroups.

The 3-dimensional ``subgroup'' (which of course is the whole $\br^3$ space)
corresponds in the 4-tuple notation to the
(whole) hyper-plane specified by

\beq \{\vec{\theta}| \sum_{i=1}^4\theta_i=0 \}. \eeq

\nin The 0-dimensional subgroup corresponds simply to the identification
lattice site that is chosen as the group identity.

\subsubsection{Group volume approximation}\label{grvolapprox}

In this approximation, which is an alternative to the monopole
approximation, we calculate  the
free energy as a function of the couplings for  each  phase  ansatz  (i.e.
each partially confining phase). The criterion for having a phase in contact
with the multiple point is that there  is  some  region  of  plaquette
action parameter space infinitesimally close to  the  multiple  point
where the corresponding free energy function is the most stable  (i.e., larger
than the free energy functions of all the other phases  that  meet  at
the multiple point). In previous work\cite{van,albu,nonabel},
we have derived an  approximate
expression\footnote{In obtaining this relation, we used Gaussian integrals
in the Lie algebra to approximate group integrals, the approximation of
independent plaquettes for the confined subgroup $H$ (i.e.,
Bianchi  identities  are  neglected), and a weak  coupling  mean  field
description  for the Coulomb phase \dofx $G/H$.} for the free energy
{\em per active link}\footnote{ For a 4-dimensional
hyper-cubic lattice, there are 3  active  links  per  site  (i.e.,  the
number of dimensions reduced by the  one  dimension  along  which  the
gauge is fixed) and 6 plaquettes per site. This  yields  2  plaquettes
per active link. So the quantity $\log Z$ per active site is the half of the
quantity $\log Z$ per plaquette.}. We used the notation
$\log Z_{H \triangleleft G}$  for the free energy function corresponding to
the phase for which $H$ is the largest confined invariant
subgroup of the gauge group $G$:

\beq (\log Z_{H\triangleleft G})_{\mbox{\tiny per active link}}=
\log\left [ \frac{(\pi/6)^{\frac{d_G}{2}}}{{\bf \beta}_G^{\frac{d_G}{2}}vol(G)}
\right ] +
\log \left [ \frac{(6\pi)^{\frac{d_H}{2}}}{{\bf \beta}_H^{\frac{d_H}{2}}vol(H)}
\right ]. \label{long1} \eeq

\nin where

\beq {\bf \beta}_{H}^{\frac{1}{2}dim\;H}\stackrel{def}{=}
\prod_i \beta_i^{\frac{1}{2}dim\;H_i} \eeq

\nin and the index $i$ runs over the Lie algebra ideals
\footnote{For example, for $H=SMG$, ${\bf \beta}_H^{\frac{1}{2}dim\;H}=
\beta_{U(1)}^{\frac{1}{2}}\beta_{SU(2)}^{\frac{3}{2}}
\beta_{SU(3)}^{\frac{8}{2}}$
and for $H=U(3)$, ${\bf \beta}_H^{\frac{1}{2}dim\;H}=
\beta_{U(1)}^{\frac{1}{2}}\beta_{SU(3)}^{\frac{8}{2}}$. Note that $vol(U(3))=
\frac{1}{3}vol(U(1))\cdot vol(SU(3))$ because $U(3)$ is obtained
by identifying the  3 elements of the
$\bz_3$ subgroup of the centre of $U(1)\times SU(3)$.}
of $H$.

Consider two partially confining phases in the case that  one  of
these is confined w.r.t to the invariant subgroup $H_I$ and the  other
is confined w.r.t. the invariant  subgroup  $H_J$.  At  any  point  in
parameter space  where  these  two  partially  confining  phases  meet
(including the multiple point of course) the condition to be satisfied
is $\log Z_{H_I\triangleleft G}= \log  Z_{H_J\triangleleft  G}$.  This
together with (\ref{long1}) leads to the following  condition  that  is
fulfilled at any point on the phase boundary separating these two phases:

\beq \log(6\pi)^{\frac{dim(H_J)-dim(H_I)}{2}}=\log
\frac{{\bf \beta}_{H_J}^{\frac{dim(H_J)}{2}}vol(H_J)}{{\bf
\beta}_{H_I}^{\frac{dim(H_I)}{2}}vol(H_I)}.
\label{long2} \eeq

\nin We  want  of  course  to  consider (\ref{long1}) in the   special
case   for   which
$G=U(1)^3$.

Using here a slightly different notation, designate by $\log Z_{H_n}$ the
free energy
per active link for  the
phase ansatz  for  which  one  of  the  above-mentioned  n-dimensional
subgroups $H_n$ of $U(1)^3$ ($dim(H_n)=n$; $n\in \{0,1,2,3\}$), is confining
and  the
factor group $U(1)^3/H_n$ behaves in a Coulomb-like way
($H_n$ could be one of the
1-dimensional      subgroups: e.g.,       $H_1=U(1)_{Peter}$
say). Let us denote by $a$ the lattice constant of the identification lattice.
Rewriting (\ref{long1}) and specialising to the case of  the  gauge
group $G=U(1)^3$ and $H_J=H_n$ reveals the dependence of  the
free energy per active link on the quantity log$a$:

\beq \log Z_{H_n}= C-(dim(U(1)^3)+dim(H_n))\log a \eeq

\nin where $dim(U(1)^3)=3$ and $dim(H_n)$ are respectively  the
dimension  of  the
gauge group (i.e., the $U(1)^3$ part of $SMG^3$) and the dimension of the
subgroup $H_n$  and  $C$  is  a
quantity that does not depend on the identification lattice constant $a$.
The slope of the various phase ans\"{a}tze is just

\beq \frac{d\log Z_{H_n}}{d\log a}=
-(dim(U(1)^3)+dim(H_n)). \label{deriv} \eeq

\nin Upon rewriting
(\ref{long2}), one  obtains  for  the  condition  defining  the  phase
boundary between the phase with confinement along the subgroup $H_n$
and the phase with confinement along $H_m$ the equation

\beq (6\pi)^{(dim(H_n)-dim(H_m))/2}=
\frac{(\frac{a^2}{2\pi^2})^{dim(H_n)/2}c_n(2\pi)^{dim(H_n)}}
{(\frac{a^2}{2\pi^2})^{dim(H_m)/2}c_m(2\pi)^{dim(H_m)}} \;\; (n,m\in
\{0,1,2,3\})
\label{cond} \eeq

\nin where the volume $vol(H_n)$ of the subgroup $H_n\subseteq U(1)^3$,
measured  in
the coordinate $\theta$, is

\beq  vol(H_n) = c_n (2\pi)^{dim(H_n)}. \label{cn}\eeq
The quantity $c_n$ is  a
factor associated with the subgroup $H_n$ that depends on the geometry
of the identification lattice.

As an example, consider  first  a  cubic  identification
lattice (actually we shall end up using an hexagonal  lattice  as  this
better satisfies the principle of multiple point criticality). For the
cubic lattice  with
$a=a_{1\;crit}\stackrel{def}{=}2\pi\sqrt{\frac{\beta_{crit}}{2}}$,  it  is
possible  to  have  the
confluence  of three  phases of the type  corresponding  to   1-dimensional
subgroups of $U(1)^3$ at a multiple point - namely those corresponding
to the 1-dimensional subgroups along the $Peter$, $Paul$, and  $Maria$
directions     of     the     lattice     having     $a_{1\;crit}=2\pi
\sqrt{\frac{\beta_{crit}}{2}}$ (the subscript ``1'' on $a_{1\;crit}$
denotes that it is a one dimensional subgroup that is critical).
Furthermore, in the case of the  cubic
identification lattice, it will be seen that phases corresponding to
all subgroups of $G=U(1)^3$
are simultaneously  critical   when   the identification  lattice
constant   $a=a_{1\;crit}=2\pi \sqrt{\frac{\beta_{crit}}{2}}$.
This  follows by observing  that  the
free energy $\log Z_n$ ($n\in \{0,1,2,3\}$) for the different ans\"{a}tze
(i.e., confinement along the various possible subgroups)
are equal for the same value of the identification lattice constant $a$
(i.e.,  for
$a=a_{1\;crit}$) because the constants $c_n$ in (\ref{cond})  are
independent of the dimension $dim(H_n)$  of  the  subgroup  (and  therefore
equal). Hence the condition (\ref{cond})  that  defines  the  boundary
between two partially confining phases is independent of dimension. This then
means that  for  the
unit cell of the cubic  identification  lattice,  all  the  quantities
$\log Z_n$ ($n\in  \{0,1,2,3\}$)  intersect  for  $a=a_{1\;crit}=2\pi
\sqrt{\frac{\beta_{crit}}{2}}=2\pi \sqrt{\frac{1.01}{2}}=4.465$. So the use
of the cubic identification lattice with $a=a_{1\;crit}$ shows that it
is possible to have a multiple point at which  8  partially  confining
phases are in contact: there  is  one  totally  confining
phase (corresponding to $H_3$), three phases corresponding to  three
2-dimensional subgroups $H_2$, three phases corresponding  to  three
1-dimensional subgroups  $H_1$,  and  a  totally  Coulomb-like  phase
corresponding to $H_0$. In particular, the coupling corresponding to
the diagonal subgroup of $U(1)^3$ (in the first approximation, this is
the coupling that we  identify  with  the continuum  $U(1)$
coupling) is down by a factor  $\sqrt{3}$  relative  to  the  critical
coupling for a $U(1)$ lattice gauge theory. This  follows  because  the
inverse of the ratio of the length of  the  diagonal  to  the critical lattice
constant  is $\sqrt{3}$. Phenomenologically, a  factor
of roughly $\sqrt{6}$ rather than $\sqrt{3}$  is  needed  so  we  must
conclude that for the $U(1)$ continuum coupling, the prediction of the
multiple point criticality  principle  using  a  cubic  identification
lattice is at odds with experiment.

However,   the  multiple  point  criticality
principle states that we should  seek the values of  the  continuum
$U(1)$ coupling at a point in parameter space at which a {\em maximum}
number of phases come together. We have already seen that
         for       a hexagonal identification  lattice  in  the  covering
group $\br^3$ of the gauge group $U(1)^3$,
we can, in terms of the 12 nearest neighbours of a site in the
hexagonal identification lattice, identify a  total
of 15 subgroups corresponding to 15 partially confining  phases. Even though
we shall discover in the sequel that 3 of these 15 \pcps - the 2-dimensional
``orthogonal'' phases given by (\ref{p1}-\ref{p3}) - are not
realistically realizable in the volume approximation inasmuch as these phases
are ``pushed'' too far away from the multiple point in the volume
approximation, there
remains 12 partially confining phases that can be made to convene at the
multiple
point. This is, in view of the multiple point criticality principle,
an improvement upon  the total of 8 phases that can be realized at the
multiple point in the case of the cubic identification lattice.

It will be  seen  that  the  price  we  must  pay  for realizing these 12
remaining partially confining phases at the multiple point in the case of
the hexagonal identification lattice instead  of  the  8  partially
confining phases of the cubic identification lattice
is that these 12 phases no longer come together exactly at a common  value  of
the identification lattice constant $a$ if we use a pure Manton action
(\ref{min}).

For the hexagonal identification lattice, the problem is that when the
lattice constant $a$ is chosen so that $a=a_{1\;crit}
\stackrel{def}{=}2\pi\sqrt{\frac{\beta_{crit}}{2}}$ corresponding to
criticality for the 1-dimensional subgroups,  this  choice  fixes  the
values of the couplings for the  2-  and  3-dimensional  subgroups  at
sub-critical values. For example, for
$a=a_{1\;crit}$, the free energy functions $\log  Z_0$  and  $\log
Z_1$ are equal corresponding to the  coexistence  of  the  totally
Coulomb-like phase and the six phases that are
confined along 1-dimensional subgroups. However, if  for  example  one
wishes to have coexistence of the totally Coulomb-like phase and the four
phases that are confined along the four 2-dimensional subgroups, it will be
seen (Table~\ref{tab2})  that $\log a$ must be decreased by
$\frac{1}{4}\log  (4/3)$.
But this reduction in $\log a$ would put the phases  corresponding to
1-dimensional subgroups into confinement.

Information about the cubic and hexagonal lattices are tabulated in
Tables~\ref{tab1} and \ref{tab2}. Table ~\ref{tab1}  pertains  to  the
cubic lattice; Table ~\ref{tab2} to  the  hexagonal
identification lattice.  The entries in the  first  four (five)
rows and  columns  of  Table~\ref{tab1} (Table~\ref{tab2}) give  the
values  of  the identification  lattice
constant $a^2$ (in terms of $a^2_{1\;crit}$) at which pairs
(corresponding to a row
and column heading) of free  energy
phase ans\"{a}tze intersect; i.e., these entries are the quantities

\beq \frac{a^2}{a^2_{1\;crit}}=\left (\frac{c_n}{c_m}\right
)^{\frac{-2}{dim(H_n)-dim(H_m)}}
\;\;\;\;(n,m \in \{0,1,2,3\}) \label{vcf} \eeq

\nin obtained by rewriting (\ref{cond}) and using that $a_{1\;crit}^2=3\pi$
(obtained from (\ref{cond}) with
$n=1$ and  $m=0$. The quantities $a$ and
$a_{1,\;crit}=2\pi \sqrt{\frac{\beta_{crit}}{2}}$ are respectively  the
identification
lattice constant and the critical value  of  the (identification)  lattice
constant. The quantities $c_n$   are  the  volume
correction factors associated with the subgroup $H_n$ ($n \in  \{0,1,2,3\}$).
These are also tabulated in the tables below. All the
volume correction factors are unity for the cubic identification lattice.
For the hexagonal lattice, $c_0$ and $c_1$ are both unity whereas
$c_2=\sqrt{3/4}$ and $c_3=\sqrt{1/2}$ corresponding respectively to the
ratio of  area of a minimal parallelogram in the hexagonal lattice to the area
of a simple plaquette in the cubic lattice  and the ratio of the volume of a
(minimal) parallelpipidum of the hexagonal lattice to the volume of a simple
cube in the cubic lattice.

\begin{table}
\caption[table1]{\label{tab1}Parameters pertaining to the cubic identification
lattice. The
entries in the first four rows and columns are all unity because phases
corresponding to all subgroups convene at the multiple point for the
critical value of the coefficient $\frac{1}{e^2_{U(1)\;crit}}$
in the Manton action; i.e., the quantity
$\frac{a^2}{a^2_{1\;crit}}=\left (\frac{c_n}{c_m}\right )^{\frac{-2}
{dim(H_n)-dim(H_m)}}$ $(n,m \in \{0,1,2,3\})$  is unity for
all $m,n\in \{0,1,2,3 \}$. The quantities in the last three columns are
as explained in Table (\ref{tab2}).} \vspace{.5cm}
\begin{tabular}{|c||c|c|c|c||c|c|c|}
\hline
CUBIC & $\log Z_{H_0}$ & $\log Z_{H_1}$ & $\log Z_{H_2}$ & $\log Z_{H_3}$ &
$\frac{d\log Z_{H_n}}{d \log a}$ & \# phases & $c_n$ \\ \hline \hline
$\log Z_{H_0}$  &
1 & 1 & 1 & 1 & -3 & 1 & 1 \\  \hline
$\log Z_{H_1}$ &   & 1 & 1 & 1 & -4 & 3 & 1 \\ \hline
$\log Z_{H_2}$ &  &  & 1 & 1 & -5 & 3 & 1 \\ \hline
$\log Z_{H_3}$ &  &  &  & 1 & -6 & 1 & 1 \\ \hline
\end{tabular}
 \end{table}
\vspace{1cm}
\begin{table}
\caption[table2]{\label{tab2}Parameters pertaining to the hexagonal
identification lattice.
As regards the five rows and first five columns, the entry in the n$th$ column
and the m$th$ row is the coefficient
$\frac{a^2}{a^2_{1\;crit}}=\left (\frac{c_n}{c_m}\right )^{\frac{-2}
{dim(H_n)-dim(H_m)}}$ $(n,m \in \{0,1,2,3\})$ .
This is the quantity by which $\frac{1}{e^2_{U(1)\;crit}}$ must be multiplied
in order that
the phases confined w.r.t. the $n$-dimensional and $m$-dimensional subgroups
can come together at the multiple point. The slope of the
$\frac{d \log Z_{H_m}}
{d \log a}$, calculated from (\ref{deriv}), is given in the sixth column.
Column
seven gives the number of phases of dimension $m$. The entries in
column eight are the ``volume'' correction factors $c_n$ (see (\ref{cn}))
in the hexagonal lattice
relative to the corresponding (unit) ``volumes'' in the cubic lattice.}
\vspace{.5cm}
\begin{tabular}{|c||c|c|c|c|c||c|c|c|}
\hline
HEXAG. & $\log Z_{H_0}$  & $\log Z_{H_1}$ & $\log Z_{H_2 orthog}$ & $\log
Z_{H_2}$ & $\log Z_{H_3}$ &
$\frac{d\log Z_{H_n}}{d \log a}$ & \# phases & $c_n$ \\ \hline \hline
$\log Z_{H_0}$  & 1
& 1 & 1 & $\sqrt{\frac{4}{3}}$ & $\sqrt[3]{2}$ & -3 & 1 & 1 \\  \hline
$\log Z_{H_1}$ &  & 1  & 1 & $\frac{4}{3}$ & $\sqrt{2}$ & -4 & 6 & 1 \\ \hline
$\log Z_{H_2 orthog}$ & & &1 & & 2 & -5 & 3 & 1 \\ \hline
$\log Z_{H_2}$ &  &  &  & 1 &  $\frac{3}{2}$ & -5 & 4 & $\sqrt{\frac{3}{4}}$ \\
\hline
$\log Z_{H_3}$ &  &  &  & & 1 & -6 & 1 & $\sqrt{\frac{1}{2}}$ \\ \hline
\end{tabular}
 \end{table}
\vspace{1cm}

However, the amount by  which
the free energy functions for these different phases fail to intersect
at a common value of the identification lattice  is hopefully
small  enough  to
be dealt with meaningfully by perturbing  the  Manton
action (using 4$th$ and 6$th$ order terms) in such a way as to  allow
12 phases to convene at a multiple point.

We therefore
replace the Manton action (containing by definition  only  second
order terms) by a more complicated action:

\[ S_{\Box,\;Manton} \rightarrow S_{\Box,\;Manton}+S_{\Box,\;h.o.} \]

\nin where $S_{\Box,\;h.o.}$ designates higher than second order terms.
In choosing the higher order terms, we want to use the lowest possible
order terms that bring together the desired phases at the  multiple  point.

The number of additional terms needed depends on how  many  phases  we
want to bring together at the multiple point. As explained  above,
we have decided to settle for the 12 phases (corresponding to
one 0-dimensional, six 1-dimensional, four 2-dimensional, and one
3-dimensional subgroups)  that have the smallest possible volume on
the  hexagonal  lattice and which are not too far from being able to convene
at the multiple point with the Manton action alone.  These 12
phases seem to exhaust the  ones  for  which  a  modification  of  the
couplings using the procedure to be explained below can  be  regarded  as  a
small perturbation; for example, the diagonal subgroup
coupling (with pure Manton action) is so far removed from the
critical couplings of the 12 hexagonal lattice phase discussed above that we
{\em a priori} give up trying to have a phase confined along the
diagonal subgroup in contact with the multiple point. The same applies
presumably to the
2-dimensional ``orthogonal'' subgroups (\ref{p1}-\ref{p3}) as
already mentioned above.

Due to the  high  degree  of  symmetry of the hexagonal lattice,  the
conditions  for  the  criticality  are  identical   for  phases
corresponding to the four 2-dimensional subgroups
and the six 1-dimensional subgroups. So the  number
of parameters we need to get all 12 phases to convene is effectively  that
for four phases (corresponding to the four possible dimensionalities of
subgroups). This requires 4-1=3 parameters. This can be compared to  the
generic number of parameters necessary for the meeting  of  12  phases:
12-1=11 parameters. The point is that the symmetry  of  the  hexagonal
identification lattice allows a non-generic multiple point in an action
parameter space spanned by just three parameters. These can be chosen
as the Manton parameter (i.e., the coefficient  to  the  second  order
term in a Taylor expansion of the action) and two parameters that  are
coefficients to respectively a  4th  and  a  6th  order  term.
These 4th and 6th order terms are to be chosen so as to have
the same symmetry as the hexagonal lattice; otherwise we lose the symmetry
that allows a non-generic multiple point. Without the symmetry, we would
in general need 11 parameters instead of 3. It is also necessary  that
these two terms contribute differently to the  different  free  energy
functions for the different types of subgroup that we want  to
bring to the multiple point. Otherwise we  could  compensate  for  the
effect of these higher order terms for all subgroups by using a single
new effective coefficient to the Manton term. In other words, we  want
our high order terms to be such that these give  {\em  different}  new
effective coefficients to  the  second  order  action  term  for  {\em
different} subgroups. The effective second order coefficient is
defined  as the coefficient in the  Manton action that would give the
same fluctuation width inside the subgroup in question as there would be
with the higher order terms in place.
To this end we use linear combinations of spherical harmonics $Y_{lm}$
with $l=4$ and $l=6$ that have the same symmetry as the  cub-octahedron
(which can be taken as the ``unit cell'' of the hexagonal identification
lattice). These linear combinations, denoted $Y_{4\;comb}$ and
$Y_{6\;comb}$, are invariant under the symmetry of the
cub-octahedron.

In using the $Y_{4\;comb}$ and
$Y_{6\;comb}$
as perturbations to the Manton action, we obtain an effective Manton inverse
squared coupling strength  that varies with the
direction $\vec{\xi}$:

\beq \frac{1}{e^2_{eff}(\vec{\xi})}. \label{invcoupeff} \eeq

\nin Here $\vec{\xi}$ denotes a vector in $\br^3$ (the covering space of
$U(1)^3$).

The desired combinations $Y_{4\;comb}$ and $Y_{6\;comb}$ that have the symmetry
of the cub-octahedron turn out, after a rather strenuous calculation, to be

\beq Y_{4\;comb}=
\frac{2}{3}\sqrt{7}Y_{40}+\frac{4}{3}\sqrt{5}(Y_{43}+Y_{4,-3})/i\sqrt{2} \eeq

\nin and

\beq Y_{6\;comb}=
(-4\sqrt{\frac{3}{35}})Y_{60}+\sqrt{\frac{11}{10}}(Y_{66}+Y_{6,-6})/\sqrt{2}+
(Y_{63}+Y_{6,-3})/i\sqrt{2}. \eeq

\nin These have been calculated in a coordinate system in which the $z$-axis
coincides with a 3-axis of symmetry of the cub-octahedron. In  Table \ref{tab3}
these combinations $Y_{4\;comb}$ and $Y_{6\;comb}$ are averaged over the
1,2 and 3-dimensional subgroups of $U(1)^3$. The fact that both combinations
vanish for $U(1)^3$ (the 3-dimensional subgroup) reflects of course the
property that spherical harmonics vanish when integrated over the surface
of a sphere. Table \ref{tab3} also gives the values of $Y_{4\;comb}$ and
$Y_{6\;comb}$ along the diagonal subgroup of $U(1)^3$.

\begin{table}
\caption[table3]{\label{tab3}The $4th$ and $6th$ order action contributions
needed to realize
12 \pcps at the multiple point. The contributions have the symmetry of the
hexagonal identification lattice.}

\beq \begin{array}{|c|c|c|} \hline \mbox{subgroup} & Y_{4\;\;comb} &
Y_{6\;\;comb} \\ \hline
\langle Y_{l\in\{4,6\}}\rangle_{3-dim} & 0 & 0 \\ \hline
\langle Y_{l\in\{4,6\}}\rangle_{2-dim} & \frac{\sqrt{7}}{4} &
\frac{5}{4}\sqrt{\frac{3}{35}} \\ \hline
        Y_{l\in\{4,6\};\;1-dim} & \frac{\sqrt{7}}{4} &
\frac{117}{32}\sqrt{\frac{3}{35}} \\ \hline
        Y_{l\in\{4,6\};\;diag subgr} & \frac{2\sqrt{7}}{3} &
-4\sqrt{\frac{3}{35}} \\ \hline
\end{array} \eeq
 \end{table}

\begin{figure}
\centerline{\epsfxsize=\textwidth \epsfbox{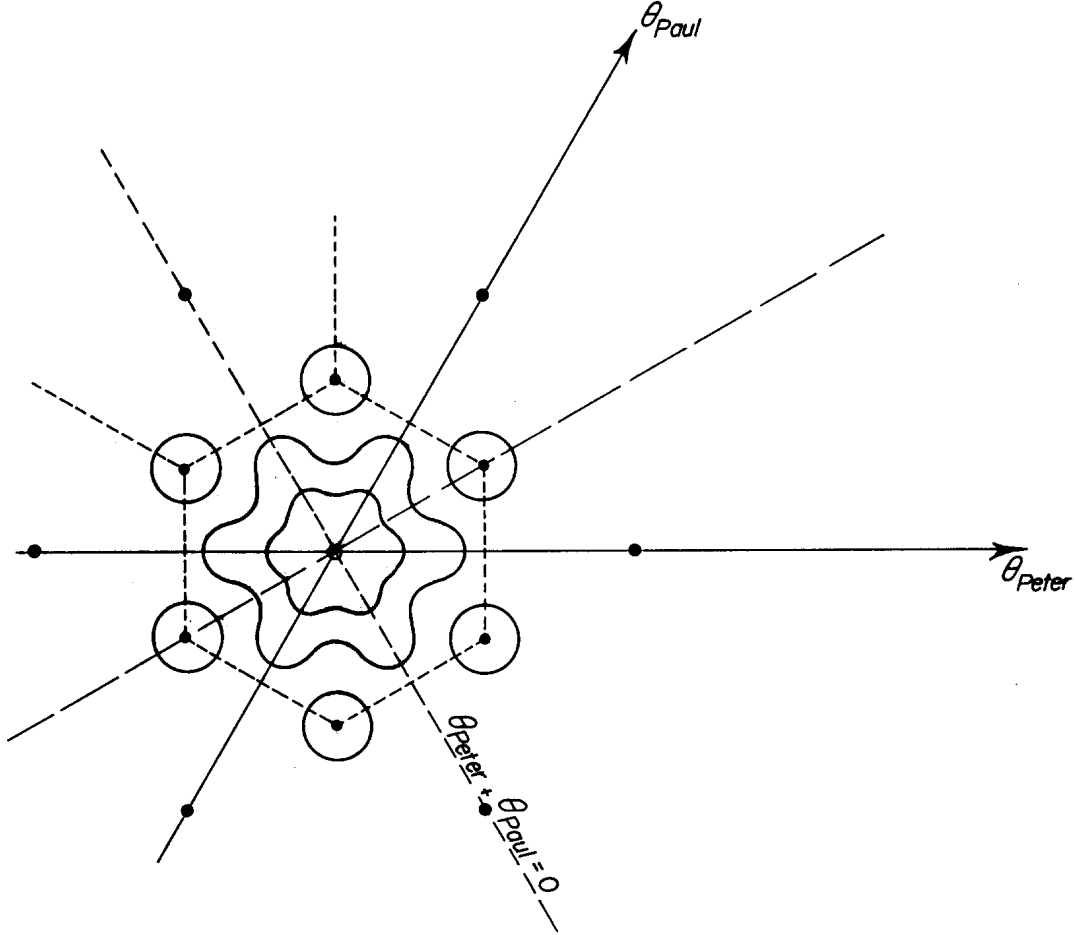}}
\caption[figwebbed]{\label{figwebbedfeet}Contours of constant perturbed Manton
action for $U(1)^2$ represented in the covering group $\br^2$ with the metric
in the plane of the paper that is identified with the Manton action metric.
The hexagonal  lattice of ``$\bullet$'' are points identified in compactifying
from $\br^2$ to $U(1)^2$.
The purpose of the correction - it is sixth order and gives  the contours
a ``webbed feet'' look - is to increase $\log Z$ for the phases with
confinement along one of the three 1-dimensional subgroups - i.e., along the
$\theta_{Peter}$ axis, the $\theta_{Paul}$ axis and along the line given by
$\theta_{Peter}+\theta_{Paul}=0$ - while disfavouring fluctuations along
directions that bisect the angles between these 1-dimensional subgroup
directions. This is accomplished by decreasing the
gradient of the action in these three subgroup directions while increasing
the gradient in
directions that bisect the above-mentioned three subgroups}
\end{figure}
Using the Tables \ref{tab1}, \ref{tab2}, and \ref{tab3}, let us now determine
the coefficients to the
2$nd$ order (i.e. Manton) as well as 4$th$ and 6$th$ order action terms
by using the requirement that averages over the
1, 2 and 3-dimensional subgroups of $U(1)^3$ are equal to the $U(1)$
critical inverse squared coupling $1/e^2_{U(1)\;crit}$ when the volume
correction factors for the hexagonal lattice are taken into consideration.
These latter are given by (\ref{vcf}). Using that $\beta=1/e^2=a^2/2\pi^2$
we can write the condition to be satisfied if the average over the subgroup
$H_n$ - i.e., $\langle 1/e^2(\vec{\xi}) \rangle_{H_n}$ -
is to have a value corresponding
to the boundary between a phase confined along $H_n$ and the totally
Coulomb phase:

\beq \langle \frac{1}{e^2(\vec{\xi})} \rangle_{H_n}=
\left(\frac{c_n}{c_0}\right)^{\frac{-2}{dim(H_n)-dim(H_0)}}
\frac{1}{e^2_{U(1)\;crit}} \label{avcoup} \eeq

\nin where $c_0$ and $H_0$ correspond to the totally Coulomb phase.
Eqn. (\ref{avcoup}) yields three equations -
one for each type of subgroup $H_n$
($n=dim(H_n)$).

For $n=3$ there are no contributions to
$\langle\frac{1}{e^2_{eff}(\vec{\xi})}\rangle_{3-dim\;subgr}$ from
$Y_{4\;comb}$ and $Y_{6\;comb}$. The second order coefficient
$\frac{1}{e_{Manton}^2}$ is therefore determined by the one equation

\beq \langle\frac{1}{e^2_{eff}(\vec{\xi})}\rangle_{3-dim\;subgr}=
\frac{1}{e_{Manton}^2}=\left(\frac{c_3}{c_0}\right)^{\frac{-2}{3-0}}
\frac{1}{e^2_{U(1)\;crit}}=2^{\frac{1}{3}}\frac{1}{e^2_{U(1)\;crit}}.\eeq

The coefficients to $Y_{4\;comb}$ and $Y_{6\;comb}$ - denoted respectively
as $B_4$ and $B_6$ - can be obtained from the equations for
$\langle\frac{1}{e^2_{eff}(\vec{\xi})}\rangle_{1-dim\;subgr}$ and
$\langle\frac{1}{e^2_{eff}(\vec{\xi})}\rangle_{2-dim\;subgr}$.
Assigning dimensionality to the strictly speaking dimensionless quantity
$1/e^2$, we use that
               [$B_4$]=[$\frac{1}{e^4}$] and [$B_6$]=[$\frac{1}{e^6}$].

For $n=1$ we have:

\beq \langle\frac{1}{e^2_{eff}(\vec{\xi})}\rangle_{1-dim\;subgr}^3=
B_6\langle Y_{6\;comb}\rangle_{1-dim\;subgr}+\left(\frac{1}{e^4_{Manton}}+
B_4\langle Y_{4\;comb} \rangle_{1-dim\;subgr}\right)^{\frac{3}{2}}= \eeq

\[
=B_6\frac{117}{32}\sqrt{\frac{3}{35}}+
\left(\frac{2^{\frac{2}{3}}}{e^4_{U(1)\;crit}}+B_4\frac{\sqrt{7}}{4}\right)^
{\frac{3}{2}}=
\left(\left(\frac{c_1}{c_0}\right)^{\frac{-2}{3-0}}
\frac{1}{e^2_{U(1)\;crit}}\right)^3=\left(1\frac{1}{e^2_{U(1)\;crit}}\right)^3
.\]

For $n=2$ we have:

\beq \langle\frac{1}{e^2_{eff}(\vec{\xi})}\rangle_{2-dim\;subgr}^3=
B_6\langle Y_{6\;comb}\rangle_{2-dim\;subgr}+\left(\frac{1}{e^4_{Manton}}+
B_4\langle Y_{4\;comb} \rangle_{2-dim\;subgr}\right)^{\frac{3}{2}}= \eeq

\[
=B_6\frac{5}{4}\sqrt{\frac{3}{35}}+
\left(\frac{2^{\frac{2}{3}}}{e^4_{U(1)\;crit}}+B_4\frac{\sqrt{7}}{4}\right)^
{\frac{3}{2}}=
\left(\left(\frac{c_2}{c_0}\right)^{\frac{-2}{2-0}}
\frac{1}{e^2_{U(1)\;crit}}\right)^3=
\left(\sqrt{\frac{4}{3}}\frac{1}{e^2_{U(1)\;crit}}\right)^3
.\]

\nin The values of the geometric factors $c_n$ are from Table \ref{tab2} and
the values of $\langle Y_{6\;comb}\rangle_{n-dim\;subgr\;H_n}$ and
$\langle Y_{4\;comb}\rangle_{n-dim\;subgr\;H_n}$ $(n=dim(H_n)\in\{0,1,2,3\})$
are taken from Table \ref{tab3}.

Solving these equations for the coefficients $B_4$ and $B_6$ yields

\beq B_4= -0.1463 \mbox{  and   } B_6=-0.7660 \eeq

We have now succeeded in fitting three coefficients of a modified Manton
(i.e. a plaquette action dominated by a second order ``Manton'' term
but having perturbative 4$th$ and 6$th$ order terms) in such a way
that 4 types of phases $H_n$ convene at a multiple point in the sense
that $\langle 1/e^2_{eff}(\vec{\xi}) \rangle_{H_n}$
($n=dim(H_n)\in\{0,1,2,3\}$) is equal to the $U(1)$
critical coupling up to a factor pertaining to the geometry of the hexagonal
identification lattice. Because the modified Manton action has the
symmetry of the hexagonal lattice, multiple point criticality for a phase
corresponding to a given dimension implies multiple point criticality for
all phases corresponding to a given dimension. For this reason we achieve
multiple point criticality for a total of 12 phases. The averaging
$\langle 1/e^2_{eff}(\vec{\xi})\rangle_{H_n}$ can be taken as an average over
all directions within the subgroup $H_n$ using a measure defined by being
invariant under rotations leaving the Manton metric invariant.

So we now have at our disposal a means of calculating a directionally
dependent effective inverse
squared coupling where the directional dependence comes from the
perturbative $4th$ and $6th$ order action terms. In a later section,
we shall want to calculate $1/e^2_{eff}$ in the direction corresponding to
the diagonal subgroup (in a chosen coordinate system).

\section{Calculation of the numerical value of the continuum coupling}
\label{calculation}

\subsection{Outline of procedure}

The aim  now  is to calculate  the  continuum  $U(1)$  standard  model  weak
hyper-charge  coupling  corresponding  to  the   ``diagonal   subgroup''
coupling at  the  multiple  point  of  the  $AGUT$  gauge  group
$SMG^3$.
In principle, the multiple point should be sought
in a very high dimensional action parameter space that is
also in contact
with a multitude of phases that are alone confined w.r.t discrete
$\bz_N$ subgroups.
In an even  more  correct search for the multiple point involving
phases with confining discrete subgroups, we should really consider Abelian
and non-Abelian groups at the same time (i.e, the full $SMG^3$ or perhaps an
even larger group) because   discrete subgroups having the characteristic of
being non-factorizable could a priori simultaneously involve Abelian subgroups
and centres of semi-simple subgroups.

\begin{figure}
\centerline{\epsfxsize=\textwidth \epsfbox{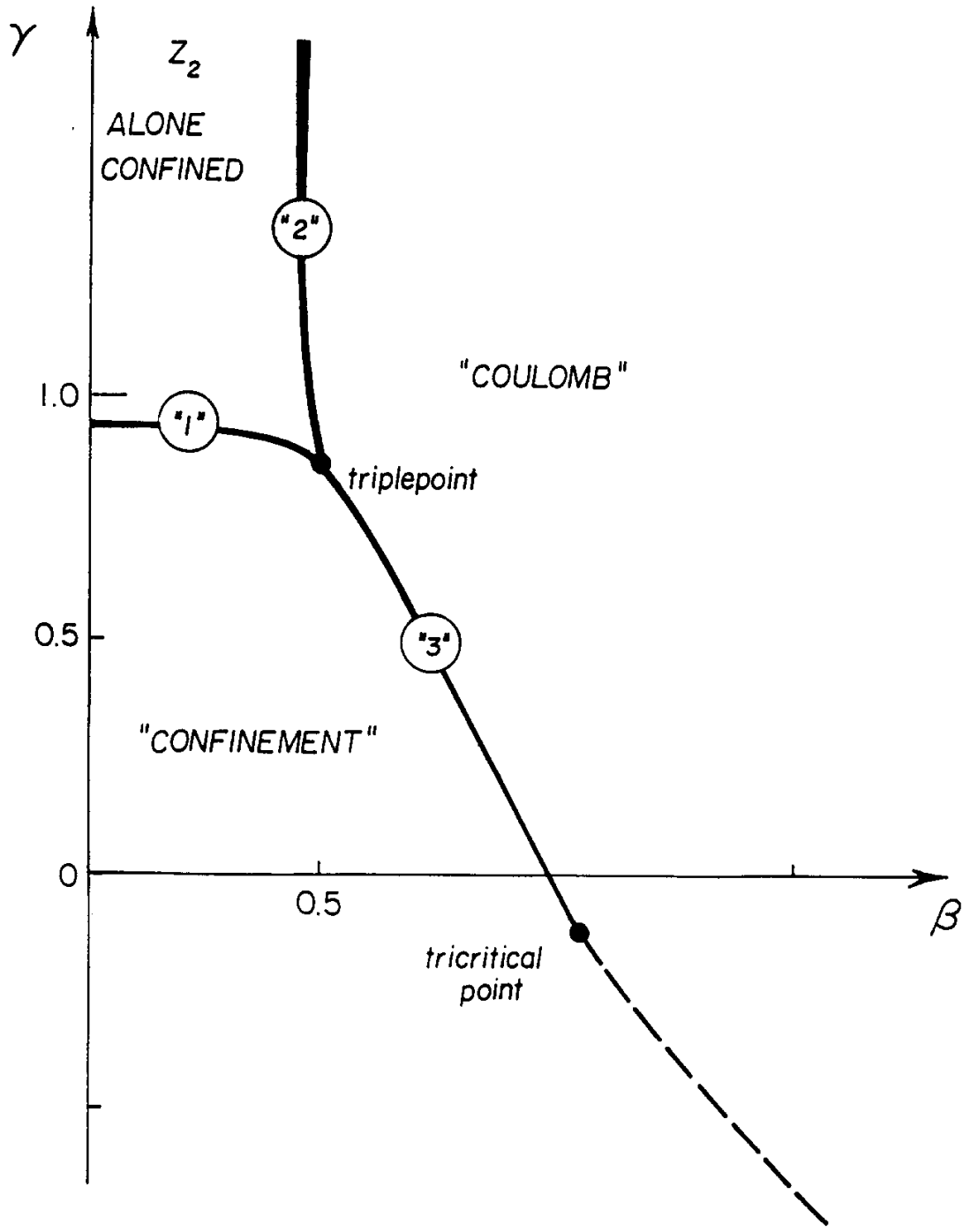}}
\caption[figurebhanot]{\label{figbhanot}The phase diagram for $U(1)$ when the
two-parameter action is used. This type of action makes it possible
to provoke the confinement of $\bz_2$ (or $\bz_3$) alone.}
\end{figure}

As a crude prototype to  a $U(1)^3$ phase diagram, we consider the (generic)
phase diagram spanned by the parameters of an action with $\cos \theta,
\cos \frac{\theta}{2}$ and $\cos \frac{\theta}{3}$ terms.
This action, which is one of the simplest generalisations of the
pure Wilson action, has been studied extensively\cite{bhanot1} and many
features of the
phase diagram (Figure~\ref{figbhanot}) are well understood. From the triple
point (TP) (which is the ``multiple point'' in this 2-dimensional phase
diagram)
emanate three characteristic phase borders: the phase border ``3''
separates the totally
confining and totally Coulomb-like phases;
the phase border ``1'' separates the totally confining phase from the phase
where only the discrete subgroup $\bz_2$ is confined;
this latter phase is separated
from the totally Coulomb-like phase by the phase border ``2''.

The calculational procedure to be used in determining the continuum $U(1)$
coupling is approximative and is done in two steps:

\begin{enumerate}

\item [A.] first  we  calculate  the
factor analogous to the factor $3=N_{gen}$  in  the  non-Abelian  case;
we call this  the enhancement factor and denote it as
$\frac{1/\alpha_{U(1)^3\;diag}}{1/\alpha_{U(1)_{crit\;TP}}}$. This factor
lies in the range 6.0 - 8.0 depending on the degree of ``first-orderness''
of the triple point (TP) transition at boundary ``2''.

\item [B.] In the second step, the continuum $U(1)$ coupling corresponding to
the multiple point
value for a single $U(1)$ is determined using an
analogy to a procedure proposed by Luck\cite{luck} and developed by
Jers\`{a}k\cite{jersak}.

\end{enumerate}

This two-step calculation can be done using more or less good approximations
as regards the extent to which the continuum $U(1)$ coupling value reflects
having phases solely confining w.r.t. discrete subgroups among the phases
that convene at the multiple point.
Let us outline the possible approximations in the order of increasing
goodness.

\begin{enumerate}

\item The roughest calculation would be to use a single parameter action with
hexagonal symmetry without regard to having phases at the triple point (TP)
that are confining solely w.r.t. the discrete subgroups $\bz_2$ and $\bz_3$
of $U(1)^3$. In this approximation, these discrete subgroups are treated as
though they were totally
confining inasmuch as it is a $U(1)$-isomorphic {\em factor group} obtained
essentially by dividing $\bz_2\times \bz_3$ out of the $U(1)$ centre of
$SMG$ that is identified with the lattice $U(1)$ critical coupling.

\item By using a two-parameter action (later a three parameter action)
leading to
the phase diagram of Figure~\ref{figbhanot}, the action
now acquires a (nontrivial) dependence on the elements within the
cosets of the factor group $U(1)/\bz_2$ (or the factor group
$U(1)/(\bz_2\times \bz_3)$ in the case of a three-parameter action)
that can reveal
how close the discrete subgroups are to being critical. However these
details are of little importance to the $U(1)$ continuum coupling; the
latter depends essentially only on a single yet to be defined parameter
$\gamma_{eff}$ the critical value of which is very nearly constant along
the phase boundary ``1''
of Figure~\ref{figbhanot}. Hence the $U(1)$ continuum coupling is also
approximately constant along this phase boundary in accord with the rule
described in the footnote on page~\pageref{rule}. The critical value
$\gamma_{eff\;crit}$ of
the parameter $\gamma_{eff}$ is expressible in terms of the critical
lattice parameters available from computer data for a lattice gauge
theory with a single $U(1)$.

\item The effect on the continuum coupling of having phases convening
at the multiple point  that are
confined solely w.r.t. $\bz_2$ and solely w.r.t. $\bz_3$
appears first when we take into account the discontinuity in $\Delta
\gamma_{eff}$ encountered in crossing the boundary ``2'' at the multiple
point. As we in both steps A. and B. above want to use the value of
$\gamma_{eff}$
corresponding to the {\em totally Coulomb-like phase} at the multiple point,
it is important for our calculation of the $U(1)$ continuum coupling to take
the ``jump'' $\Delta \gamma_{eff}$ into account. Inasmuch as the continuum
subgroup \dofx are in the same phase on both sides of boundary ``2'',
this discontinuity $\Delta \gamma_{eff}$ is entirely due to a phase
transition for the discrete subgroup(s). Moreover, the presence of a
discontinuity presumably reflects the degree of first-orderness
of the triple point transition at border ``2'' inherited from a pure $\bz_2$
and $\bz_3$ transition (i.e., for $\gamma >> 1$ in Figure~\ref{figbhanot}).

\item The discrete subgroups  $\bz_2$ and $\bz_3$ contribute
differently to the ``jump'' $\Delta \gamma_{eff}$ in crossing the boundary
``2'' due to the fact that
$\bz_2^3$ does not inherit the hexagonal symmetry of $U(1)^3$ while
$\bz_3^3$ is more likely to do so.
This is discussed at the end of Section~\ref{sec-portray}.

\end{enumerate}

It is important to recall that the normalisation of the $U(1)$
that we have argued for is implemented by the identification of the
$U(1)$ lattice critical coupling with the ($U(1)$-isomorphic)
{\em factor-group} $= SMG/(SU(2)\times SU(3))$ for  some  one  of
the Cartesian product factors say $SMG_{Peter}$. Since we have  argued
or assumed that phases with genuine discrete subgroups of this  $U(1)$
factor-group are not to be in contact with the multiple point chosen by
Nature, the only discrete subgroups that are to be taken  into  account
are     discrete     subgroups          of      the      $U(1)$
{\em subgroup} of $SMG$. The  relation  between  the  $U(1)$
subgroup  and  the  factor-group  $SMG/(SU(2)\times  SU(3))$   can   be
described  as  $U(1)_{factorgr}=U(1)_{subgr}/{\bf  Z}_6$.

In using $U(1)/\bz_6$ as the factor group to be identified with the lattice
critical $U(1)$, we identify the elements of $\bz_6$ and thereby ``hide''
any differences that there might be in the probabilities for being at
different elements of $\bz_6$ when a one-parameter action is used
(approximation 1 in list above). But the details of how the heights
of the peaks in probability at different elements of $\bz_6$ differ are
important if we want to arrange that the discrete subgroups of $\bz_6$ are
by themselves to be confined in phases convening at the triple point.
However such details become visible again if the (one-parameter)
Wilson action (roughest approximation {\bf 1} in the list above)
is replaced by the
(two-parameter) ``mixed'' fundamental-adjoint action
(approximation {\bf 2} and {\bf 3} in the list above).
By introducing an additional parameter in this way, we
render the group elements identified in the factor groups
$U(1)_{subgr}/\bz_2$ and  $U(1)_{subgr}/\bz_3$ inequivalent (i.e.,
the action acquires a dependence on the elements {\em within} the
cosets of these factor groups). So in
effect, by going from the Wilson action to the two-parameter action we
lift the factor group up into a kind of covering space. The result is that by
replacing the  $U(1)_{factorgr}$
critical coupling by the triple point
coupling  for the $U(1)_{subgr}$ of $SMG$,
we essentially arrange that the subgroups ${\bf Z}_2$ and ${\bf Z}_3$
can confine individually in phases that convene at the triple point (TP).

In both steps A. and B. of the calculation of the continuum $U(1)$ coupling,
we make use of the ``jump'' $\Delta \gamma_{eff}$
in the quantity $\gamma_{eff}$ that
in Section \ref{secgamef} below will be argued to be an effective coupling
in the sense that in the region of the phase
diagram near the phase border ``1'' in Figure~\ref{figbhanot}
(i.e., on both sides of ``1'') it
is to a good approximation valid that the phase realized (i.e.,
the totally confined or the phase with
only $\bz_2$ confined) is determined by the value of this one variable
$\gamma_{eff}$
($\gamma_{eff}$ is a certain combination of the parameters
$\gamma$ and $\beta$ of the two-parameter action
(see Figure~\ref{figbhanot})).
Consequently, the variable $\gamma_{eff}$ is necessarily constant along the
phase boundary ``1'' and we can also assume that the corresponding
continuum coupling has a constant
value along this boundary. The change in $\gamma_{eff}$ -
i.e. $\Delta \gamma_{eff}$ -
comes first at the boundary ``2'' in going into the totally Coulomb-like phase.
The value of  $\Delta \gamma_{eff}$ (calculated in Section~\ref{deltagam})
depends on the degree of ``first-orderness'' that at the
multiple point ($\gamma\approx 1$) is inherited from the pure $\bz_2$ and
$\bz_3$ transitions at $\gamma \rightarrow \infty$. Without
the correction for discrete subgroups embodied by $\Delta \gamma_{eff}$, the
multiple point coupling is obtained as if the discrete
subgroups were totally confining.

In the step A., the quantity  $\Delta \gamma_{eff}$, which
reflects
the degree of first-orderness inherited from the pure $\bz_2$ and $\bz_3$
transition in crossing boundary ``2'' at the multiple point, is
used to interpolate between the enhancement factor of about 8  obtained
with the
volume approximation and the enhancement factor of 6
obtained with the independent monopole approximation.
These approximations are most suitable for respectively first and second order
transformations. The calculation of the enhancement factor is done in
Section~\ref{enhance}.

In  step B. of the calculation (performed in Section~\ref{contcoup}),
the quantity $\Delta\gamma_{eff}$ is again used - this time in the
combination $\gamma_{eff}+\Delta \gamma_{eff}$ - to calculate the  $U(1)$
continuum coupling corresponding to
the triple point  values of a (single) $U(1)$ lattice gauge theory.
We seek the continuum coupling in the corner of the totally {\em Coulomb-like}
phase (necessary if Planck-scale confinement of observed fermions is to be
avoided)
that lies at the
triple point - that is, in the ``corner'' formed by the phase borders ``2''
and ``3''. According to the above argumentation, we know that the
continuum coupling at any position along the border ``1'': it is just equal
to the value at $\gamma=\gamma_{crit}$ and $\beta=0$. In particular, this is
true at the multiple point in the phase with only $\bz_2$ confining (i.e.,
in the ``corner'' formed by the phase boundaries ``1'' and ``2''). But as
argued above, we want the coupling corresponding to the Coulomb phase
``corner'' formed by borders ``2'' and ``3''. This requires a correction
that accounts for going from the multiple point corner formed by
``1'' and ``2'' to the multiple point corner formed by borders ``2''
and ``3'' (and in principle also a small correction from crossing
border ``1''). It  is this
transition, corresponding to the transition from a
phase with solely $\bz_2$
confining to a totally Coulomb-like phase that is accompanied by the ``jump''
denoted by $\Delta \gamma_{eff}$.

\subsection{The  approximation  of  an  effective   $\gamma$ \label{secgamef}}

In the literature (\cite{bhanot1,bhanot2}) we find the phase diagram for a
$U(1)$
group with a mixed lattice action having a $\gamma$  term  defined  on
the factor-group $U(1)/{\bf Z}_2$:

\beq
S_{\Box}(\theta)= \gamma cos(2\theta) + \beta cos(\theta). \label{mix}
\eeq

\nin With this action it is easy to provoke confinement of the whole
group as well as the totally Coulomb phase and phase confined solely w.r.t.
a discrete subgroup isomorphic to $\bz_2$ for a judicious choice of
the action parameters $\gamma$ and $\beta$ that span the phase diagram
in Figure~\ref{figbhanot}.
Indeed the phase diagram of Figure~\ref{figbhanot} clearly reveals a triple
point
common to three phases.
The interpretation of these phases as  the  three referred to above
is confirmed by  rough  mean  field
estimates for the phase borders.
In the case of the non-Abelian subgroups $SU(2)$ and $SU(3)$, two of the
phases in Figure~\ref{figbhanot}  are  actually  connected,  because one of
the  phase
borders ends at a tri-critical point.
However, this does not of course preclude the  existence
of a multiple (i.e., triple) point.

Before proceeding, it is useful to change  notation  by
scaling the variable $\theta$ down by a  factor  two
inasmuch as it is recalled (see
\ref{onlyz6}) that we want to normalise relative to the factor  group
\footnote{The motivation is that the  normalisation  that  we  use  to
define the coupling $\alpha_1$  (i.e.,  $\alpha_{1,\;Peter}$,  etc.)  is
relative  to  the  factor  group  rather  than  the  subgroup   (i.e.,
$U(1)_{factorgr}  =  SMG/(SU(2)\times  SU(3))  \sim  U(1)_{subgr}/{\bf
Z}_6$ rather than $U(1)_{subgr}$).  Instead  of  $U(1)/{\bf  Z}_6$,  we
consider for illustrative purposes  the  analogous  situation
$U(1)/{\bf  Z}_2$. This
case is also comparable with readily available results to be found   in   the
literature\cite{bhanot1}.} $U(1)/\bz_2$.

\beq
S_{\Box}(\theta)= \gamma \cos(\theta) + \beta \cos(\theta/2).
\label{newbhanot}\eeq

\nin Note that with this notational convention, Bianchi identities are
fulfilled modulo $4\pi$. In discussions of monopoles dealt with
in later sections, we shall have occasion to distinguish ``full''
$4\pi$-monopoles and ``minimal strength'' $\frac{2\pi}{4\pi}$-monopoles.
The latter
will be seen to correspond to the
``length'' of the factor group $U(1)/\bz_2$. These remarks first become
relevant and more transparent when, in a later section, we explain the idea
of ``minimal strength'' monopoles. In the case of $\bz_2$, such monopoles
are referred to as $\frac{2\pi}{4\pi}$-monopoles. These will be seen
to be the monopoles present relative
to a $\bz_2$ background field.

In order to obtain numerical results, the multiple
point coupling in this diagram will in a later Section be related to the point
at which
$\beta=0,\gamma=\gamma_{critical}$
inasmuch as we have a procedure for relating this point  to the continuum
coupling at the triple point (hereafter ``TP'') ``corner'' formed by
the phase boundaries ``1'' and
``2'' and subsequently at the for us interesting TP ``corner'' formed
 by the phase boundaries ``2'' and ``3'' (i.e., the totally Coulomb phase at
 the TP - see Figure~\ref{figbhanot}).
We shall actually argue that to
a very good approximation the continuum coupling does not vary  along
the phase border ``1'' (separating  the  ``total  confinement''  and  the
``phase with only ${\bf Z}_2$ confining'') in going from
$(\beta,\gamma)=(0,\gamma_{crit})$
to the corner at the TP formed by the phase boundaries ``1'' and ``2''.
It is first upon crossing the phase boundary ``2'' into the totally Coulomb
phase at the  TP that there is a change - a jump $\Delta \gamma_{eff}$ -
in the quantity $\gamma_{eff}$ that  immediately below will be seen to be an
effective coupling. This jump $\Delta \gamma_{eff}$ comes from a jump in
the  relative
probability of finding the plaquette variable at the(a)  non-trivial element of
$Z_2(\bz_3)$ upon making the transition at boundary ``2'' separating the
totally
Coulomb-like phase from the phase solely confined w.r.t. $\bz_2$ or $\bz_3$ in
Figure~\ref{figbhanot}. As the continuum \dofx are in the same phase on both
sides
of the boundary ``2'', the discontinuity $\Delta \gamma_{eff}$ must be entirely
due to the discrete subgroup transition which inherits a considerable
degree of the first order nature of the pure
(i.e., for $\gamma\rightarrow \infty$) discrete group transition.

In order to see how the effective coupling $\gamma_{eff}$ comes about, we
consider the
partition function for the action (\ref{newbhanot})

\beq Z=\int {\cal D}\theta(\link) \exp(\sum_{\Box}(\gamma cos(\theta)  +
\beta cos (\theta/2))). \eeq

\nin It can  be rewritten as

\beq  Z=  \int  {\cal  D}\hat{\theta}(\link)   \exp(\sum_{\Box}(\gamma
\cos(\hat{\theta})+\log(\cosh(\beta(\cos(\hat{\theta}/2)-1)))+\log(1+\langle
\sigma \rangle_{\hat{\theta}}(\tanh(\beta(\cos\hat{\theta}/2-1))))
\label{zexact}
\eeq

\nin  where

\beq \sigma= \mbox{sign}\cos(\theta/2) \eeq

\nin and where the variable  $\hat{\theta}$, which takes values on the interval
$0\leq \hat{\theta}\leq 2\pi$, is related to $\theta$ by

\beq \hat{\theta}=\left \{ \begin{array}{c} \theta \mbox{ for } \sigma=+1
\\     \theta\pm     2\pi     \mbox{ for }     \sigma=-1      \end{array}
\right \}\;(\mbox{mod } 4\pi) \eeq

\nin and

\beq    \langle    \sigma    \rangle_{\hat{\theta}}=\langle     \sigma
\rangle_{\mbox{\tiny with restriction }\theta (\Box_{\tiny A}\;) = \hat{\theta}
\;
(\mbox{\tiny mod }2\pi)}=
\eeq

\beq     =     \frac{\int     {\cal     D}\theta(\link)e^{S}      \delta
(\theta(\Box_{\tiny                   A})-\hat{\theta}\;(\mbox{mod}
2\pi))\sigma(\Box_{\tiny A}\;)}{\int {\cal  D}\theta(\link)e^{S}  \delta
(\theta(\Box_{\tiny A}\;)-\hat{\theta}\;(\mbox{mod} 2\pi))} \eeq

\nin where $\Box_{\tiny A}$ is some fixed plaquette (that due to
long distance translational invariance can be arbitrarily chosen).
Up to now, this (rather formal) treatment has been exact.

The effective coupling is defined by requiring equality of
averages of the second
derivatives of two expressions for the action: namely the action
$\gamma_{eff}\cos\theta$ and the action appearing as the exponent of
(\ref{zexact}); that is,

\beq \langle \frac{d^2}{d\theta^2}(\gamma_{eff}\cos\theta )\rangle=
\label{2act}\eeq

\[ \langle \frac{d^2}{d\theta^2}(\gamma
\cos(\hat{\theta})\;\;+\;\;\log(\cosh(\beta(\cos(\hat{\theta}/2)-1)))
\;\;+\;\;\log(1+\langle
\sigma \rangle_{\hat{\theta}}(\tanh(\beta(\cos\hat{\theta}/2-1)))
\rangle.  \]

\nin Before taking the derivative on the right-hand side of (\ref{2act}),
we expand the second and third terms of the action in the exponent of
(\ref{zexact}) in the small
quantity $\beta(\cos(\hat{\theta}/2)-1)$. To leading order, the second term
in the exponent of (\ref{zexact}) is

\beq                        \log(cosh(\beta(\cos(\hat{\theta}/2)-1)))=
\frac{1}{2}(\beta(\cos(\hat{\theta}/2)-1))^2          +          \dots
\approx \frac{1}{2}(\beta(-(\frac{\hat{\theta}^2}{8}))^2)
\cdots \label{secondterm}\eeq

\nin while the third term to leading order in
$\beta(\cos(\hat{\theta}/2)-1)$ is

\beq                       \log(1+\langle                       \sigma
\rangle_{\hat{\theta}}(tanh(\beta(\cos(\hat{\theta}/2)-1)))) \approx
\langle                       \sigma
\rangle_{\hat{\theta}}(\beta(\cos(\hat{\theta}/2)-1)).
\label{hop} \eeq

\nin Performing the derivatives in (\ref{2act}) yields

\beq \langle \gamma_{eff}\cos\hat{\theta}\rangle=\langle\gamma\cos\hat{\theta}+
\frac{\langle\sigma\rangle_{\hat{\theta}}\beta}{4}\cos(\hat{\theta}/2)
\rangle  \label{derivative}\eeq

\nin where on the right-hand side the term with $\cos\hat{\theta/2}$
arises as the second derivative of the leading term in (\ref{hop}):

\beq   \langle\sigma\rangle_{\hat{\theta}}\beta(\cos(\hat{\theta}/2)-
1))\eeq

\nin which is of degree one in $\beta(\cos(\hat{\theta}/2)-1)$.
In the approximation used, the contribution from the leading
term in (\ref{secondterm}) is neglected as this term is of second degree in
$\beta(\cos(\hat{\theta}/2)-1)$.

\nin Rewriting $\cos(\hat{\theta}/2)$ as $\frac{\cos(\hat{\theta}/2)}
{\cos\hat{\theta}}\cos\hat{\theta}$ on the right-hand side of
(\ref{derivative}), we can extract the effective coupling $\gamma_{eff}$
as

\beq \gamma_{eff}=
\gamma+\frac{\langle\sigma\rangle_{\hat{\theta}}\beta}{4}
\langle\frac{\cos(\hat{\theta}/2)}
{\cos\hat{\theta}}\rangle \approx
\gamma+\frac{\langle\sigma\rangle_{\hat{\theta}}\beta}{4}
\langle \cos^{-\frac{3}{4}}\hat{\theta}\rangle  \label{gammaeffect}
\eeq

In the roughest approximation, we take
$\langle \cos^{-\frac{3}{4}}\hat{\theta}\rangle=1$
in (\ref{gammaeffect}) and thereby obtain $\gamma_{eff}$ as

\beq \gamma_{eff}=\gamma +
\langle     \sigma      \rangle_{\hat{\theta}}\frac{\beta}{4}
\approx \gamma +
\langle     \sigma      \rangle\frac{\beta}{4}
\label{actz2}  \eeq

\nin where in the last step, $\langle     \sigma      \rangle_{\hat{\theta}}$
has been replaced by  $\langle \sigma \rangle$
inasmuch as $\langle     \sigma      \rangle_{\hat{\theta}}$ is
to a good approximation independent of $\hat{\theta}$. The reason is that the
region in $\hat{\theta}$ over which we shall average is not very large - even
for critical $\gamma$. This combined with the fact that
$\langle \sigma\rangle_
{\hat{\theta}}$ depends (for symmetry reasons) to lowest order on
$\hat{\theta}^2$ allows us to ignore the dependence of $\langle \sigma\rangle$
on $\hat{\theta}$.

Near the boundary ``1'' separating the totally confining phase from the phase
where $\bz_2$ alone is confined, it is claimed that
the physics is quite accurately described by a particular single
combination  of  the
two  lattice  action  parameters  $\beta$  and  $\gamma$  that can be
used as a replacement for the dependence on both parameters.
That this is a rather good approximation has to do
with the fact that  fluctuations
in the ${\bf Z}_2$ degrees of freedom are strong all the way along the
phase border ``1'' because $\bz_2$ is confined on both sides of this
boundary. This gives rise to a very effective averaging over the distribution
at
$\theta$ and $\theta+2\pi$; this combined with the argument
that the dependence of the distribution on $\hat{\theta}$ is small
means that
the  information content in both  $\gamma$  and
$\beta$ that is  relevant is manifested essentially as a  single  parameter
$\gamma_{eff}$.

In particular, both the continuum coupling and the question of
which phase is realized (i.e.
the position of the phase boundary ``1'')
should,  in  the  region  where this approximation is valid,
only depend the single  parameter  $\gamma_{eff}$.  Hence
the continuum coupling will not vary along  this  phase  border.  This
implies that $\gamma_{eff}$  will have the same value at the triple point
(TP)
as for $\beta=0$.
At the TP, there are three corners because three phases meet here;
each
has it own continuum  coupling provided the phase transitions are first order.
The  above  argument  leads  to  the
conclusion that the continuum coupling at the multiple point in
the corner of the phase with alone ${\bf  Z}_2$  confined
equals the value of this coupling in the same phase but where
$\beta=0$ and where  $\gamma$  is  infinitesimally  above  $\gamma_{ crit}$.
Analogously, the continuum coupling in the totally confining phase (to
the extent that this makes sense) is the same at the multiple point
corner and the
point in this phase where $\beta =0$  and  $\gamma$  is  infinitesimally
below the critical value.

If we want to be able to provoke confinement solely  along other
discrete subgroups
than $\bz_2$ (e.g., along $\bz_3$), an action more general than (\ref{mix})
is needed.
Such a more general action would be

\beq
S=\gamma\cos \theta +\beta_2\cos\frac{\theta}{2}+\beta_3\cos\frac{\theta}{3}
+\beta_6\cos\frac{\theta}{6} \label{fullact} \eeq

\nin Taking the second derivative of $S$:

\beq
-S^{\prime\prime} =\gamma\cos \theta +\frac{\beta_2}{4}\cos\frac{\theta}{2}
+\frac{\beta_3}{9}\cos\frac{\theta}{3}+\frac{\beta_6}{36}\cos\frac{\theta}{6}
\label{fullactder}  \eeq

\nin Assume that $\gamma$ is large compared to $\beta_2$, $\beta_3$, and
$\beta_6$. We can then write

\beq \gamma_{eff}=(-S^{\prime\prime}(0))P_0+(-S^{\prime\prime}(2\pi)P_2+
(-S^{\prime\prime}(4\pi))P_4+(-S^{\prime\prime}(6\pi))P_6
+(-S^{\prime\prime}(8\pi))P_8+(-S^{\prime\prime}(10\pi))P_{10} =\eeq

\beq
=\left(\gamma+\frac{\beta_2}{4}+\frac{\beta_3}{9}+\frac{\beta_6}{36}\right)P_0+
\left(\gamma-\frac{\beta_2}{4}-\frac{\beta_3}{18}+\frac{\beta_6}{72}\right)P_2+
\left(\gamma+\frac{\beta_2}{4}-\frac{\beta_3}{18}-\frac{\beta_6}{72}\right)P_4+
\eeq

\[+\left(\gamma-\frac{\beta_2}{4}+\frac{\beta_3}{9}-
\frac{\beta_6}{36}\right)P_6+
\left(\gamma+\frac{\beta_2}{4}-\frac{\beta_3}{18}-\frac{\beta_6}{72}\right)P_8
\left(\gamma-\frac{\beta_2}{4}-\frac{\beta_3}{18}+
\frac{\beta_6}{72}\right)P_{10}  \]

\nin where $P_0,P_2,P_4,P_6,P_8$ and $P_{10}$ are the probabilities
that a plaquette takes a value near (in the corresponding
sequence) $0,2\pi,4\pi,6\pi,8\pi$ and $10\pi$.
Regrouping, we have

\beq \gamma_{eff}=\underbrace{P_0+P_2+P_4+P_6+P_8+P_{10})}_{=1}\gamma + \eeq

\[+\frac{\beta_2}{4}\underbrace{(P_0(1)+P_2(-1)+P_4(1)+P_6(-1)+P_8(1)
+P_{10}(-1))}_{\langle \sigma_{\sbz_2} \rangle}+ \]
\[+
\frac{\beta_3}{9}\underbrace{(P_0(1)+P_2(-\frac{1}{2})+
P_4(-\frac{1}{2})P_6(1)+P_8(-\frac{1}{2})+
P_{10}(-\frac{1}{2}))}_{\langle \sigma_{\sbz_3} \rangle}+ \]
\[
+\frac{\beta_6}{36}\underbrace{(P_0(1)+P_2(\frac{1}{2})+
P_4(-1)+P_6(-1)+P_8(-\frac{1}{2})+
P_{10}(\frac{1}{2}))}_{\langle \sigma_{\sbz_6} \rangle}
=\]

\beq =\gamma + \frac{\beta_2}{4}\langle\sigma_{\bz_{2}}\rangle+
               \frac{\beta_3}{9}\langle\sigma_{\bz_{3}}\rangle+
              \frac{\beta_6}{36}\langle\sigma_{\bz_{6}}\rangle
               \label{gamefz23}\eeq

\nin where

\beq \sigma_{\sbz_2}=\mbox{ sign }\cos(\theta/2) \eeq

\[ \sigma_{\sbz_3}=\mbox{ sign }\cos(\theta/3) \]

\[ \sigma_{\sbz_6}=\mbox{ sign }\cos(\theta/6) \]

Equation (\ref{gamefz23}) contains (\ref{actz2}) as a special case; the
more detailed derivation of (\ref{actz2}) was included for illustrative
purposes.

Note that with the action (\ref{fullact}), Bianchi identities are now
fulfilled modulo $12\pi$. The analogy to the remarks pertaining to monopoles
immediately following (\ref{newbhanot}) are for the action (\ref{fullact})
that ``full'' monopoles correspond to charge $12\pi$ and ``minimal strength''
monopoles - denoted $\frac{2\pi}{12\pi}$ - to the ``length'' of the
factor group $U(1)/\bz_6$. These ``minimal strength'' monopoles will be
described as monopoles relative to a $\bz_6$ background field or
alternatively as monopoles modulo a $\bz_6$ background. These remarks
become more relevant in the following section where we consider the
effect of including phases at the multiple point that are critical w.r.t.
$\bz_2$ and $\bz_3$.

\subsection{Estimating the degree of ``first orderness'' in the
transition from the $\bz_2$ confining phase to the totally Coulomb
phase at the triple point} \label{deltagam}

In the limit
of very large $\gamma$ values, the phase transition at border ``2'' becomes
a pure $\bz_2$ transition inasmuch as all the probability is concentrated
at a $\bz_2$ subgroup of $U(1)$. We want to use known results for $\bz_2$
to estimate the degree of ``first-orderness'' of the transition in crossing
the boundary ``2'' at the multiple point.
A proper $\bz_2$ transition corresponds to infinite
$\gamma$ whereas $\gamma$ at the multiple point is of the order unity. However,
we expect the phase transition in crossing the border ``2'' at the triple
point to inherit to some extent the properties (i.e., a degree of
first-orderness) of a $\bz_2$ phase transition even though $\gamma$ at the
triple point is only of order unity. The reason is that, also at the
triple point, the transition at the border ``2'' (from the phase with
$\bz_2$ alone confining to the totally Coulomb phase) really only involves
the $\bz_2$ \dof. That the transition in crossing border ``2''
at the triple point presumably does not have the full degree of first-orderness
of a
pure $\bz_2$ transition is due to the importance of group elements of $U(1)$
that
depart
slightly (and continuously) from the elements of $\bz_2\subset U(1)$.
Having such
elements  make
possible ``$\frac{2\pi}{4\pi}$-monopoles'' the density of which increases
as $\gamma$ becomes smaller.
What are ``$\frac{2\pi}{4\pi}$-monopoles''? Here we make connection with
the remarks immediately following (\ref{newbhanot}) and, more generally,
the remarks in the last paragraph of the preceding section.
Think of the six plaquettes bounding a 3-cube.
In the phase with $\bz_2$ alone confining (and with $\gamma$ large but
not infinite), plaquette configurations of a 3-cube
can involve an odd number of
plaquettes that have plaquette variable
values near the nontrivial element of $\bz_2$
(in the notation of (\ref{newbhanot}) in which Bianchi identities are
fulfilled
modulo $4\pi$, the nontrivial element of $\bz_2$ corresponds to $2\pi$ so
$\bz_2=\{0,2\pi\}\subset U(1)$) in combination with small deviations from
$\bz_2$ (the deviations lie along $U(1)$ in which of course $\bz_2$ is
embedded)
such that together the six plaquette values of a 3-cube sum to zero
(mod $4\pi$ in the notation of (\ref{newbhanot})).
We can regard the flux through such a configuration as that coming from a
``$\frac{2\pi}{4\pi}$-monopole'' relative to a $2\pi$ ``background'' flux
coming from the general abundance of plaquettes having the value near
the (nontrivial) element $2\pi\in \bz_2\subset U(1)$.

If one considers an isolated $\bz_2$ theory (i.e., a $\bz_2$ that is not
embedded in a $U(1)$ as is the case for  infinite $\gamma$),
there can be no monopoles because there is for
$\bz_2$ no way to have 6 ``small'' elements that sum up to a circumnavigation
of the whole group.
However, for finite $\gamma$, the distribution of group elements
accessible due to quantum fluctuations spreads out slightly from $\bz_2$ to
$U(1)$
elements ``close to $\bz_2$'' with the result that
it is possible to have $\frac{2\pi}{4\pi}$-monopoles in the
sense introduced above. In other words,
in the phase with only $\bz_2$ confining, it is possible to have monopoles
modulo a $\bz_2$ background (i.e.,$\frac{2\pi}{4\pi}$-monopoles)
if $\gamma$ is not so large as to
preclude
continuous plaquette variable deviations from $\bz_2$ along $U(1)$ of
sufficient magnitude so that these deviations from $\bz_2$
for plaquette values of a 3-cube can
add up to the length of the factor group $U(1)/\bz_2$. When
Bianchi identities are satisfied modulo $4\pi$ by such configurations,
we can say that we get half (i.e., $\frac{2\pi}{4\pi}$) of the way to $0$
(mod $4\pi$)
using $\frac{2\pi}{4\pi}$-monopoles; the other half of the way to $0$
(mod $4\pi$)
is provided by the $2\pi$ background field having as the source an
odd number of plaquettes with values near the nontrivial element of
$\bz_2\subset U(1)$.

\begin{figure}
\centerline{\epsfxsize=\textwidth \epsfbox{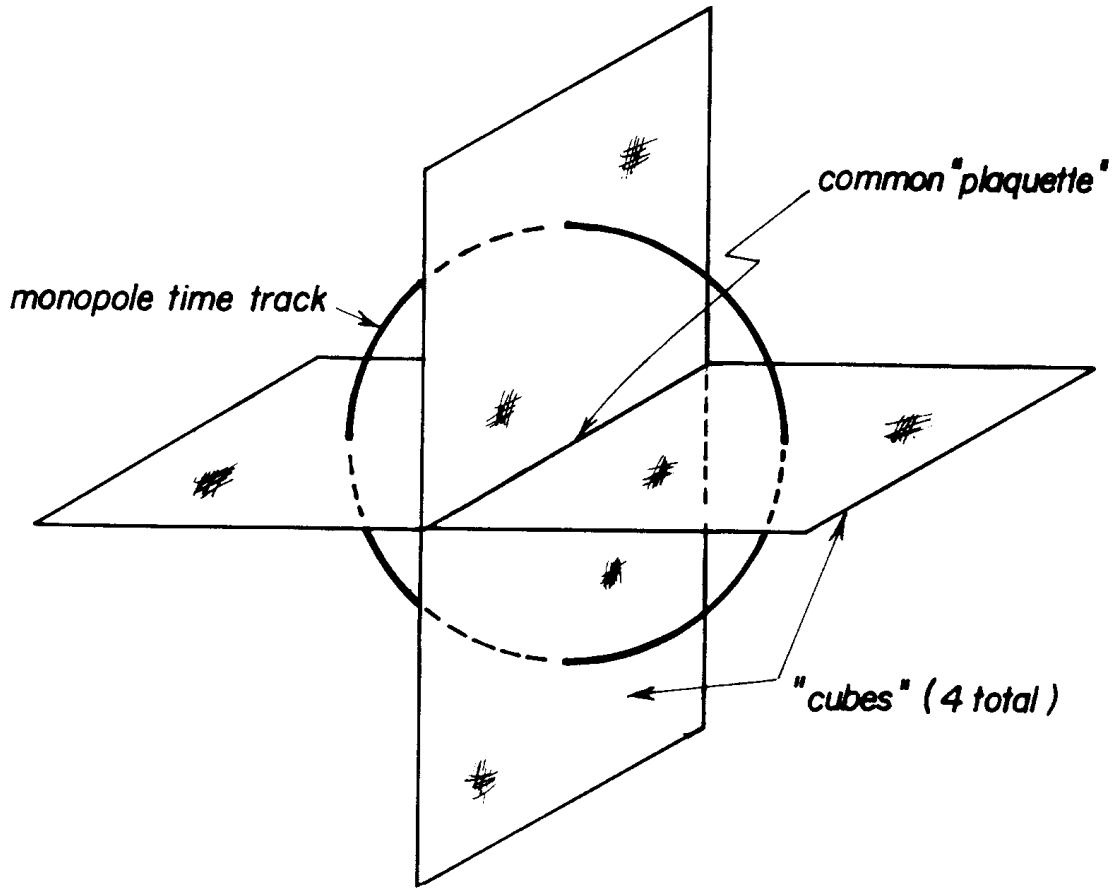}}
\caption[The dominant configuration for monopoles]{\label{domconfig}
The important monopoles are expected to be of minimal strength and to be
found essentially only in the dominant configuration of four cubes
surrounding a common plaquette. The dominant configuration is illustrated
above in a picture having one dimension less than the actual (4-dimensional)
dominant configuration. The actual dominant configuration - i.e.,
a plaquette
common to four 3-cubes has in the above dimensionally reduced picture
become a link common to four plaquettes.}
\end{figure}

In the sequel, we shall restrict our attention to ``minimal strength''
monopoles\footnote{A ``minimal strength'' $\bz_N$ monopole is a
configuration of 6 plaquettes surrounding a 3-cube such that the sum of
continuous deviations from elements of $\bz_N$ add up to the length of the
factor group $U(1)/Z_N$.} (i.e.,
$\frac{2\pi}{4\pi}$-monopoles in the case of the action (\ref{mix}))
inasmuch as such ``minimal strength'' monopoles in the
dominant configuration in which a foursome of 3-cubes encircles a common
plaquette. This dominant configuration which is illustrated in
Figure~\ref{domconfig}
can be expected
to constitute the
vast majority of the monopoles present. In the case of the
action~(\ref{newbhanot}), the dominant monopoles
are the  $\frac{2\pi}{4\pi}$
monopoles (These are the only possible only less than full strength monopoles)
In the case of the action~(\ref{fullact}), minimal strength (and
presumably dominant)
monopoles are $\frac{2\pi}{12\pi}$ monopoles; in principle there could also be
monopoles of strength $4\frac{4\pi}{12\pi}$ and $\frac{6\pi}{12\pi}$.

We claim that as $\gamma \rightarrow \infty$, the
probability of having such a dominant configuration monopole decreases
exponentially;
accordingly
there is only a thin population of minimal ``strength monopoles'' (and an
even much thinner population of monopoles other than the ``minimal strength''
type). Hence it is
presumably a very good approximation to describe the presence of
monopoles as due solely to the dominant configuration of ``minimal strength''
monopoles.

In the case of the action (\ref{newbhanot}), this means
four 3-cube $\frac{2\pi}{4\pi}$-monopoles that encircle a common
plaquette having a value corresponding to the nontrivial element of $\bz_2$.
Consider by way of example the case where each of the four
3-cubes in this dominant configuration have
the value $2\pi/5$
on  five plaquettes (with the sixth ``encircled'' common plaquette
having the value $\pm 2\pi$). Such a 3-cube configuration would, relative to a
$\bz_2$ background flux (expected for large $\gamma$ and small
$\beta$'s), behave as a $\frac{2\pi}{4\pi}$-monopole with a flux of $2\pi/5$
emanating from each of five plaquettes.

The dominant-configuration $\frac{2\pi}{4\pi}$-monopoles
can be expected to occur with some
low but nonzero density in the lattice near the phase border ``2'' even
for large (but not too large) $\gamma$ values. Our suspicion, confirmed by
calculations below, is that the degree of ``first-orderness'' of the phase
transition at the boundary ``2'' is greater the smaller the chance that
small deviations from $\bz_2$ (lying in $U(1)$) can, for the six
plaquettes of a 3-cube, add up to a
$\frac{2\pi}{4\pi}$-monopole (or, stated equivalently, add up to
the length of the factor group $U(1)/\bz_2$).

As $\gamma$ decreases,
an increasing number of $\frac{2\pi}{4\pi}$-monopoles is encountered.
At the triple point (TP), where $\gamma\approx 1$,
the presence of a larger number of $\frac{2\pi}{4\pi}$-monopoles than for
very large $\gamma$
mitigates but does not eliminate the high degree of ``first-orderness''
characteristic of pure $\bz_2$ transitions (for which the deviations
from $\bz_2$ (along $U(1)$)
of six 3-cube
plaquette variable values cannot sum to the length of the whole $U(1)/\bz_2$
due to $\gamma$ being too large).

In order to deal quantitatively with the effect of
$\frac{2\pi}{4\pi}$-monopoles, and
thereby with the question of how much of the behaviour of a pure
$\bz_2$ transition is inherited by the phase transition at border ``2''
at the triple point, it is
useful to define {\em two} new variables $\on$ and $\B$:

\beq \on\stackrel{def}{=} \left \{ \begin{array}{l}
+1 \mbox{ if $U(\Box)$ closest to $e^{i0}\in \bz_2$} \\
-1 \mbox{ if $U(\Box)$ closest to $e^{i\pi}\in \bz_2$} \end{array} \right.
\eeq

\nin The other new variable $\B$ (the subscript ``$BIO$''
is an acronym for \underline{B}ianchi \underline{I}dentity
\underline{O}beying) is defined as follows:

\beq \B\stackrel{def}{=} \on \cdot \eeq

\[\cdot\left \{ \begin{array}{l}
+1 \mbox{ \footnotesize for ordinary $\Box$
(i.e. {\em not} the encircled $\Box$ in the dominant config.)} \\
-1 \mbox{ \footnotesize  for $\Box$ encircled by four 3-cube
$\frac{2\pi}{4\pi}$-monopoles in the dominant config.} \end{array} \right.
\]

\nin The variable $\B$ differs from
the variable $\on$ only by a sign change of $\on$ in the case where
the plaquette $\Box$ coincides with the ``encircled'' plaquette. The
``encircled''
plaquette is
always present in the four 3-cube
$\frac{2\pi}{4\pi}$-monopoles of the dominant monopole
configuration.

Let us make the observation that the values assigned by the variable
$\on$ to the plaquettes of a 3-cube satisfy the $\bz_2$-Bianchi identity if the
3-cube is not a $\frac{2\pi}{4\pi}$-monopole;
i.e., in our approximation,  not one of the four 3-cube
$\frac{2\pi}{4\pi}$-monopoles encircling
a common plaquette
in the dominant $\frac{2\pi}{4\pi}$-monopole configuration. Note, however,
that the Bianchi
identity {\em is} violated by the values assigned by the variable $\on$ to the
plaquettes of a 3-cube when there is a $\frac{2\pi}{4\pi}$-monopole.
For instance, it is readily seen that
for the very special $\frac{2\pi}{4\pi}$-monopole example given above
($U(\Box)=e^{i\frac{2\pi}{4\pi}\cdot 2\pi}=-1$ on the ``encircled''
plaquette; $U(\Box)=e^{i2\pi/5}$ on the remaining 5 plaquettes of the
$\frac{2\pi}{4\pi}$-monopole
3-cube), the Bianchi identity is violated:

\beq \prod_{\Box\in \partial(\frac{2\pi}{4\pi}-monopole\;3-cube)}\on=
(-1)\cdot 1^5=-1  \neq 1 \eeq

\nin inasmuch as $\on=-1$ for $U(\Box)=-1$ and $\on=1$ for
$U(\Box)=e^{i2\pi/5}$.

More generally, a $\frac{2\pi}{4\pi}$-monopole (which really just means a
monopole
modulo a $\bz_2$ background) consists of a configuration of plaquette
variable values of a 3-cube that deviate continuously from elements of
$\bz_2$ in such a
way that the total sum of continuous deviations (lying in $U(1)$)
from $\bz_2$ equals, modulo
$4\pi$, $2\pi$ multiplied by the number of plaquettes  for which the continuous
deviations are centred at the nontrivial element of $\bz_2$. Note that in
order to have a monopole, an odd number of the six plaquettes of a three
cube must be near the nontrivial element (i.e., $2\pi$) of $\bz_2$

Even more generally, we have for a monopole modulo a $\bz_N$ background (i.e.,
a monopole for which the continuous $U(1)$ deviations from $Z_N\subset U(1)$
add up to a multiple of the length of the factor group $U(1)/\bz_N$):

\beq
\prod_{\mbox{\tiny $\Box \in$ 3-cube}}( U(\Box)g_{_{\tiny
nearest}}(U(\Box))^{-1}) =
\prod_{\mbox{\tiny $\Box \in$ 3-cube}} g_{\tiny nearest}(U(\Box))
\;\;\;\;\;\;\;
(g_{\tiny nearest}(U(\Box))\in \bz_N) \eeq

\nin where $g_{\tiny nearest}(U(\Box))$ is defined as that element of $\bz_N$
which is nearest to $U(\Box)$:

\beq dist^2(U(\Box),g_{\tiny\; nearest}(U(\Box)))\stackrel{def}{=}
inf\{ dist^2(U(\Box),g^{\prime})\} \;\;\;\;(g^{\prime}\in \bz_N) \eeq

\nin where $dist^2(U(\Box),g^{\prime})$ denotes the squared distance from a
plaquette variable value $U(\Box)$ and an element
$g^{\prime}\in \bz_N$.
We are really interested in $\bz_6=\bz_2\times \bz_3$ inasmuch
as we are also interested in the modification of first-orderness due to
an increasing number of monopoles modulo $\bz_3$ in going from large $\gamma$
to the triple point.
However, for the purpose of exposition, we continue to use the
example of monopoles modulo $\bz_2$.

With the modification of the variable $\on$ that defines the variable
$\B$, we have in $\B$ a variable that, for sufficiently large $\gamma$,
assigns values to
configurations of plaquettes that respects the $\bz_2$
Bianchi identities - also for $\frac{2\pi}{4\pi}$-monopole configurations
(when the monopoles are of the dominant configuration type which is the only
type for which the variable $\B$ is defined).

Note that the variable $\B$ differs from the variable $\on$ only if there
are $\frac{2\pi}{4\pi}$-monopoles. For $\gamma \rightarrow \infty$ such
monopoles disappear and $\B=\on\in \bz_2=\{+1,-1\}$.
In going to smaller values of $\gamma$
in the phase with only $Z_2$ confining, an increasing range of
fluctuations along $U(1)$ centred at the elements of
$\{0,2\pi\} =\bz_2 \subset U(1)$ provide  alternative
(Bianchi identity-obeying)
configurations that supplement the essentially discrete
group-valued plaquettes characteristic of large $\gamma$ configurations.

We want now to determine approximately the ($\gamma$ dependent) relation
between
the distributions of the two variables $\on$ and $\B$. The average value of
$\on$ is estimated
using the identity

\beq \langle \on \rangle = P(\B=+1)\langle\on \rangle_{(\B=+1)}+ \eeq
                       \[    +P(\B=-1)\langle\on \rangle_{(\B=-1)} \]

\nin where $P(\B=+1)$ and $P(\B=-1)$ denote respectively the probabilities
that $\B=+1$ and $\B=-1$ while $\langle\on \rangle_{\B=+1}$ and
$\langle\on \rangle_{\B=-1}$ denote averages of $\on$ subject respectively
to the constraints that $\B=+1$ and $\B=-1$.

Denoting by $\xi=\xi(\gamma)$ the
($\gamma$ dependent) probability that a plaquette coincides with the
``encircled plaquette'' of the dominant $\frac{2\pi}{4\pi}$-monopole
configuration,
there obtains

\beq \langle\on \rangle_{\B=+1}
=\frac{e^{\beta}\cdot 1 +\xi e^{-\beta}\cdot(-1)}{e^{\beta}+\xi e^{-\beta}}
\eeq

\nin and

\beq \langle\on \rangle_{\B=-1}
=\frac{e^{-\beta}\cdot (-1) +\xi e^{\beta}\cdot(+1)}{e^{-\beta}+\xi e^{\beta}}
\eeq

Using

\beq P(\B=+1)=\frac{1}{2}+\frac{1}{2}\langle \B \rangle \eeq

\nin and

\beq P(\B=-1)=\frac{1}{2}-\frac{1}{2}\langle \B \rangle     \eeq

\nin we have

\beq \langle \on \rangle=\frac{1}{2}\left(\frac{e^{\beta}-\xi e^{-\beta}}
{e^{\beta}+\xi e^{-\beta}}+
\frac{\xi e^{\beta}-e^{-\beta}}{\xi e^{\beta}+e^{-\beta}}\right)+
\frac{1}{2}\left(\frac{e^{\beta}-\xi e^{-\beta}}{e^{\beta}+\xi e^{-\beta}}-
\frac{\xi e^{\beta}-e^{-\beta}}{\xi e^{\beta}+e^{-\beta}}\right)
\langle \B \rangle \eeq

\beq \approx \langle \B \rangle(1-2\xi\cosh 2\beta)+2\xi\sinh 2\beta
\label{on} \eeq

\nin where in the last step we have used that $\xi$ is assumed to be small.

We want now to calculate the jump in (\ref{actz2}) in going from the
phase with only $\bz_2$ confining to the totally Coulomb phase at the
the triple point. That is, we want $\Delta \gamma_{eff}$ along the boundary
``2'' in Figure~\ref{figbhanot} as a function of $\gamma$:

\beq \Delta \gamma_{eff}=\Delta(\gamma+\langle \sigma \rangle
\frac{\beta_{crit}(\gamma)}{4})=
\Delta\langle \sigma \rangle \frac{\beta_{crit}(\gamma)}{4}=
\frac{\beta_{crit}(\gamma)}{4}\Delta\langle \on \rangle.
\label{hm}\eeq

\nin where we have made the identification
$\langle \sigma \rangle=\langle \on \rangle$. Substituting (\ref{on})
into (\ref{hm}) we get

\beq  \frac{\beta_{crit}(\gamma)}{4}\Delta\langle \on \rangle=
\frac{\beta_{crit}(\gamma)}{4}\Delta\langle \B \rangle
(1-2\xi\cosh 2\beta(\gamma)) \label{hmm} \eeq

\nin In our approximative procedure we identify $\Delta\langle \B \rangle$
with the jump in $\Delta \langle S_{\Box} \rangle$ for a $\bz_2$ gauge theory
since the phase transition ``2'' at the triple point is determined from
the phase of $\bz_2$.

Let us define a parameter $\beta_{BIO}$ as the action parameter $\beta$ in
a $Z_2$ gauge theory which optimally reproduces the distribution of the
variables $\B$ in the $U(1)$ theory (with the mixed action (\ref{newbhanot}))
by using an action of the form

\beq S=\beta_{BIO}\sum_{\Box}\B. \eeq

\nin In other words, $\beta_{BIO}$ is defined such that

\beq \langle\B  \rangle_{\mbox{\tiny in }U(1) \mbox{\tiny theory with }
S=S(\beta,\gamma)}=
\langle\B  \rangle_{\sbz_2 \mbox{ \tiny theory with}
S=\beta_{BIO}\sum_{\Box}\B } \eeq

We now want to obtain $\beta_{BIO}$ as a function of $\beta$ and $\xi$ (and
hereby $\gamma$ inasmuch as $\xi=\xi(\gamma)$) by equating the ratio
of the probabilities

\beq \frac{P(\B=1)}{P(\B=-1)} \label{probratio}\eeq

\nin for the two actions $S=S(\beta,\gamma)$ and $S_{BIO}=
\beta_{BIO}\sum_{\Box}\B$:

\beq
\frac{\overbrace{e^{\beta}}^{\on=+1}+\overbrace{\xi e^{-\beta}}^{\on=-1}}
{\underbrace{e^{-\beta}}_{\on=-1}+\underbrace{\xi e^{\beta}}_{\on=+1}}=
\frac{e^{\beta_{BIO}}}
{e^{-\beta_{BIO}}}.
\label{onto}\eeq

\nin This procedure for estimating $\beta_{BIO}$ is somewhat errant in that
Bianchi identities are ignored on both sides of equation (\ref{onto}) in
various ways: first in the calculation of the ratio (\ref{probratio}) and
second, and presumably less importantly, in the simulation-by a-$\bz_2$
theory that defines $\beta_{BIO}$. The hope is that these error roughly
cancel inasmuch as the same error is present on both sides of the equation.

Taking the logarithm of both
sides of (\ref{onto}) and solving for $\beta_{BIO}$ yields

\beq \beta_{BIO}=\beta+
\frac{1}{2}\log\frac{1+\xi e^{-2\beta}}{1+\xi e^{2\beta}} \label{betabio} \eeq

\nin We want to use (\ref{betabio}) to relate $\beta_{BIO}$ and
$\beta_{crit}(\gamma)$ along the boundary ``2'' in Figure~\ref{figbhanot}.
Using that  $\xi << e^{\beta},\;1$ there obtains

\beq \beta_{crit\;BIO}\approx
\beta_{crit}(\gamma)+\frac{1}{2}\xi(e^{-2\beta}-e^{2\beta})
=\beta_{crit}(\gamma)-\xi \sinh 2\beta. \label{betabioa} \eeq

\nin Substituting (\ref{betabioa}) for $\beta_{crit}(\gamma)$ on the right-hand
side
of (\ref{hmm}) yields

\beq
\Delta \gamma_{eff}=\frac{\beta_{crit}(\gamma)}{4}\Delta\langle \on \rangle=
\frac{1}{4}\beta_{crit\;BIO}\left(1+\frac{\xi\sinh 2\beta}{\beta_{BIO}}\right)
\Delta \langle \B \rangle(1-2\xi \cosh 2\beta) = \eeq

\beq = \frac{1}{4}\beta_{crit\;BIO}\Delta\langle \B \rangle
\left(1+\xi\left(\frac{\sinh 2\beta}{\beta_{crit\;BIO}}-2\cosh
2\beta\right)\right).
\label{just} \eeq

\nin Solving (\ref{betabioa}) for $\xi$ and substituting into (\ref{just})
yields

\beq  \Delta \gamma_{eff}=
\frac{1}{4}\beta_{crit\;BIO}\Delta\langle \B \rangle
\left(1+(\beta_{crit}(\gamma)-
\beta_{crit\;BIO})\left(\frac{1}{\beta_{crit\;BIO}}-
\frac{2}{\tanh 2\beta}\right)\right). \eeq

{}From the literature \cite{creutz1}
we have values for $\langle S_{\Box} \rangle_
{\sbz_2}=
\Delta\langle \B \rangle$ and $\beta_{crit\;BIO}$. The quantity
$\beta_{crit}(\gamma)-\beta_{crit\;BIO}$ is estimated graphically using
a $U(1)$ phase diagram found in the literature\cite{bhanot1}  corresponding
to the action (\ref{newbhanot}). It is
now finally possible to calculate $\Delta\gamma_{eff}$ at the triple point for
the transition from the phase with only $\bz_2$ confining to the totally
Coulomb-like phase.

It is indeed fortunate that the subtraction $\frac{1}{\beta_{BIO}}-
\frac{2}{\tanh 2\beta}$ almost cancels thereby rendering our calculation
of $\Delta \gamma_{eff}$ rather insensitive to the large uncertainty in the
graphical estimate of $\beta_{crit}(\gamma)-\beta_{BIO}$. This means that the
major contribution  to the change in first-orderness in going from very large
$\gamma$ to $\gamma \approx 1$ at the triple point is achieved simply by
determining  $\beta_{crit}(\gamma)$ by the condition that
$\beta_{crit\;BIO}=\beta_{crit}(\gamma=\infty)$. This makes it possible to
perform an analogous correction to the first-orderness in going from a
pure $\bz_3$ theory to the triple point for an action $\gamma\cos \theta +
\beta_3 \cos \frac{\theta}{3}$ without having access to the phase diagram
for the $U(1)$ theory with an action of this form
                  (that we need for the graphical estimate of
$\beta_{TP\;crit}(\gamma)-\beta_{\sbz_3\;BIO}$).

In subsequent calculations, we shall make use of the fact
that the probability $\xi$
of having a
$\bz_2$ and a $\bz_3$ monopole must be roughly equal.
The argument goes as follows: we can assume
that essentially all monopoles present will be of the ``minimal strength''
type. In the case of the $SMG$, this means  monopoles
modulo a $\bz_6$ background: i.e., $\frac{2\pi}{12\pi}$-monopoles. These
are built up of $U(1)$ elements close to $\bz_6$ such that the deviations
from $\bz_6$ of six 3-cube plaquette variables add up to the full extent of
$U(1)/\bz_6$.
Of course it is still assumed that these ``minimal strength'' monopoles
essentially
only are found in
dominant configuration of four 3-cubes that encircle a common plaquette
But such a
``minimal strength'' monopole is a superposition of a $\bz_2$ and a $\bz_3$
monopole:

\beq \frac{2\pi}{12\pi}=\frac{6\pi}{12\pi}-\frac{4\pi}{12\pi} \eeq

Assuming a rarity of
$\pm \frac{2\pi}{4\pi}$-monopoles (i.e.,
$\pm \frac{6\pi}{12\pi}$-monopoles in the $12\pi$ normalisation)
as well as $\frac{\pm 4\pi}{12\pi}$-monopoles  (i.e.,
monopoles corresponding to the strength of a nontrivial element
of the $\bz_3$ subgroup),
monopoles are for all practical purposes exclusively of
the $\frac{2\pi}{12\pi}$ type. And each of these ``minimal strength''
monopoles is formally a linear combination
of exactly one $\bz_2$-monopole and one $\bz_3$-monopole. Hence these latter
monopole types are ``present'' in essentially equal numbers.

As we would like to include not only the degree of first orderness
inherited from $\bz_2$ at the triple point, but also that inherited from
$\bz_3$, we need to generalise (\ref{hm})
and (\ref{just})  which were
derived for $\bz_2$ alone. The generalisation of
(\ref{hm}) is obtained by varying (\ref{gamefz23}):

\beq \Delta \gamma_{eff}=\frac{\beta_2}{4}\Delta(\langle
\sigma_{\sbz_2}\rangle)+
                         \frac{\beta_3}{9}\Delta(\langle
\sigma_{\sbz_3}\rangle)+
                         \frac{\beta_6}{36}\Delta(\langle
\sigma_{\sbz_6}\rangle).
\label{delgamun}                         \eeq

\nin where the notation has been changed such that $\langle \sigma \rangle
\stackrel{def}{=}\langle \on \rangle$
in (\ref{hm}) is in (\ref{delgamun}) denoted by
$\langle \sigma_{\sbz_2} \rangle$. For $\bz_3$ the analogous
quantity is denoted by  $\langle \sigma_{\sbz_3} \rangle$
                                      in (\ref{delgamun}).
Moreover, we have the notational change
$\beta_{crit\;BIO}\rightarrow \beta_{crit\;\sbz_2}$ in going from (\ref{hm}) to
(\ref{delgamun}). In (\ref{delgamun}) the analogous
quantities for $\bz_3$ and $\bz_6$ are  denoted respectively as
$\beta_{crit\;\sbz_3}$ and $\beta_{crit\;\sbz_6}$.
We have taken the $\beta_6$ term in (\ref{delgamun}) as being zero. This is
presumably justified by the smallness of the $\bz_6$ ``jump'' contribution
when treated (incorrectly) as being independent of $\bz_2$ and
$\bz_3$.

In going to the  new notation,  (\ref{just}) becomes (for $\bz_2$)

\beq \Delta \gamma_{eff}=\frac{1}{4}\beta_{crit\;\sbz_2}\Delta\langle S_{\Box}
\rangle_{\sbz_2}
\left(1+\xi\left(\frac{\sinh 2\beta}{\beta_{crit\;\sbz_2}}-2\cosh
2\beta\right)\right)
\eeq

\nin The generalisation that also includes the discontinuity
inherited from $\bz_3$ that contributes to
$\Delta\gamma_{eff}$
at the triple point transition from the
phase with just the discrete subgroups $\bz_2$ and $\bz_3$
confining to the totally Coulomb-like phase is

\beq
\Delta \gamma_{eff}=\sum_{N\in\{2,3\}}
\frac{1}{N^2}\beta_{crit\;\sbz_N}\Delta\langle S_{\Box}\rangle_{\sbz_N}
\left(1+\xi\left(\frac{\sinh 2\beta}{\beta_{crit\;\sbz_N}}-2\cosh
2\beta\right)\right)
 \label{final}. \eeq

\nin From the argumentation above, we know  that $\xi$ is expected to have the
same value in both terms of (\ref{final}).

In (\ref{final}) it is seen that the subgroups $\bz_2$ and $\bz_3$
both contribute a term to $\Delta\gamma_{eff}$ at the triple point.
Presumably it is a good
approximation to calculate $\Delta\gamma_{eff}$ as if contributions from
$\bz_2$ and $\bz_3$ are mutually independent inasmuch as these subgroups
factorize at the multiple point. However, even in this approximation there
will still be an indirect interaction between these subgroups via the continuum
\dofx in $U(1)$ and via the encircled plaquette in the dominant monopole
configuration.
Using (\ref{final}), the contributions from $\bz_2$ and  $\bz_3$
to $\Delta\gamma_{eff}$ are calculated and tabulated in
Table \ref{tab4}.

\begin{table}
\caption[table4]{\label{tab4}The quantity $\Delta\gamma_{eff}$ calculated
using the appropriate terms in (\ref{final}).
In the last row, the quantities for $\bz_6$ are calculated (incorrectly)
in a manner analogous to that used for $\bz_2$ and $\bz_3$. This procedure
presumably overestimates the effect of $\bz_6$ contributions.}
\beq
\begin{array}{|l|l|l|l|l|}\hline
  &\beta_{crit\;\sbz_N} & \Delta\langle S_{\Box}\rangle_{\sbz_N} & \xi &
\Delta\gamma_{eff} \\
  \hline
\bz_2 & 0.44 & 0.44 & 0.04 & 0.047_3 \\
   \hline
\bz_3 & 0.67 & 0.56 & 0.04 & 0.039_3 \\
   \hline
(\bz_6) & (1.00) & (0.13) & (0.0437) & (0.0033) \\
\hline\end{array}
\eeq
 \end{table}

\subsection{Calculating the enhancement factor for $1/\alpha_{U(1)}$
corresponding to the Planck scale breakdown of $U(1)^3$ to the diagonal
subgroup} \label{enhance}

The two approximations that we have developed in order to gain an insight
into the phase diagram for the group $U(1)^3$ - the independent monopole
approximation and the group volume approximation - are more or less
suitable according to whether the phase transitions are second or
first order.

To determine the correct enhancement factor, we interpolate between the
independent monopole approximation that gives this factor as 6 and the
volume approximation that puts this factor at about 8. This interpolation
is done by calculating the jump $\Delta W_{\Box\;``3"}$ in the Wilson
operator at the  boundary ``3'' transition at the TP
(see Figure~\ref{figbhanot}) that reflects the degree of first-orderness
inherited
at this transition from pure $\bz_2$ and $\bz_3$ transitions. As
$\Delta \gamma_{eff}$ expresses the degree of
first-orderness at the TP in going into the totally Coulomb-like phase,
$\Delta W_{\Box\;``3"}$ is calculated using the assumption that it
depends essentially on $\Delta \gamma_{eff}$.

The first approximation is the monopole  condensate  approximation  in
which the relevant quantity for which phase is realized is
the amount of fluctuation  in  the  convolution  of  the  6  plaquette
variables enclosing a 3-cube.

In the second approximation - based on the group volume approximation -
it turns out that to attain the multiple point in the hexagonal symmetry
scheme, it is necessary to introduce
additional  parameters  in  the  form  of
coefficients to 4$th$ and 6$th$ order perturbations to the Manton  action.
These additional parameters  are  used  to   get  the  free  energy
functions (corresponding to different phases) to coincide in parameter
space at a point - ``the'' multiple point. This point is shared by what
we expect is a maximum number of phases.

If, for example,  the Coulomb to confining phase transition for a
$Peter$-$U(1)$
subgroup is purely
second order,  this phase transition would not be expected to
cause any change in at what value of the distance along another subgroup axis
(e.g., the $Paul$-axis)  the
first identification-lattice point is encountered.
The reason is that there is no discontinuous change in the degree of
fluctuation in the $Peter$-plaquette variable in making the transition.
In this case we expect the independent monopole approximation
to work well.

On the other hand, if the phase transition is
very strongly first order so that the fluctuations along the $Peter$-subgroup
become discontinuously larger upon
passing into the $Peter$ confinement phase, this can be expected to affect
the threshold at which other subgroups go into confinement in a sort of
``interaction effect''. In this situation the volume-approximation
can be useful because it can take into account (and actually overestimates)
the influence that fluctuations along different directions in the group
can have on each other. The independent monopole approximation
tends to ignore
this effect.

Because the group volume approximation accounts for the interaction effect
between
fluctuations along different subgroups, it was necessary to use
$4th$ and $6th$ order action terms in order to
get a multiple point at which 12 phases convene (corresponding to continuous
invariant subgroups; we neglect an infinity of discrete subgroups in this
approximation). The effect of
the higher order terms is the preferential enhancement of quantum
fluctuations along the one dimensional (nearest neighbour) subgroup
directions of
the identification lattice thereby effectively eliminating the influence
that fluctuations along one subgroup have on the fluctuations along
another subgroup and vice versa.

In fact, the volume approximation effectively replaces the gauge group $G$
by its factor group $G/H$ when $H$ has confinement-like behaviour.
This amounts to treating the fluctuations along the component of the group
lying within the cosets $gH\;\;(g\in G)$
as being so large that, as far as Bianchi identities are concerned,
we can regard the distribution of elements within the cosets of $H$ as
essentially being that of the Haar measure
\footnote{We are interested in whether or not Bianchi identities introduce
correlations between plaquette variables that are sufficiently coherent
so as to lead to spontaneous breakdown of gauge symmetry under
transformations of the type (\ref{lin}). If the distribution along cosets
of $H$ is effectively the Haar measure,
all elements within a coset are accessed with equal probability and there
is not spontaneous breakdown under transformations of the type (\ref{lin})
as far the \dofx corresponding to the invariant subgroup $H$ are concerned.
Hence, the fulfilment of Bianchi identities in the case of the \dofx for which
we may not forget about them (i.e., when these identities can introduce
coherent correlations between plaquettes) is insured
by the more lenient requirement that Bianchi identities only need be fulfilled
after mapping the $U(\Box)\in G$ into the factor group $G/H$.
This is consistent with our
definition of confinement, which is that correlations between
values of different plaquette variables that are imposed by
Bianchi identities effectively disappear when a subgroup goes into the
confining phase. In the volume approximation, we can for calculational
purposes therefore assume the Haar
measure for the distribution of plaquette variables. Recall from earlier
sections that this is really not the case. Rather, going into confinement
at a first order phase
transition is accompanied by a discontinuous broadening of the  width of
the distribution
of elements within the cosets of the confined subgroup. But this is
sufficient to
suddenly allow the fulfilment of Bianchi identities by having  the sum
of plaquette variables add up to a nonzero multiples of $2\pi$ which
in turn reduces the effectiveness of Bianchi identities in introducing
coherent correlations between plaquettes which again allows larger plaquette
variable fluctuations
which again makes it even easier to avoid correlations from Bianchi
identities in a sort of self-perpetuating chain of events.}.

\subsubsection{The   independent   monopole
condensate approximation - the calculation}

In the  independent  monopole  approximation,  we  can
reach the multiple point using the  Manton  action alone (i.e.,  no
higher order terms). The diagonal $U(1)$  subgroup  to  be  identified
with the $U(1)$ of the $SMG$ is that given by $\theta(1,1,1)$  in  the
coordinate choice (\ref{coord3}).

The first identification lattice point  met by this diagonal
subgroup occurs for $\theta=2\pi$; i.e., the point $2\pi(1,1,1)$.
Hence the quantisation rule
$y/2 \in{ \bf Z}$ (for  particles  not  carrying  non-Abelian  gauge  coupling)
is achieved by the
naive        continuum        limit         identification         \beq
\exp(i\theta(-))=\exp(iag_1A_{\mu}y/2)\;\;\;\mbox{for}\;\;\;     y/2=1.
\eeq

\nin For $y/2=1$ (corresponding to $e^+_L$), the  covariant  derivative
is

\beq
D_{\mu}=
\partial_{\mu}
-ig_1A_{\mu}.
\eeq

The equation analogues to (\ref{dist1dim}) for the diagonal subgroup
(on the 3-dimensional identification lattice) is
\beq
\frac{\beta_{diag}}{2}(2\pi)^2=length(2\pi(1,1,1))= \label{diag6} \eeq

\[ =(2\pi)^2(1,1,1)
\frac{\beta_{crit}}{2}\left( \begin{array}{ccc} 1 & \frac{1}{2} & \frac{1}{2}
\\ \frac{1}{2}
& 1 & \frac{1}{2} \\ \frac{1}{2} & \frac{1}{2} & 1 \end{array} \right)
\left(   \begin{array}{c} 1 \\ 1 \\ 1 \end{array} \right)=(2\pi)^2\frac{
\beta_{crit}}{2} \cdot 6 \]

\nin for the multiple point. Contrary to the case of  the
non-Abelian couplings that are weakened  by  a  factor  $N_{gen}=3$  in
going to the diagonal subgroup of $U(1)^3$, the $U(1)$ coupling at  the
multiple
point is weakened by a factor 6 in going to the diagonal  subgroup  of
$U(1)^3$. In general, the weakening factor in the  hexagonal  case  in
going from $U(1)^{N_{gen}}$ to the diagonal subgroup $U(1)$ along  the
direction $(1,1,\cdots,1)$ is $N_{gen}+ \left(\begin{array}{c} N_{gen}
\\ 2 \end{array}\right)=N_{gen}(N_{gen}+1)/2$:

\nin so that
\beq g^2_{diag}=\frac{g^2_{crit}}{N_{gen}(N_{gen}+1)/2}.  \eeq

\subsubsection{The volume of groups scheme}

In the earlier section~\ref{grvolapprox}, we have developed a  means for
calculating
an effective inverse squared coupling having a directional dependence on
$4th$ and $6th$ order action terms. We now calculate
the effective inverse squared coupling (\ref{invcoupeff}) along the diagonal
subgroup:

\beq \frac{1}{e^2_{eff}(\vec{\xi})} \mbox{  (for $\xi=(1,1,1)$)  } = \eeq

\[
\left(B_6Y_{6\;\;comb}(diag)+\left(\frac{1}{e^2_{Manton}}+
B_4Y_{4\;\;comb}(diag)\right)^{\frac{3}{2}}\right)^{\frac{1}{3}} =\]

\[ =\left(\frac{-0.766}{e^6_{U(1)\;\;crit}}(-4\sqrt{\frac{3}{35}})+
\left(\frac{2^{\frac{2}{3}}}{e^4_{U(1)\;\;crit}}+
\frac{-0.146}{e^4_{U(1)\;\;crit}}\frac{2}{3}\sqrt{7}\right)^{\frac{3}{2}}
\right)^{\frac{1}{3}}= \]

\[ =1.34\cdot\frac{1}{e^2_{U(1)\;\;crit}} \;\;\;\mbox{  (in vol. approx.)}\]

\nin From (\ref{diag6}) we have that the inverse squared coupling
corresponding to the diagonal subgroup of $U(1)^3$ is
                                       a factor 6 larger than
$\frac{1}{e^2_{U(1)\;\;crit}}|_{\xi = (1,1,1)}\;$:

\beq \frac{1}{e^2(diag)}=6\cdot\frac{1}{e^2_{eff}(\vec{\xi})}\left |_{
\vec{\xi}=(1,1,1)}\right.=6\cdot 1.34=8.04 \eeq

\subsubsection{The calculation of the enhancement factor}

We have seen that the enhancement factor
$\frac{1/\alpha_{U(1)^3}}{1/\alpha_{U(1)}}$
has respectively the values 6.0 and $1.34\cdot 6.0$ according to whether the
``independent monopole'' or the ``volume'' approximation is used. These
approximations tend respectively to ignore and to overestimate the
dependence
that fluctuations in one subgroup can have on which phase is
realized along other subgroups or factor-groups.
This interaction effect depends on the
degree of first-orderness of the phase transition; this degree of
first-orderness is used in
our procedure
to determine to what extent the pure ``monopole  approximation'' should
be ``pushed'' towards the ``volume approximation''.
We seek a combination of these two approximations - with the relative weight
determined by the degree of first-orderness -
that is to be embodied in the value of $\Delta \gamma_{eff}$ that subsequently
is used in both steps of the calculation of the $U(1)$ continuum coupling.
In this section, we use $\Delta \gamma_{eff}$ to determine the $U(1)$ coupling
at the multiple point of a phase diagram for a $U(1)^3$ gauge group.

The correction
for the degree of first-orderness will be implemented by choosing the ``hop''
$\Delta W=\Delta\langle \cos \theta\rangle$ in the Wilson operator at the
TP transition to the totally Coulomb-like phase in such a way that it
reflects the residual first-orderness.
This transition
obviously has to separate confinement-like and Coulomb-like phases for the
{\em continuum} degrees of freedom.
There are two possibilities -
namely the TP transition at border ``1'' and the TP transition at border
``3'' corresponding let us say to respectively the jumps
$\Delta W_{\Box\;TP\;``1"}$
and $\Delta W_{\Box\;TP\;``3"}$ in the Wilson operator.

But we now
argue that $\Delta W_{\Box\;TP\;``1"}$ is not what we want because it
doesn't
reflect the degree of ``residual'' first-orderness (at the TP) that is due
to the $Z_N$ transition\footnote{In fact by using a trick of changing variables
(more on this below), we
can actually show that $\Delta W_{\Box\;TP\;``1"}\approx
\Delta W_{\Box\;\gamma=0\;``3"}$.
The latter reflects (less pronounced) residual first-orderness of the $Z_N$
transition quite far removed from the TP - namely that for $\gamma=0$
which is not so far from the tri-critical point at $\gamma=-0.11$ where
all remnants of the $Z_N$ transition disappear and the transition at
boundary ``3''continues for $\gamma<-0.11$ as a pure second order transition.}
The reason has to do with  $Z_N$ ($N=2,3$) being in the same
phase on both sides of the
border ``1'' at the TP. Accordingly, $\Delta W_{\Box\;TP\;``1"}$ cannot
reflect the discrete group transition.

So it is the discontinuity  $\Delta W_{\Box\;TP \;``3"}$ that we want to
use to interpolate between the  ``independent monopole'' and the ``volume''
approximation so as to obtain the
enhancement factor $\frac{\alpha_{U(1)^3\;diag}}{\alpha_{U(1)\;crit}}$ that
reflects the appropriate degree of first-orderness for the TP transition
in going from confinement to Coulomb-like behaviour for the continuum \dof.

In order to estimate the residual ``first-orderness'' present at the multiple
point in making the transition to the totally Coulomb-like phase from the
phase(s)
with confinement solely w.r.t to discrete subgroup(s), we shall
use the already proposed scenario in which we speculate that the increased
frequency of minimal strength monopoles (i.e., $\frac{2\pi}{4\pi}$
monopoles in the $4\pi$ normalisation implicit in
(\ref{newbhanot})) is
related
to the fact
that the phase transition along the border ``2'' in Figure~\ref{figbhanot}
becomes less and
less strongly
first order as $\gamma$ decreases. That is, we speculate that the increasing
role of minimal strength monopoles (in the $4\pi$ normalisation, the minimal
strength $\frac{2\pi}{4\pi}$ monopoles are the only monopoles; in the $12\pi$
normalisation, there are, in addition to minimal and most abundant
$\frac{2\pi}{12\pi}$ monopoles, also (less common) $\frac{6\pi}{12\pi}$-
and $\frac{4\pi}{12\pi}$-monopoles)
in typical plaquette
configurations
is the reason that the transitions to the totally Coulomb-like phase at border
``2'' and subsequently, also at border ``3'' in Figure~\ref{figbhanot}
becomes less and less first order as $\gamma$ is diminished.

As mentioned just above, it is well
known that, for $U(1)$, the phase transition at border ``3'' becomes  second
order at the tri-critical point (at a slightly negative value of $\gamma$)
and continues as a second order phase transition for $\gamma$ values less
than the tri-critical value $\gamma_{TCP}$.
The above picture is not inconsistent with the results of numerical studies
that clearly reveal even a pure $U(1)$  gauge theory with a Wilson action
(i.e., a theory with $\gamma=0$) as having a
weakly first order phase transition as evidenced by a ``jump''
$\Delta W_{\Box}$ in the Wilson operator $W_{\Box}$.
Indeed one finds in the work of Jers\`{a}k\cite{tricrit} {\em et al} fits that
relate the ``jump'' $\Delta W_{\Box}$ in the Wilson operator $W_{\Box}
\stackrel{def}{=} \langle \cos (\theta_{\Box}) \rangle$ to $\Delta \gamma
\stackrel{def}{=}\gamma-\gamma_{TCP}$ where $\gamma_{TCP}$ denotes the
value of $\gamma$ in the tri-critical point:

\beq  \Delta W_{\Box}=A(\gamma-\gamma_{TCP})^{\beta_{\mu}} \label{gamnul}\eeq

\nin The values for $\gamma_{TCP}$ and $\beta_{\mu}$ are given respectively as
$\gamma_{TCP}=-0.11\pm 0.05$ and $\beta_{\mu}=1.7\pm 0.2$ while the constant
$A$ is deduced to be $A=0.68_{35}$. For $\gamma=0$ (corresponding to a
Wilson action), there obtains $\Delta W_{\Box}=0.68(0.11)^{1.7}=0.016$.

Actually this latter discontinuity will be seen to be of interest to us
because it can be shown
that this jump is to a good approximation the jump $\Delta W_{\Box,\;``1"}$
encountered in crossing border ``1'' near the multiple point. The reasoning
is as follows: the jump $\Delta W_{\Box,\;``1"}$ is to a good approximation
constant
along the phase border ``1''; consequently, $\Delta W_{\Box,\;``1"}$ near the
multiple point is essentially the same as that at $\gamma=1.01$ and $\beta=0$
which, in turn, is,
by a simple change of  notation, identical with the discontinuity
$\Delta W_{\Box}$ at $\gamma=0$, $\beta=1.01$ that using (\ref{gamnul}) was
found to have the value  $\Delta W_{\Box}=0.016$.

So what is wanted for the purpose of calculating the enhancement factor is the
jump $\Delta W_{\Box,\;``3"}$ encountered at the multiple point in
traversing border ``3" separating the totally Coulomb-like and totally
confinement-like phases. What we have is a way to calculate
$\Delta W_{\Box;\;``2"}$: this procedure
relates $\Delta W_{\Box,\;``2"}$ to the cubic root\cite{lautrup,moriarty}
of the quantity
$\Delta \gamma_{eff}$ (see Section~\ref{deltagam}) encountered in crossing
the border ``2'' at the multiple point. Were
it not that the transition at border ``1'' is (weakly) first order but
instead second order, then we would have had
$\Delta W_{\Box,\;``1"}=0 $
and

\beq  \Delta W_{\Box,\;``2"}=\Delta W_{\Box,\;``3"}=
A(\Delta \gamma_{eff})^{\frac{1}{3}}\;\;\;\mbox{ (when
$\Delta W_{\Box,\;``1"}=0$)}
\label{2ndord}\eeq

\nin where $A=0.252$.
However, having argued that $\Delta W_{\Box,\;``1"}=0.16\neq 0 $
corresponding to a weakly first order transition in crossing border ``1''
 in the
vicinity of the multiple point, we conclude  on the grounds of
continuity that this jump must be the difference in the ``jumps''
$\Delta W_{\Box,\;``2"}$ and $\Delta W_{\Box,\;``3"}$ in crossing
respectively
the borders ``2'' and ``3'' at the multiple point (see Figure~\ref{figbhanot}).
Recall that
these jumps, observed in crossing the borders ``2'' and ``3'' near the
multiple point
are essentially assumed to be the residual effects of first-order pure
discrete subgroup transitions at
large $\gamma$.
So in principle at least, the ``jump'' $\Delta W_{\Box;\;``3"}$ is obtained by
correcting  \footnote{The reason that we do the calculation in this
circuitous way - instead of trying to directly estimate
$\Delta W_{\Box\;``3"}$
by first calculating the ``$\Delta \gamma_{eff}$'' at boundary ``3'' - is
that it is not clear what this latter $\Delta \gamma_{eff}$ means. The reason
that we calculate $\Delta \gamma_{eff}$ at boundary ``2'' is that the
phases on both sides of this boundary are very similar w.r.t the
continuum \dof. This allows us to conclude that our $\Delta \gamma_{eff}$ at
boundary ``2'' can be associated essentially alone with the discrete
subgroup transition.} $\Delta W_{\Box\;``2"}$ (calculated by using
(\ref{2ndord})) by the
amount of the ``jump'' $\Delta W_{\Box,\;``1"}$
in crossing border ``1''.
Using that $\Delta W_{\Box,\;``1"}$ is small, we
make this correction in an approximate way by increasing $\Delta \gamma_
{eff}$ in (\ref{2ndord}) by the corrective quantity

\beq \Delta \gamma_{corr\;``1"}\stackrel {def}{=}
(\frac{\Delta W_{\Box,\;``1"}}{A})^3  \label{corr} \eeq

\nin obtained by inverting (\ref{2ndord}). In this approximation, we obtain

\beq \Delta W_{\Box,\;``3"}\approx A(\Delta \gamma_{eff}+
\Delta \gamma_{corr\;``1"})^{\frac{1}{3}}= \label{jump3}\eeq

\[ = A(\Delta \gamma_{eff}+ (\frac{0.016}{0.252})^3)^{\frac{1}{3}}  \]

\nin where we have used that $\Delta W_{\Box,\;``1"}=0.016$ in (\ref{corr})
which in turn has been used in (\ref{jump3}). Strictly speaking, it is
inconsistent to assume additivity in the ``jumps'' $\Delta W_{\Box,\;``1"}$,
$\Delta W_{\Box,\;``2"}$, and $\Delta W_{\Box,\;``3"}$ (essential because
of continuity requirements) {\em and} at the same time that both
$\Delta W_{\Box,\;``2"}$ and $\Delta W_{\Box,\;``3"}$ are related to an
appropriate $\gamma_{eff}$ by a cubic root law. Consistency requires
$\Delta W_{\Box,\;``1"}=0$ corresponding to a second order transition. For
small $\Delta W_{\Box,\;``1"}$, this inconsistency is not bothersome
and the approximation (\ref{jump3}) is good. In fact the corrective term
$\Delta \gamma_{corr\;``1"}$ is  so small so as not to yield a difference
in $\Delta W_{\Box,\;``2"}$ and $\Delta W_{\Box,\;``3"}$ that is discernible
to within the calculational accuracy.

Equation (\ref{jump3}) provides a way of calculating
the for us interesting $\Delta W_{\Box,\;``3"}$ at the multiple point.
Various values of $\Delta W_{\Box,\;``3"}$ are tabulated in Table \ref{tab5}.
These are calculated for different values of $\Delta \gamma_{eff}$ that in
turn are obtained as combinations of the $\Delta \gamma_{eff}$ in
Table \ref{tab4} calculated
for the $\bz_2$, $\bz_3$, and $\bz_6$ discrete subgroups of $U(1)$.

Before we use these various $\Delta W_{\Box,\;``3"}$ values to calculate the
enhancement factor
$\frac{\alpha_{U(1)^3\;diag}}{\alpha_{U(1)\;crit}}$, we need to
develop a way of using the $\Delta W_{\Box,\;``3"}$ to
interpolate between the ``pure monopole''
and the ``volume'' approximation. We now do this for the general case of any
discontinuity $\Delta W_{\Box}$.
In general, when there is a ``jump'' $\Delta W_{\Box}$,
we estimate that we get the most
correct enhancement factor
$\frac{1/\alpha_{U(1)^3\;diag}}{1/\alpha_{U(1)}\;crit}$
by linearly interpolating between the enhancement factor ``6'' corresponding
to the independent monopole approximation and the enhancement factor
$1.34\cdot 6=8.04$ corresponding to the volume approximation. That is,
the enhancement factor is calculated as

\[ \left(\frac{1/\alpha_{U(1)^3\;diag}}{1/\alpha_{U(1)}}\right)_{actual}=
\left(\frac{1/\alpha_{U(1)^3\;diag}}{1/\alpha_{U(1)}}\right)_{ind\;mono}+ \]
\[ +\frac{\eta}{\tau}
\left[\left(\frac{1/\alpha_{U(1)^3\;diag}}{1/\alpha_{U(1)}}\right)_{vol}-
\left(\frac{1/\alpha_{U(1)^3\;diag}}{1/\alpha_{U(1)}}\right)_{ind\;mono}\right]
\]

\beq= 6+\frac{\eta}{\tau}[6(1.34-1)] \label{intpol}\eeq

\nin where $\frac{\eta}{\tau}$ is given by

\beq
\frac{\eta}{\tau}=
\frac{\left(\frac{Coul\;fluc}{conf\;fluc}\right)^2_{ind\;\;mono}-
     \left(\frac{Coul\;fluc}{conf\;fluc}\right)^2_{actual}}
     {\left(\frac{Coul\;fluc}{conf\;fluc}\right)^2_{ind\;\;mono}-
     \left(\frac{Coul\;fluc}{conf\;fluc}\right)^2_{vol}} \label{etatau}\eeq

\nin and $\eta$ is defined as the numerator while $\tau$ the denominator on
the right hand side of (\ref{etatau}).  Write

\beq \left(\frac{Coul\;fluc}{conf\;fluc}\right)^2\stackrel{def}{=}
     \frac{1-\langle \cos \theta\rangle_{Coul}}
     {1-\langle \cos \theta\rangle_{conf}} \label{coulconf}
=1-\frac{\Delta W_{\Box}}{1-\langle \cos \theta \rangle_{conf}} \label{yes}
\eeq

\nin where in the last step we have used that
$\langle \cos\theta \rangle_{Coul}=\langle \cos\theta \rangle_{conf}+
\Delta W_{\Box}$.

Using that
$\left(\frac{Coul\;fluc}{conf\;fluc}\right)^2_{ind\;mono}=1$ essentially
by definition, we have using (\ref{etatau}) and (\ref{yes}) that

\beq \eta=\frac{\Delta W_{\Box}}
{1-\langle \cos \theta \rangle_{conf}}= \Delta W_{\Box}/0.377 \label{eta} \eeq

\nin where in (\ref{eta}) we have used $\langle \cos\theta \rangle_{conf\;ph}
=0.623$.

Various values of $\eta$ are tabulated in Table \ref{tab5} corresponding
to the values of $\Delta W_{\Box,\;``3''}$ that are also tabulated
in the same Table.

The quantity $\frac{\eta}{\tau}$ are used to obtain the values for the
enhancement factors $\frac{1/\alpha_{U(1)^3\;diag}}{1/\alpha_{U(1)}\;crit}$
tabulated in the final two columns of Table \ref{tab5}.
The two columns correspond to
$\frac{1/\alpha_{U(1)^3\;diag}}{1/\alpha_{U(1)}\;crit}$ for
two different values of $\tau$. The first, corresponding
to the roughest approximation, is for $\tau=1$ inasmuch as we make the
approximation
$\left(\frac{Coul\;fluc}{conf\;fluc}\right)^2_{vol}
\approx 0$. The enhancement factors in the column at the extreme right hand
side are obtained using a  better estimate\footnote{$\tau=
1-\left(\frac{Coul\;fluc}{conf\;fluc}\right)^2_{vol}=
1-\frac{2(1-\langle \cos \theta \rangle)_{Coul/\; crit}}
{\langle \theta^2\rangle_{conf}}=1-\frac{2(1-0.65)}{\pi^2/3}\approx 1-0.21
=0.79.$
Here $\langle \theta^2 \rangle_{conf}$ is calculated as though one had the
ideal Haar measure distribution which is the distribution used in effect in
our volume approximation.}
of $\tau$:
\beq \tau=1-\left(\frac{Coul\;fluc}{conf\;fluc}\right)^2_{vol}=1-0.21=0.79.\eeq

The values of $\frac{1/\alpha_{U(1)^3\;diag}}{1/\alpha_{U(1)\;crit}}$ in the
last column of Table \ref{tab5} will appear in Table \ref{tab6} in
conjunction with the calculation of the Planck scale value of the continuum
$U(1)$ fine-structure constant $1/\alpha_{U(1)\;Pl.\;scale}$.

\begin{footnotesize}
\begin{table}
\caption[table5]{\label{tab5}Enhancement factors given in the
last four columns on the right are given for two ways of calculating
$\tau$ as well as with and without $\Delta\gamma_{corr``1"}$ included
in the calculation of $\Delta W_{\Box}$ in (\ref{jump3}).
For the quantity $\tau\stackrel{def}{=}
1-\left(\frac{Coul\;fluc}{conf\;fluc}\right)^2$
we have $\tau=1$ when confinement fluctuations are taken as infinite and
$\tau=0.79$  when confinement fluctuations are taken as finite.
The second and third columns contain
$\Delta W_{\Box}$ calculated respectively with and without the quantity
$\Delta\gamma_{corr``1"}$ in (\ref{jump3}).
The values for $\Delta\gamma_{eff}$ in the first column are taken from
Table~\ref{tab4}. The
quantity $\eta$ in the fourth and fifth columns is defined in (\ref{etatau})
and calculated according
to (\ref{eta})
with and without the quantity
$\Delta\gamma_{corr``1"}$ in the expression (\ref{jump3}) for
$\Delta W_{\Box}$. }
\vspace{.7cm}
\[\!\!\!\!\!\!\!\!\!\!\!\!\!\!\!\!\!\!\!\!\mbox{
\begin{tabular}{|c||c|c|c|c|c|c|c|c|c|}\hline
Procedure & $\Delta \gamma_{eff}$ & \mc{2}{c|}{$\Delta W_{\Box ,``3"}$ from
(\ref{jump3})} &
    \multicolumn{2}{c|}{$\eta$ from (\ref{eta})} &
    \multicolumn{4}{c|}{$\frac{1/\alpha_{{U(1)}^3_{``diag"}}}
{1/\alpha_{{U(1)}_{``crit"}}}$ from (\ref{intpol})} \\
    \cline{3-10} & & with  & $\Delta \gamma_{corr``1"}$  &
    with & $\Delta \gamma_{corr``1"}$ &
 \mc{2}{c|}{with $\Delta \gamma_{corr``1"}$} & \mc{2}{c|}{$\Delta
\gamma_{corr``1"}=0$}\\
\cline{7-10} & & $\Delta \gamma_{corr``1"}$ & $=0$ & $\Delta \gamma_{corr``1"}$
& $=0$ &  $\tau=0.79$ & $\tau=1$ & $\tau=0.79$ & $\tau=1$ \\ \hline \hline
Vol Approx: &&\mc{2}{c|}{}&\mc{2}{c|}{}&\mc{2}{c|}{}&& \\
Haar (ideal)& &\mc{2}{c|}{{\tiny ($\langle \cos \theta \rangle_{Coul}=0.65$)}}
&\mc{2}{c|}{1} &\mc{2}{c|}{} &  & 8.04 \\
Haar, compact & &\mc{2}{c|}{{\tiny ($\langle \cos \theta \rangle_{Coul}=0.65$)}
} & \mc{2}{c|}{0.79}  & \mc{2}{c|}{} & 8.04 &  \\
Mean field & &\mc{2}{c|}{}&\mc{2}{c|}{$\frac{1}{2}$} & \mc{2}{c|}{} &
7.29 & 7.02 \\ \hline
Ideal ind mono &0&\mc{2}{c|}{0} & \mc{2}{c|}{1} &\mc{2}{c|}{6} &\mc{2}{c|}{6}
\\ \hline
No discrete &0 & 0.016 &0 & 0.042$_4$  & 0  & 6.11$_0$ & 6.08$_7$  & 6 & 6 \\
subgroups  &&&&&&&&& \\ \hline
Using  & & & & & & & & &  \\
$\bz_2$ only: & 0.047$_3$ & 0.091$_3$ & 0.091$_1$ & 0.24$_2$ & 0.24$_2$ &
6.62$_5$ & 6.49$_4$ & 6.62$_4$  & 6.49$_3$ \\  \hline
Using & 0.047$_3$+ & & & & & & & & \\
$\bz_2+\bz_3$: & 0.039$_3$ & 0.11$_2$ & 0.11$_1$ & 0.29$_6$ & 0.29$_6$ &
6.76$_4$ &6.60$_4$ & 6.76$_4$  & 6.60$_3$ \\ \hline
Using & $\frac{0.047_3}{2}+$ & & & & & & & & \\
$\frac{1}{2}(\bz_2+\bz_3)$: & $+\frac{0.039_3}{2}$ & 0.088$_7$ & 0.088$_5$ &
0.23$_5$ & 0.23$_5$ &  6.60$_7$ & 6.48$_0$ & 6.60$_6$  & 6.47$_9$ \\ \hline
Using & $\frac{0.047_3}{2}$ & & & & & & & & \\
$\frac{1}{2}\bz_2+\bz_3$: & $+0.039_3$ & 0.10$_0$ & 0.10$_0$ & 0.26$_6$ &
0.26$_6$ & 6.68$_8$ & 6.54$_3$ & 6.68$_7$  & 6.54$_2$  \\ \hline
\end{tabular}}\]
\end{table}
\end{footnotesize}

\subsection{Continuum  critical
coupling from critical $U(1)$ lattice coupling}\label{contcoup}

The Planck scale prediction for the $U(1)$ fine-structure constant is to be
obtained as the product of the enhancement factor and the continuum critical
coupling that corresponds to the lattice critical coupling.

We have the enhancement factor in Table~\ref{tab5} (calculated using different
approximations) but we have yet to translate
the lattice $U(1)$ critical coupling
into a continuum one. This is the purpose of this section.

We use a procedure analogous to that used by Jers\`{a}k et al\cite{jersak}.
In this work the continuum coupling is calculated
numerically. Using Monte Carlo methods on the lattice, the Coulomb potential
is computed  and fitted to the formula proposed by Luck (\cite{luck}).
In the Coulomb phase with  the Wilson action, the fit yields

\beq
Wilson:\;\;\;   \alpha(\beta) =0.20 - 0.24 (\frac{\beta
-\beta_{crit}}{\beta})^{0.39}\;= \;
0.20 - 0.24 (1-\frac{1.0106}{\beta})^{0.39}. \label{contfor}
\eeq
For the Villain action (in the Coulomb phase) the analogous result is
\beq
Villain:\;\;\;   \alpha(\beta) = 0.20 - 0.33 (1-\frac{0.643 }{\beta})^{0.52}.
\label{vil} \eeq

It is of course our intention to substitute $\gamma_{eff}$ for what
Jersak et al. designates as $\beta$. This is justified in as much as
(\ref{contfor}) is valid for $\beta \geq\beta_{crit}$; i.e., for $\beta$ lying
{\em within the  Coulomb phase}. The replacement of
\beq\beta -\beta_{crit}\eeq

\nin  by

\beq \gamma_{eff\;tot\;Coul\;ph}-\gamma_{eff.\;only\;\sbz_2\;conf}
\stackrel{def}{=} \Delta \gamma_{eff} \eeq

\nin is valid
inasmuch as the phases separated by the phase boundary ``2'' in
Figure~\ref{figbhanot} are both in the Coulomb phase as far as the
{\em continuum}
\dofx are concerned. Values obtained for $\alpha$ using (\ref{contfor})
with $\beta$ replaced by $\gamma_{eff}$ and $\beta-\beta_{crit}$ by
$\Delta \gamma_{eff}$ are tabulated for various values of the latter
in the third column of Table~\ref{tab6}.

\begin{footnotesize}
\begin{table}
\caption[table6]{\label{tab6}Our Planck scale prediction for the $U(1)$
fine-structure constant
is obtained as the product of the enhancement factor (from the last four
columns of Table~(\ref{tab5})) and the value of
$1/\alpha_{cont}$ obtained from the critical value of the lattice parameter
$\gamma_{eff}$ and the ``jump'' $\Delta \gamma_{eff}$ in this same quantity
in crossing the phase border ``3'' (see Figure~\ref{figbhanot}) at the multiple
point.
We list a number of
combinations that differ according to how the discrete subgroups are treated
w.r.t. whether the discrete subgroups are large enough to have the symmetry
of the hexagonal identification lattice
and how $\tau$ (see Table~\ref{tab5}) is calculated as an indication of the
sensitivity of our
prediction to such details. The prediction marked with ``$\bullet$'' indicates
the predicted value calculated in what we regard as the most correct manner.
Also included are results for the Villian action where (\ref{vil}) has been
used to calculate $\alpha(\beta)$. In this table, we use the same
$\Delta\gamma_{eff}$
in both (\ref{contfor}) and (\ref{vil}) (this is incorrect; see next table).
In the Villian case, the $\Delta W_{\Box\;``3"}$  used in calculating the
enhancement factor is calculated as $\Delta W_{\Box\;``3"}=
0.16(\Delta\gamma_{eff})^{0.29}$ (this is the counterpart of (\ref{jump3})
for the Wilson action)
with $\Delta_{``1"\;corr}=0$). The coefficient ``0.16'' is estimated
from
Monte Carlo data in (\cite{jersak}) and is rather uncertain\footnotemark.}
\beq \!\!\!\!\!\!\!\!\!\!\!\!
\begin{array}{|l|l|l|l|l|l|l|l|}\hline
\multicolumn{8}{|c|}{ \mbox{single $U(1)$}}  \\ \hline
\mbox{Procedure} &\Delta\gamma_{eff} & \alpha_{cont.} & 1/\alpha_{cont.}
&\mc{2}{c|}{\mbox{enh. fac Haar}}&
\mc{2}{c|}{\mbox{prediction $1/\alpha_{Pl. scale}$}} \\
\cline{5-8}  & & & & \tau=0.79 & \tau=1 & \tau=.79 & \tau=1 \\ \hline
\mbox{Wilson action (using (\ref{contfor}))}&&&&&&& \\ \hline
\mbox{$\bz_2$ only with $\Delta\gamma_{corr``1"}$ in (\ref{jump3})}
& 0.047_3 & 0.128_6 & 7.77_8 & 6.62_5 & 6.49_4 & 51.5 & 50.5 \\ \hline
\mbox{$\bz_2$ only without $\Delta\gamma_{corr``1"}$ in (\ref{jump3})}
& 0.047_3 & 0.128_6 & 7.77_8 & 6.62_4 & 6.49_3 & 51.5 & 50.5 \\ \hline
\mbox{$\bz_2+\bz_3$ with $\Delta\gamma_{corr``1"}$ in (\ref{jump3})}
& 0.086_6 & 0.110_8 & 9.02_1 &  6.76_4 & 6.60_4 & 61.0 & 59.6 \\ \hline
\mbox{$\bz_2+\bz_3$ without $\Delta\gamma_{corr``1"}$ in (\ref{jump3})}
& 0.086_6 & 0.110_8 & 9.02_1 & 6.76_4 & 6.60_3 & 61.0 & 59.6 \\ \hline
\mbox{$\frac{1}{2}(\bz_2+\bz_3)$ with $\Delta\gamma_{corr``1"}$ in (\ref{jump3}
)} & 0.043_3 & 0.130_9 & 7.64_0 & 6.60_7 & 6.48_0 & 50.5 & 49.5 \\ \hline
\mbox{$\frac{1}{2}(\bz_2+\bz_3)$ without $\Delta\gamma_{corr``1"}$ in
(\ref{jump3})} & 0.043_3 & 0.130_9 & 7.64_0 & 6.60_6 & 6.47_9 & 50.5 & 49.5 \\
\hline
\mbox{$\frac{1}{2}\bz_2+\bz_3$ with $\Delta\gamma_{corr``1"}$ in (\ref{jump3})}
 & 0.063_0 & 0.120_6 & 8.29_2 & 6.68_9 & 6.54_3 & 55.5\;\;\bullet & 54.3 \\
\hline
\mbox{$\frac{1}{2}\bz_2+\bz_3$ without $\Delta\gamma_{corr``1"}$ in
(\ref{jump3})} & 0.063_0 & 0.120_6 & 8.29_2 & 6.68_7 & 6.54_2 & 55.5\;\;\bullet
& 54.3 \\ \hline\hline
\mbox{Villian action (using (\ref{vil}))}&&&&&&& \\ \hline
\mbox{$\bz_2$ only}   &0.047_3 &0.11_8 &8.46_5 & 6.45_2  & 6.35_7 & 54.6 & 53.8
 \\ \hline
\mbox{$\bz_2+\bz_3$} &0.086_6 &0.091_1 &10.9_8 & 6.53_9 & 6.42_6 & 71.8 & 70.6
\\ \hline
\mbox{$\frac{1}{2}(\bz_2+\bz_3)$}&0.043_3&0.12_2 &8.22_6 & 6.44_1  & 6.34_8
& 53.0\;\;\bullet & 52.2 \\ \hline
\mbox{$\frac{1}{2}\bz_2+\bz_3$} & 0.063_0 & 0.10_6 & 9.42_4 &6.49_1 & 6.38_8
& 61.2\;\;\bullet & 60.2 \\ \hline
\end{array}
\eeq
\end{table}
\end{footnotesize}
\footnotetext{
Making that the assumption that the phase transitions for both the Wilson
and Villian actions are second order, we take the  difference
$\langle \theta^2 \rangle -\langle \theta^2 \rangle_{crit.}$
as being the same when the string tension is the same for both action
types. Using figure 4a in Jersak et al: Nucl. Phys. B251, 1985, 299,
we obtain the coefficient 0.23 as the coefficient of $(\Delta \gamma_{eff})
^{.29}$. Allowing for the fact that the transitions are not strictly
second order gives rise to a correction that results in a coefficient
of 0.16 instead of the 0.23.}
The $\gamma_{eff}$ used in Table~\ref{tab6} is that  calculated in
(\ref{actz2})
to lowest order in $\hat{\theta}$. We want now to go to next order in
$\hat{\theta}$. For the Wilson action suitable for having a phase confined
solely
w.r.t. $\bz_2$, the appropriate effective coupling is
given by (\ref{gammaeffect}). We denote this improved effective
coupling by $\gamma_{eff\;corr}$:

\beq \gamma_{eff\;corr}=
\gamma+\frac{\langle\sigma_{\bz_{2}}\rangle_{\hat{\theta}}\beta}{4}
\langle\frac{\cos(\hat{\theta}/2)}
{\cos\hat{\theta}}\rangle \approx
\gamma+\frac{\langle\sigma_{\bz_{2}}\rangle_{\hat{\theta}}\beta}{4}
\langle \cos^{-\frac{3}{4}}\hat{\theta}\rangle \mbox{      (Wilson action)}.
\label{gameffectwil} \eeq

\nin The analogous improved $\gamma_{eff\;corr}$ for the
Villian action case has the $\cos\hat{\theta}$ in the denominator in the
average on the left-hand side
of (\ref{gameffectwil}) removed
corresponding to the Villian action being approximately a Manton action
(having a second derivative that is $\hat{\theta}$-independent) instead
of being
equal to $\cos\hat{\theta}$ as in the Wilson case. So for the Villian action
we have

\beq \gamma_{eff\;corr}= \gamma+\frac{\beta}{4}\langle\sigma_{\bz_{2}}\rangle
\cdot\langle\cos(\hat{\theta}/2) \rangle \approx
\gamma+\frac{\beta}{4}\langle\sigma_{\bz_{2}}\rangle
\cdot\langle\cos^{\frac{1}{4}}\hat{\theta} \rangle
\;\;\;\mbox{(Villian action)}. \label{gameffectvil} \eeq

The effective couplings (\ref{gameffectwil}) and (\ref{gameffectvil})
are for respectively the Wilson and
Villian actions. In both cases there can be a phase confined solely w.r.t.
$\bz_2$. The analogous couplings for the Wilson and Villian actions in
the case where there is a  phase confined solely w.r.t. $\bz_3$ are given
respectively by (\ref{gameffwilz3}) and (\ref{gameffvilz3}) below; i.e.,
by

\beq \gamma_{eff\;corr}=
\gamma+\frac{\beta}{9}\langle\sigma_{\bz_{3}}\rangle
\langle \cos^{-\frac{8}{9}}\hat{\theta}\rangle \mbox{      (Wilson action)}
\label{gameffwilz3} \eeq

\nin and

\beq \gamma_{eff\;corr}=
\gamma+\frac{\beta}{9}\langle\sigma_{\bz_{3}}\rangle
\cdot\langle\cos^{\frac{1}{9}}\hat{\theta} \rangle
\;\;\;\mbox{(Villian action)}.
\label{gameffvilz3} \eeq

\begin{footnotesize}
\begin{table}
\caption[table6a]{\label{tab7}Here we use the improved effective couplings
$\gamma_{eff\;corr}$ in (\ref{gameffectwil}) and (\ref{gameffectvil})
corresponding respectively to Wilson and
Villian actions for which there is a phase confined solely w.r.t. $\bz_2$.
The analogous improved effective couplings (\ref{gameffwilz3}) and
(\ref{gameffvilz3}) are used respectively
for the Wilson and
Villian actions that can provoke phases confined solely w.r.t. $\bz_3$.
Strictly speaking, for the improved calculation of $\Delta \gamma_{eff}$ -
i.e., $\Delta \gamma_{eff\;corr}$ - we should (for say the Wilson action
in the case where we have a phase confined solely w.r.t $\bz_2$)
calculate as follows:
$ \Delta\gamma_{eff\;corr}= \frac{\beta}{4}\left(
\langle \sigma \rangle_{Coul}\left\langle\frac{\cos\frac{\hat{\theta}}{2}}
{\cos\hat{\theta}}\right\rangle_{Coul}-
\langle \sigma \rangle_{conf}\left\langle\frac{\cos\frac{\hat{\theta}}{2}}
{\cos\hat{\theta}}\right\rangle_{conf}\right)$
but because
$\langle \sigma \rangle_{conf} << \langle \sigma \rangle_{Coul}$ we have
$\Delta\gamma_{eff\;corr}
\approx
\frac{\beta}{4}\left(
\langle \sigma \rangle_{Coul}\left\langle\frac{\cos\frac{\hat{\theta}}{2}}
{\cos\hat{\theta}}\right\rangle_{Coul}\right).$
We calculate $\Delta \gamma_{eff\;corr}$ iteratively inasmuch as the latter
is needed to get $\Delta W$ which is needed to get $\langle \cos\hat{\theta}
\rangle$ which in turn is needed to calculate $\Delta \gamma_{eff\;corr}$.
The $\Delta W$ obtained iteratively using $\Delta \gamma_{eff\;corr}$ is also
used in calculating the enhancement factor in Table~\ref{tab7}.
In the case of the Villian action, the cos$\hat{\theta}$ in the denominator of
$\left\langle\frac{\cos\frac{\hat{\theta}}{2}}
{\cos\hat{\theta}}\right\rangle$ is removed. The case having a
phase confined solely w.r.t. $\bz_3$ is calculated in a way analogous to
that for $\bz_2$ for respectively
the Wilson and Villian action cases.}
\beq \!\!\!\!\!\!\!\!\!\!\!\!\!\!\!
\begin{array}{|l|l|l|l|l|l|l|l|}\hline
\multicolumn{8}{|c|}{ \mbox{single corrected $U(1)$ }}  \\ \hline
\mbox{Procedure} &\Delta\gamma_{eff,\;corr} & \alpha_{cont.} & 1/\alpha_{cont.}
 &\mc{2}{c|}{\mbox{enh. factor Haar}}&
\mc{2}{c|}{\mbox{prediction $1/\alpha_{Pl. scale}$}} \\
\cline{5-8}  & & & & \tau=0.79 & \tau=1 & \tau=0.79 & \tau=1 \\ \hline
\mbox{Wilson action (using (\ref{contfor}))}&&&&&&& \\ \hline
\mbox{$\bz_2$ only with $\Delta\gamma_{corr``1"}$ in (\ref{jump3})} &
0.060_0 & 0.122_0 & 8.19_6 & 6.67_7 & 6.53_5 & 54.7 & 53.6 \\ \hline
\mbox{$\bz_2+\bz_3$ with $\Delta\gamma_{corr``1"}$ in (\ref{jump3})} &
0.109_4 & 0.103_1 & 9.69_7 & 6.82_6 & 6.65_3 & 66.2 & 64.5 \\ \hline
\mbox{$\frac{1}{2}(\bz_2+\bz_3)$ with $\Delta\gamma_{corr``1"}$ in (\ref{jump3}
)}&0.0561_5 & 0.123_9 & 8.07_2 & 6.66_2 & 6.52_3 & 53.8\;\;\bullet & 52.7 \\
\hline
\mbox{$\frac{1}{2}\bz_2+\bz_3$ with $\Delta\gamma_{corr``1"}$ in (\ref{jump3})}
&0.0810_8 & 0.112_9 & 8.85_4 & 6.74_8 & 6.59_1 & 59.7\;\;\bullet & 58.4 \\
\hline \hline
\mbox{Villian  action (using (\ref{vil}))}&&&&&&& \\ \hline
\mbox{$\bz_2$ only} &0.0431_8 &0.121_7&8.21_9 & 6.44_1 & 6.34_8 &52.9 & 52.2
\\
\hline
\mbox{$\bz_2+\bz_3$} &0.0811_9&0.0942_4&10.6_1&6.52_9&6.41_8&69.3&68.1 \\
\hline
\mbox{$\frac{1}{2}(\bz_2+\bz_3)$}&0.0404_4&0.124_1 &8.05_5&6.43_2&6.34_1&
51.8\;\;\bullet&51.1 \\ \hline
\mbox{$\frac{1}{2}\bz_2+\bz_3$}
&0.0594_1&0.108_7&9.20_4&6.48_3&6.38_2&59.7\;\;\bullet&58.7
\\ \hline
\end{array}
\eeq
\end{table}
\end{footnotesize}

\subsection{Results: Comparison of MPCP predictions with experimental
values of fine-structure constants}

Our Planck scale predictions for the gauge coupling constants come about as the
product of the appropriate enhancement factor in going from the multiple
point of $SMG^3$ to the diagonal subgroup {\em and} the continuum value of
the lattice critical coupling.

In the case of the non-Abelian gauge couplings, the enhancement factor is
just $N_{gen}=3$  whereas for the $U(1)$ coupling the enhancement factor
is more than twice as large as in the non-Abelian case.
Had the phase transition at the multiple point been purely second order,
we would expect an enhancement factor of $\frac{1}{2}N_{gen}(N_{gen}-1)=6$
(instead of $N_{gen}=3$ as in the non-Abelian case) due to interaction terms
of the type $F_{\mu\nu\;Peter}F^{\mu\nu}_{Paul}$ where the indices
$Peter, Paul,\cdots$ label the various $SMG$ factors of $SMG^{N_{gen}}$
(of which there are $N_{gen}=3$). However, the fact that transitions between
phases solely confined w.r.t. discrete subgroups and the totally
Coulomb-like phase inherit a residual first-orderness of the pure discrete
subgroup transitions leads to an enhancement factor larger than
$\frac{1}{2}N_{gen}(N_{gen}-1)=6$.
The enhancement factor
for $U(1)$ is calculated
using different approximations the result of which are tabulated in
Table~\ref{tab5}

The values we have calculated for the $U(1)$
gauge coupling (i.e., the values for the diagonal subgroup of $SMG^3$ at
the multiple point of $SMG^3$) and the values calculated for
the non-Abelian couplings
are predicted to coincide with experimental values that have been extrapolated
to the Planck scale using the assumption of a minimal standard model.
In the renormalization group extrapolation procedure\cite{amaldi} used,
we accordingly
assume a desert with just a single Higgs ($N_{Higgs}=1$). The number
of generations (families) is of course taken to be 3.

In doing the
renormalization group extrapolation of experimental values to Planck
scale, we start the running
at the scale of $M_Z=91.176\pm 0.023$ using
values from LEP experiments\cite{kim}. We also extrapolate the other way:
we extrapolate our Planck scale predictions down to the scale of $M_Z$
so as these can be directly compared with experimental values of
fine-structure constants. Predicted and experimental\cite{amaldi} values
of the three
fine-structure constants are compared at both the Planck scale and the scale
of $M_Z$ are compared in Table~\ref{tab8}. We have included predicted
values obtained using several different variations in some details of our
model. For the non-Abelian fine-structure constants, the naive continuum
limit and the continuum-corrected continuum limit values are taken
from our earlier work\cite{nonabel}.

\begin{table}
\caption[Table 8]{\label{tab8} Our predictions using slightly different
calculational methods (approximations) and assumptions; these are
compared with experimental  values  (Delphi
results) extrapolated using the renormalization group  to  the  Planck
scale. The minimal Standard Model has been assumed in doing the
extrapolation. The predicted values for $U(1)$ in the last eight rows
are taken from Table~\ref{tab7} (with $\tau=0.79$).}

\vspace{.4cm}

\begin{tabular}{||c|c|c||}
\hline\hline {\bf SU(3)} & $\alpha^{-1}(\mu_{Pl.})$  & $\alpha^{-1}(M_Z)$   \\
\hline
{\bf Experimental values} & {\bf 53.6} & {\bf 9.25}$\pm 0.43$ \\ \hline
Continuum corrected continuum limit& $56.7\;\bullet$  &  12.$_8 \;\bullet$
      \\ \hline
Monopole correction & $56\pm 6\;\bullet$    & $12._1\pm 6 \;\bullet$ \\
\hline
Naive continuum limit & $80._1$    & $36._2$    \\ \hline \hline
{\bf SU(2)} & $\alpha^{-1}(\mu_{Pl.})$  & $\alpha^{-1}(M_Z)$   \\ \hline
{\bf Experimental values} & {\bf 49.2} & {\bf 30.10}$\pm 0.23$ \\ \hline
Continuum corrected continuum limit & $49.5\;\bullet$  &  29.$_8\; \bullet$
       \\ \hline
Monopole correction     & $48._3\pm 6\;\bullet$  & $28._5\pm 6\; \bullet $   \\
  \hline
Naive continuum limit & $65._1$   &  $45._3$     \\ \hline \hline
{\bf U(1)} & $\alpha^{-1}(\mu_{Pl.})$  & $\alpha^{-1}(M_Z)$   \\
& {\tiny ($SU(5)$ norm. in parenthesis)} & {\tiny ($SU(5)$ normalisation in
parenthesis)} \\ \hline
{\bf Experimental values}: & {\bf 54.}$_8$ (32.9) & {\bf 98.70}$\pm 0.21$
($59.22\pm 0.13$) \\ \hline
Continuum corrected continuum limit & 66 ($39._6$):  &  109.1 (65.5)       \\
\hline
Naive continuum limit (w. enh. 6.8): & 84.6(50.8)    &127.7    (76.6)  \\
\hline
Independent monopole approx.  &     30 (18) & 73 (44)   \\ \hline
$\bz_2$ (Wilson action): & 54.7 (32.8) & 97.8 (58.7)\\ \hline
$\bz_2$ (Villian action): & 52.9 (31.7) & 96.0 (57.6) \\ \hline
$\bz_2+\bz_3$ (Wilson action): & 66.2 (39.7) & 109.3 (65.6)\\ \hline
$\bz_2+\bz_3$ (Villian action): & 69.3 (41.6) & 112.4 (67.5)\\ \hline
$\frac{1}{2}(\bz_2+\bz_3)$ (Wilson action): & 53.8 (32.3)$\;\; \bullet$ & 96.9
(58.2)$\;\;\bullet$ \\ \hline
$\frac{1}{2}(\bz_2+\bz_3)$ (Villian action): & 51.8 (31.1)$\;\;\bullet$ & 94.9
(57.0)$\;\;\bullet$\\ \hline
$\frac{1}{2}\bz_2+\bz_3$ (Wilson action):& 59.7 (35.8)$\;\;\bullet$ & 102.8
(61.7)$\;\;\bullet$ \\ \hline
$\frac{1}{2}\bz_2+\bz_3$ (Villian action): & 59.7 (35.8)$\;\;\bullet$ & 102.8
(61.7)$\;\;\bullet$ \\ \hline
\end{tabular}
\end{table}

\section{Conclusion}\label{conclusion}

We use the principle of multiple point criticality to calculate the values
of the three standard model gauge couplings. These agree with experiment
to well within the calculational
accuracy of 5 to 10\%. In the context used here, the principle states that
Nature seeks out the action parameter values in the phase diagram of a lattice
gauge theory that correspond to the multiple point. At this point, a maximum
number of phases convene.
The gauge group is taken as
the $N_{gen}$-fold Cartesian product of the standard model group:
$SMG^{N_{gen}}$ where $N_{gen}=3$ is the number of fermion generations. So
there is a $SMG$ factor for each family of quarks and leptons.
This gauge group is referred to as the
\underline{A}nti \underline{G}rand
\underline{U}nified \underline{T}heory $(AGUT)$ gauge group.
 At the Planck scale,
the gauge couplings are predicted to have the multiple point values
corresponding to
the diagonal subgroup of $SMG^{N_{gen}}$. The diagonal subgroup, which is
isomorphic to the usual standard model group, arises as that surviving the
Planck scale breakdown of the more fundamental $SMG^{N_{gen}}$ under
automorphic symmetry operations.

In order to provoke the many phase that should convene at the multiple
point -
including those corresponding to confinement solely of discrete
subgroups of the gauge group - we need a rather general action
the parameters of which span a multidimensional phase-diagram space.
In many cases, such phases
would be called lattice artifacts because the boundary between such
lattice-scale phases disappears in going to long wavelengths and what
is distinguishable as a Coulomb-like phase at lattice scales becomes
indistinguishable from a confining phase at large distances. Such
phases are usually regarded as not being of physical significance
because they depend on the presence of a lattice which has been
introduced only as a calculational regulator that must leave no
trace of its presence upon taking a continuum limit.

Our point of view is that a Planck scale lattice is one way of implementing
the fundamental necessity of having a truly existing regulator at roughly
the Planck scale. We would claim that field theories are intrinsicly
inconsistent without the assumption of a fundamental regulator. While the
lattice seems to play a fundamental role in our model, it is really only
a way of manifesting the necessity of a fundamental regulator. We would
of course hope that critical behaviour for any field theory
formulated using other regulators (e.g., strings) would lead to approximately
the same critical values for the coupling constants so that $MPC$ predictions
based on the assumption that Nature had chosen a different regulator would not
yield very different values of coupling than those obtained by using a lattice
regulator. Obtaining the same values of couplings when using different
regulators would suggest that the principle of multiple point
criticality has a validity that transcends the particulars of the regulator.

Our claim is then that even the presence of phases that are only
distinguishable on a Planck scale lattice can have profound consequences
for physics. And this is so despite the fact that such phases can
- even though quantitatively distinguishable at the lattice
scale (e.g., two phases with different finite correlation lengths) -
become qualitatively indistinguishable at long distances.
This situation is not unfamiliar in other situations. For
example, at the triple point of water, three different phases can be accessed
by suitable changes in intensive parameters by just a small amount. However,
two of the three phases are not qualitatively distinct: at the tri-critical
point, the distinction between liquid and vapour disappears. This however does
not change the fact that all three phases are important in defining the
triple point values of temperature and pressure.

The new result in the present paper is that we calculate the $U(1)$ gauge
coupling and thereby now have  a prediction for all three gauge couplings
inasmuch as we have calculated the non-Abelian couplings in the earlier work.

The main difference between the Abelian and non-Abelian case is that
the diagonal subgroup couplings squared for $U(1)$ are  a factor $N_{gen}+
\left (\begin{array}{c} N_{gen}\\ 2 \end{array} \right )=(N_{gen}+1)
N_{gen}/2=6$
weaker than the critical values from
Monte Carlo data rather than the naively expected weakening factor $N_{gen}=3$
that is found for the non-Abelian couplings in going to the diagonal
subgroup.
The reason for the difference in the weakening factor in going to the
diagonal subgroup of $SMG^3$  is that in the case of $U(1)$ there is the
possibility of interaction terms $F_{\mu\nu\; Peter}F^{\mu\nu}_{Paul}$ in
the Lagrangian.

In trying to estimate the uncertainty in our calculation of the
$U(1)$ gauge coupling, two points of
view can be taken:

a) we could take the viewpoint that we do not really know which of the phases
characterised by being solely
confined w.r.t. discrete subgroups should also convene at the multiple point
in certain  cases. In particular, we could claim that we do not
know to what extent that $\bz_2-$ and $\bz_3$-like subgroups, in analogy
to the U(1)-continuum,
give rise to a hexagonal phase system at the multiple point. If this is the
case, we
have to let our lack of knowledge about such details of the phase diagram (and
the multiple point chosen by Nature) be included in the uncertainty in our
prediction.

b) we could take the standpoint that our choice of procedure for
including the effects of having solely confining
$\bz_2-$ and $\bz_3$-like subgroups at the multiple point is correct
and that we accordingly can do our calculations based on a correct
picture of the pattern of phases that convene
at the multiple point,
also w.r.t. solely confining  discrete
subgroups. In this case,  uncertainties in our results are assumed to be
due only to uncertainties in the Monte Carlo procedures used and in the
approximations we
use in our corrections of Monte Carlo data in order to get our
predictions.

In the case a) we must regard the differences in predictions arising when
$\bz_2-$ and $\bz_3$-like subgroups are taken into account
in different ways as being a measure of the uncertainty. For the predicted
$U(1)$
coupling at the Planck scale, this viewpoint leads to an estimated uncertainty
of about 5\%. We implement this point of view in Table~\ref{tab9} by averaging
all combinations in which there is a $\frac{1}{2}\bz_2$ contribution. This
results in an average of the combinations having $\bz_3$ and those having
$\frac{1}{2}\bz_3$ as the contributions from $\bz_3$. This reflects our
lack of certainty as to how the $\bz_3$ contribution should be treated.

In addition to this uncertainty, there will of course be the uncertainties
in the Monte Carlo results which we have used which
may be taken as 5\%. Also, our corrections are presumably not performed to
better than some 4\%, so it is unlikely that the
uncertainty in our prediction in case b) is less than 6.4\%.
In case a) we should rather
take the  uncertainty as being 8\%. These percent-wise uncertainties
concern the squared couplings  referred to the Planck scale. These correspond
to absolute Planck scale uncertainties of $4._5$ and $ 3._5$ in the inverse
fine-structure constant in
respectively the cases a) and b).
But since the renormalization group correction
consists basically of adding a rather well-determined constant to the inverse
fine-structure constants, the absolute uncertainty
in the $1/\alpha$'s is the same at all scales.

It is remarkable that in spite of these uncertainties being rather modest
we have agreement with experiment within them!

It is interesting to formulate our predictions as a number that can be
compared with the famous $\alpha^{-1}=137.036\dots$. From Table~\ref{tab9}
we deduce that the phenomenologically observed value of
$\alpha^{-1}$ decreases by $8.2\pm 0.5$ in going from low
energies to that of $M_Z$:

\beq 137.036-(\alpha_1^{-1}(M_Z)+\alpha_2^{-1}(M_Z))=
137.036-(98.70\pm 0.23 + 30.10\pm 0.23)=8.2\pm 0.3
 \eeq

\nin Our theoretical prediction for the famous $\alpha^{-1}=137.036$
is in the case a)

\beq \alpha^{-1}_1(M_Z)\;+\;\alpha^{-1}_2(M_Z)\;+\;8.2\pm 0.5=   \eeq

\[ = 99.4\pm 5 \;+\;  29.2 \pm 6 \pm 3.5\; +\; 8.2\pm 0.3= 136._8\pm 9 \]

\nin and in the case b)

\beq \alpha^{-1}_1(M_Z)\;+\;\alpha^{-1}_2(M_Z)\;+\;8.2\pm 0.5= \eeq
\[ =102.8\pm 3.5\;+\;29.2\pm 6 \pm 3.5\; +\; 8.2\pm 0.3= 140._2 \pm 8. \]

\begin{table}
\caption[Table9]{\label{tab9}The predicted values of
$\alpha^{-1}(M_Z)$ for $SU(3)$ and $SU(2)$,
are obtained as the average of several calculational procedures. The first
set of uncertainties comes from Monte Carlo data and from
the approximation procedure that we used to get our predictions from the
Monte Carlo critical couplings. The second set of uncertainties are the RMS
deviations from the average value of $\alpha^{-1}(M_Z)$ using the several
different calculational procedures. The predicted $\alpha^{-1}(M_Z)$ values
for $U(1)$ and uncertainties arise as the result of the implementing
the viewpoints a) and b) elaborated upon immediately above.}

\beq
\begin{array}{|l|l|l|} \hline
  & \alpha^{-1}(M_Z) & \alpha^{-1}(M_Z)\\
  & predicted        & experimental     \\ \hline
SU(3) & 12.4\pm 6 \pm 6 & 9.25\pm 0.43     \\ \hline
SU(2) & 29.2 \pm 6 \pm 3.5 & 30.10\pm 0.23  \\ \hline
U(1)&\begin{array}{cc} a) & 99.4 \pm 5 \\ b) & 102.8 \pm 3._5 \end{array}
& 98.70\pm 0.23       \\ \hline
\end{array}
\eeq
\end{table}

Since $\alpha_s^{-1}$ is rather small at experimental scales,
the absolute uncertainty  is percent-wise large at these scales.
But really
it is probably best to see our $\alpha_s$-prediction (at Planck scale)
as a prediction of the logarithm of the ratio
of the strong interaction scale to the Planck scale which then
allows only a crude prediction of $\alpha_s(M_Z)$.
Note that the strong scale to Planck scale ratio
is actually one of Dirac's surprising $10^{20}$ factors! So this
``large number'' is found here as an exponential of an order one number
that is proportional to the number of generations ($\pi^2$ in
the denominator of the $\beta$-functions leads to couplings that walk
slowly with scale).

Assuming the coexistence of more than one phase separated by
transitions that are first order
is roughly equivalent to assuming
the principle of multiple point criticality. 
This principle offers the hope of a general explanation for the occurrence
of fine-tuned intensive quantities in
Nature. Indeed, the conspicuous values taken by a number of physical constants
 - e.g., the vanishing effective
cosmological constant, the fine-structure constants, $\Theta_{QCD}$ -
have values that coincide
with values obtained if it is assumed that Nature seeks out
multiple point values for intensive parameters\footnote{The
smallness of the Higgs mass relative to (say) the Planck scale is also
a conspicuous quantity
that could have been expected to be explainable as a multiple point value.
It is interesting that recent work\cite{cdfhbn} indicates that the high
value of
the top quark mass
precludes an explanation of the lightness of the  Weinberg-Salam Higgs as
a multiple point. However, the assumption that Nature has multiple point(s)
 together with the requirement that the phase
transition
between degenerate phases at the multiple point is {\em maximally} first order
leads to strikingly impressive predictions for the mass of the top
quark and the expected mass of the Weinberg-Salam Higgs.}.

As mentioned above, multiple point values of intensive parameters
occur in the presence of coexisting phases separated by first order
transitions.
Such coexistence could be enforced by having fixed but not fine-tuned amounts
of extensive quantities. We have shown in recent work\cite{nl1,nl2} that the
enforced coexistence of extensive quantities in spacetime is tantamount to
having long range nonlocal interactions of a special type: namely
interactions that are identical between fields at all spacetime points
regardless
of the spacetime distance between them. Such omnipresent nonlocal interactions,
which can be described by a very general form of a reparameterization invariant
action, would not be perceived as non-locality but rather most likely
absorbed into physical constants. Even still, the presence of nonlocal
interactions opens the possibility for having contradictions of a type
reminiscent of the ``grandfather paradox'' naively encountered in ``time
machines''. However, we can show\cite{bomb} that generically there is a
``compromise''
that averts paradoxes. It is interesting that this solution coincides with
multiple point values of intensive quantities such as fine-structure
constants and the cosmological constant.
Hence one can speculate that it is a mild form of non-locality,
intrinsic to fundamental physics, that is the underlying explanation
of Nature's affinity for the multiple point.

\section{Acknowledgements}

We would like to express a special thanks to L. Laperashvili and
Kasper Olsen for useful and stimulating discussions.
We are very grateful for useful and stimulating interactions with
Colin Froggatt and Christian Surlykke. Sven Erik Rugh and Karen
Ter-Martirosyan have faithfully provided us with healthy skepticism.
Also many thanks to L. Laperashvili for providing us with many useful
references. Financial support from EEC Fund CHRX-CT94-0621 and
INTAS Grant 93-3316 is gratefully acknowledged.

\bibliography{rpred}
\end{document}